\documentclass[sigconf]{acmart}

\usepackage{color,soul}
\usepackage{booktabs}
\usepackage{multirow}
\usepackage{threeparttable, tablefootnote}

\usepackage{algorithm}
\usepackage{listings}
\usepackage[noend]{algpseudocode}
\usepackage{sansmath}
\usepackage{xurl}
\usepackage{balance}
\usepackage{longtable}
\usepackage{makecell}
\usepackage{pseudo}
\usepackage[listings]{tcolorbox}
\tcbuselibrary{listings,theorems}

\definecolor{arsenic}{rgb}{0.33, 0.33, 0.33}
\newtcbtheorem[number within=section]{mytheo}{}{colback=green!5,colframe=green!35!black,fonttitle=\bfseries}{th}
\newtcbtheorem[]{procedure}{}{colback=white,colframe=arsenic}{th}

\definecolor{gainsboro}{rgb}{0.86, 0.86, 0.86}

\usepackage{pgf-umlsd}

\usepackage{listings}
\definecolor{mGreen}{rgb}{0,0.6,0}
\definecolor{mGray}{rgb}{0.5,0.5,0.5}
\definecolor{mPurple}{rgb}{0.58,0,0.82}
\definecolor{backgroundColour}{rgb}{0.95,0.95,0.92}
\lstdefinestyle{CStyle}{
	backgroundcolor=\color{backgroundColour},
	commentstyle=\color{mGreen},
	keywordstyle=\color{magenta},
	numberstyle=\tiny\color{mGray},
	stringstyle=\color{mPurple},
	basicstyle=\footnotesize,
	breakatwhitespace=false,
	breaklines=true,
	captionpos=b,
	keepspaces=true,
	numbers=left,
	numbersep=5pt,
	showspaces=false,
	showstringspaces=false,
	showtabs=false,
	tabsize=2,
	language=C
}

\usepackage{amsmath}
\usepackage{filecontents}
\usepackage{caption}
\usepackage{subcaption}

\usepackage{tikz}
\newcommand*\emptycirc[1][1ex]{No}
\newcommand*\halfcirc[1][1ex]{
	partially}
\newcommand*\fullcirc[1][1ex]{Yes}
\usetikzlibrary{matrix,shapes,arrows,positioning,chains, calc}

\newcommand{\our}{\textsc{CloneBuster}}
\newcommand{\DL}{TTP}

\newcommand{\picscalingvalue}{0.45}

\usepackage[]{graphicx}
\usepackage{subcaption}

\setcopyright{none}
\renewcommand\footnotetextcopyrightpermission[1]{}

\begin{document}
	
	\title{No Forking Way: Detecting Cloning Attacks on Intel SGX Applications}
	
	\author{Samira Briongos}
	\email{samira.briongos@neclab.eu}
	\affiliation{
		\institution{NEC Laboratories Europe}
		\country{Germany}
	}

	\author{Ghassan Karame}
	\email{ghassan@karame.org}
	\affiliation{
		\institution{Ruhr-Universit\"{a}t Bochum}
		\country{Germany}
	}

	\author{Claudio Soriente}
	\email{claudio.soriente@neclab.eu}
	\affiliation{
		\institution{NEC Laboratories Europe}
		\country{Spain}
	}
	
	\author{Annika Wilde}
	\email{annika.wilde@rub.de}
	\affiliation{
		\institution{Ruhr-Universit\"{a}t Bochum}
		\country{Germany}
	}
	
	\renewcommand{\shortauthors}{Briongos et al.}
	
	\begin{abstract}
		Forking attacks against TEEs like Intel SGX can be carried out either by rolling back the application to a previous state, or by cloning the application and by partitioning its inputs across the cloned instances. Current solutions to forking attacks require  Trusted Third Parties (TTP) that are hard to find in real-world deployments. In the absence of a TTP, many TEE applications rely on monotonic counters to mitigate forking attacks based on rollbacks; however, they have no protection mechanism against forking attack based on cloning. In this paper, we analyze 72 SGX applications and show that approximately 20\% of those are vulnerable to forking attacks based on cloning---including those that rely on monotonic counters.
		
		To address this problem, we present \our, the first practical clone-detection mechanism for Intel SGX that does not rely on a TTP and, as such, can be used directly to protect existing applications. \our{} allows enclaves to (self-) detect whether another enclave with the same binary is running on the same platform. To do so, \our{} relies on a cache-based covert channel for enclaves to signal their presence to (and detect the presence of) clones on the same machine. We show that \our{} is robust despite a malicious OS, only incurs a marginal impact on the application performance, and adds approximately 800 LoC to the TCB. When used in conjunction with monotonic counters, \our{} allows applications to benefit from a comprehensive protection against forking attacks.
	\end{abstract}

	\maketitle
	\pagestyle{plain}

	\section{Introduction}\label{sec:intro}
	
	Trusted Execution Environments (TEE), such as Intel SGX, enable user processes to run in isolation (i.e., in so-called enclaves) from other software on the same platform, including the OS. Intel SGX applications are, however, susceptible to so-called \emph{forking attacks}, where the adversary partitions the set of clients and provides them with different views of the system. Forking attacks may be mounted either by cloning an enclave or by rolling back its state~\cite{brandenburger17dsn}. Rollback attacks exploit the fact that the sealing functionality of Intel SGX lacks freshness guarantees. This opens the door for a malicious OS to feed a victim enclave with stale state, whenever the enclave requests to unseal its state from storage---thereby ``rolling back'' the enclave to a previous state.
	Cloning attacks leverage the fact that Intel SGX does not provide means to control the number of enclaves, with the same binary, that a malicious OS can launch on the same machine.
	
	Forking attacks against enclaves---either by rollback or by cloning---result in serious consequences in a number of applications ranging from digital payments~\cite{lind2017teechain} to password-based authentication~\cite{strackx16usenix}. For example, in a password manager application, forking attacks may allow an adversary to brute-force a password despite rate-limiting measures adopted by the application. Similarly, in a payment application, an adversary could spend the same coins in multiple payments by reverting the state of its account balance.
	
	\vspace{0.25 em}\noindent \textbf{Problem.} A comprehensive solution to thwart forking attacks requires a centralized trusted third party (TTP)~\cite{DBLP:conf/ccs/DijkRSD07} or  a distributed one~\cite{DBLP:conf/ndss/Kaptchuk0M19,brandenburger17dsn,ROTE2017,NARRATOR2022}. Unfortunately, in most real-world applications, TTPs are hard to find. Moreover, some TTP-based solutions might themselves be subject to cloning attacks during the initialization process, unless the initialization involves yet another TTP (e.g., a trusted administrator~\cite{ROTE2017} or a blockchain~\cite{NARRATOR2022}). Without TTPs, most applications can mitigate forking attacks based on rollbacks by means of hardware-based monotonic counters~\cite{strackx16usenix}. However, an application that uses monotonic counters can still be cloned---making it still susceptible to forking attacks.
	To confirm this intuition, we thoroughly analyzed the security of 72 SGX-based proposals listed in~\cite{awesome_sgx,sgx_papers} with respect to forking attacks. Our findings show that 14 of those applications (i.e., roughly 20\%) are vulnerable to forking attacks based on cloning. Among those vulnerable proposals, only 3 rely on monotonic counters to counter rollback attacks, but can still be forked by cloning.
	A notable (production-ready) application that is vulnerable to forking by cloning is BI-SGX~\cite{Bisgx}. Previous work has shown that BI-SGX is vulnerable to forking attacks based on rollbacks~\cite{jangid2021towards}; the authors of~\cite{jangid2021towards} propose to fix the vulnerability using monotonic counters.
	We show that relying on monotonic counters is not enough to prevent forking attacks and report a forking attack based on cloning against the fixed version of BI-SGX that uses monotonic counters.
	
	\vspace{0.25 em}\noindent \textbf{Research question.} \emph{Can we design an anti-cloning solution that is practical, efficient, and does not require a TTP?} To the best of our knowledge, no such solution exists at the moment.
	
	\vspace{0.25 em}\noindent \textbf{Concrete solution.} To address this question, we propose \our, the first practical clone detection mechanism for SGX enclaves that does not rely on any external party. \our{} provides enclaves with the ability to (self-) detect whether other enclaves with the same binary are running on the same platform---without relying on a TTP. More precisely, we show how to leverage cache-based covert channels as a signaling mechanism for enclaves. Intuitively, if each enclave running on a machine uses the same channel to signal its presence to (and detect the presence of) other enclaves loaded with the same binary, cloning attacks can be promptly detected. \our{} ensures robust detection of clones despite noise on the channel---due to other benign applications polluting the cache---and even \emph{when the OS is malicious}. When used in conjunction with monotonic counters, \our{} enables enclaves to benefit from a comprehensive protection against all types of forking attacks (including rollback attacks) without relying on an external trusted party. Moreover, we show that \our{} could be equally used in solutions like ROTE~\cite{ROTE2017} or NARRATOR~\cite{NARRATOR2022} to avoid the use of yet another TTP when the system is being initialized. We summarize our contributions as follows:
	
	\begin{description}
		\item[Impact of cloning on SGX applications:] We thoroughly analyze the vulnerability of 72 SGX-based applications against forking attacks (cf. Section~\ref{sec:motivation}). We show that 14 applications either do not account for any protection mechanism against forking or simply prevent forking attacks based on rollbacks by means of monotonic counters---these remain vulnerable to forking attacks based on cloning. Inspired by these findings, we discuss in details how to mount a forking attack based on cloning against such applications. We also describe and implement an attack against a production-ready open-source application.
		\item[\our:]
		We introduce a practical, novel clone-detection mechanism, dubbed \our, that does not rely on a TTP (cf. Section~\ref{sec:solution}). We analyze the security of \our{} and show that it can effectively detect clones in spite of a malicious OS (cf. Section~\ref{sec:security}).
		We also show how to extend \our{} to allow the detection of up to $n>1$ clones and therefore cater for the cases where the application logic accommodates for up to $n-1$ clones.
		Finally, we show that \our{} can be equally used to secure the initialization process of existing TTP-based forking solutions, such as ROTE~\cite{ROTE2017} and NARRATOR~\cite{NARRATOR2022}.
		\item[Prototype implementation \& evaluation] We implemented a prototype of \our{} and evaluated it under realistic workloads (cf. Section~\ref{sec:eval}). We additionally report the performance of \our{} when used to detect forking attacks on an open-source production-ready SGX application. Our evaluation results show that \our{} achieves high detection (F1 score up to 0.999), with a maximum performance penalty of 4\%; the TCB increase is only 800 LoC. The code of our prototype is available at~\cite{fullversion}.
	\end{description}

	\section{Background}
	\label{sec:background}
	
	\subsection{Intel SGX}
	Intel Software Guard Extensions (SGX) is an x86 instruction set extension that offers hardware-based isolation to trusted applications that run in so-called \emph{enclaves}~\cite{IntelArchitecture}.
	Enclave isolation leverages dedicated, hardware-protected memory called Enclave Page Cache (EPC). The OS is responsible for loading the enclave's software in the EPC while the processor keeps track of deployed enclaves and their memory pages in the Enclave Page Cache Map (EPCM). By leveraging EPCM, the hardware restricts access to EPC from processes running at higher privilege levels, including the OS or the hypervisor. In particular, the Memory Management Unit (MMU) uses the EPCM to abort any attempt to access the enclave memory that has not been issued by the enclave itself or that does not comply with the specified read/write/execute permissions.
	
	Attestation allows the platform to issue publicly verifiable statements of the software configuration of an enclave. In particular, each application enclave has two identities: one called MRENCLAVE computed as the hash of the enclave binary loaded into memory; the other identity is called MRSIGNER and identifies the enclave developer. During attestation, a designated system enclave outputs a signature over both identities to certify that the application runs in an enclave on an SGX-enabled platform.
	
	Intel SGX also allows enclaves to store encrypted data on disk. This is achieved via a ``sealing'' interface that uses hardware-managed cryptographic keys. Sealed data is encrypted and authenticated using keys that are dependent on the platform and on one of the enclave identities. Sealing data against the MRENCLAVE identity ensures that only enclaves loaded with the same binary on the same platform can unseal it; on the other hand, sealing data against the MRSIGNER identity ensures that all enclaves running on the same platform and issued by the same developer (hence, with the same MRSIGNER) can unseal it.
	
	We note that Intel SGX has been deprecated in last-generation CPUs for desktops but will still be available for server-grade platforms~\cite{intelproducts}, which fits the confidential computing in the cloud paradigm.
	Further, the size of the EPC will increase up to 1TB on multi-socket systems.
	
	\subsection{Cloning SGX Enclaves}
	
	Cloning an application (irrespective of whether it resides within an enclave) may or may not include its runtime memory. ``Live'' cloning consists of creating a copy of a running process, that includes also the runtime memory of the original process. In contrast, a ``non-live'' cloning operation creates a clone by only copying the code and the persistent state.
	
	We note that Intel SGX limits live cloning of enclaves ``by design''. In particular, EPC encrypted memory and hardware-managed EPCM prevent live cloning of enclaves: in a nutshell, an encrypted memory page assigned to a given enclave, cannot be copied and assigned to another one.
	
	With respect to non-live cloning, we note that the sealing functionality used to persist state information to disk prevents cross-platform cloning. In particular, cryptographic keys that Intel SGX uses for sealing enclave data, depend on the host where the enclave is running. Therefore, state sealed by an enclave on a given host cannot be unsealed on a different host.
	
	Nevertheless, Intel SGX does not prevent non-live cloning of an enclave on the same platform, nor does it provide a mechanism to distinguish two such clones. In particular, the number of enclaves that can be set up on a given host and executed at the same time---regardless of the loaded binary---is only limited by system resources. Thus, little prevents an adversary, that controls the OS on a given host, to launch a number of enclaves with the same binary. In case one of those enclaves seals data to disk, all other enclaves with the same binary have access to that data---since sealing keys on a given host only depend on the enclave identity. As a result, if one enclave is attested and provisioned with a secret, all clones will have access to the same secret. Intel acknowledges that there is no mechanism to distinguish enclaves loaded with the same binary on the same platform, since they all share the same identities (i.e., MRSIGNER and MRENCLAVE).\footnote{\url{https://intel.ly/3uprwdh}}
	
	\begin{table*}[!htbp]
		\centering
		\footnotesize
		\noindent\scalebox{0.99}{
			\begin{tabular}{|l|c|l|l|l|l|c|l|l|}
				
				\cline{1-4} \cline{6-9}
				\multirow{2}{*}{\textbf{Project}} & \textbf{Source code} &  \multicolumn{2}{c|}{{\textbf{Vulnerable to}}} & & \multirow{2}{*}{\textbf{Project}} & \textbf{Source code} &  \multicolumn{2}{c|}{{\textbf{Vulnerable to}}} \\ \cline{3-4} \cline{8-9}
				& \textbf{available} &  \textbf{Rollback} & \textbf{Cloning} & & & \textbf{available} & \textbf{Rollback} & \textbf{Cloning} \\
				
				\cline{1-4} \cline{6-9}
				\multicolumn{4}{|c|}{\textbf{Encrypted Databases and Key-value Stores}} 						 				 & &  \cellcolor{gainsboro}\textbf{X-Search} \cite{xsearch_paper,xsearch_code} $^{ap}$ & \cellcolor{gainsboro}Yes & \cellcolor{gainsboro}N/A & \cellcolor{gainsboro}Yes (C) \\\cline{1-4} \cline{6-9}
				\cellcolor{gainsboro}\textbf{Aria} \cite{aria} $^{p}$ & \cellcolor{gainsboro}No & \cellcolor{gainsboro}N/A & \cellcolor{gainsboro}Yes (A)    									 				 & & \multicolumn{4}{c|}{\textbf{Blockchains}} \\ \cline{6-9}
				\cellcolor{gainsboro}\textbf{Avocado} \cite{avocado_paper,avocado_code} $^{a}$ & \cellcolor{gainsboro}Yes & \cellcolor{gainsboro}N/A & \cellcolor{gainsboro}Yes (A)    			 				 & & \textbf{BITE} \cite{bite} $^{p}$ & No  & No (MC) & N/A \\
				\cellcolor{gainsboro}\textbf{Enclage} \cite{enclage} $^{p}$ & \cellcolor{gainsboro}No & \cellcolor{gainsboro}N/A & \cellcolor{gainsboro}Yes (A)    								 				 & & \textbf{BLOXY} \cite{bloxy} $^{p}$ & No  & N/A & N/A \\
				\cellcolor{gainsboro}\textbf{EnclaveCache} \cite{enclavecache} $^{p}$ & \cellcolor{gainsboro}No & \cellcolor{gainsboro}Yes & \cellcolor{gainsboro}Yes (B)    					 				 & & \textbf{Ekiden} \cite{ekiden_paper,ekiden_code} $^{a}$ & Yes  & No (\DL) & No (\DL) \\
				\textbf{EnclaveDB} \cite{EnclaveDB} $^{p}$ & No  & No (MC) & No (\DL)    						 				 & & \textbf{Hybrids on Steroids} \cite{troxy} $^{p}$ & No  & No (MC+\DL) & No (\DL) \\
				\textbf{HardIDX} \cite{hardidx} $^{p}$ & No  & Yes & N/A    									 				 & & \textbf{MobileCoin} \cite{mobilecoin_paper,mobilecoin_code} $^{a}$ & Yes  & No (\DL)& No (\DL) \\
				\cellcolor{gainsboro}\textbf{NeXUS} \cite{nexus_paper,nexus_code} $^{p}$ & \cellcolor{gainsboro}Yes & \cellcolor{gainsboro}No (MC) & \cellcolor{gainsboro}Yes (B)    			 				 & & \textbf{Oasis} \cite{oasis} $^{a}$ & Yes  & No (\DL)& No (\DL)\\
				\cellcolor{gainsboro}\textbf{ObliDB} \cite{oblidb_paper,oblidb_code} $^{p}$ & \cellcolor{gainsboro}Yes & \cellcolor{gainsboro}N/A & \cellcolor{gainsboro}Yes (A)    				 				 & & \textbf{Obscuro} \cite{obscuro_paper,obscuro_code} $^{a}$ & Yes  & N/A & N/A \\
				\textbf{PESOS} \cite{pesos} $^{p}$ & No  & N/A & N/A    										 				 & & \textbf{Phala Network} \cite{phala_paper,phala_code} $^{a}$ & Yes  & No (\DL) & No (\DL) \\
				\textbf{SeGShare} \cite{segshare} $^{p}$ & No  & No (MC) & N/A    								 				 & & \textbf{Private Chaincode} \cite{fabric_paper,fabric_code} $^{a}$ & Yes  & N/A & N/A \\
				\cellcolor{gainsboro}\textbf{ShieldStore} \cite{shieldstore_paper,shieldstore_code} $^{ap}$ & \cellcolor{gainsboro}Yes  & \cellcolor{gainsboro}No (MC) & \cellcolor{gainsboro}Yes (B)  			 & & \textbf{Private Data Objects} \cite{pdo_paper,pdo_code} $^{a}$ & Yes  & No (\DL) & No (\DL) \\
				\textbf{SPEICHER} \cite{speicher_paper,speicher_code} $^{a}$ & Yes  & No (MC) & No (\DL)    	 				 & & \textbf{Proof of Luck} \cite{pol_paper,pol_code} $^{a}$ & Yes  & N/A & No ($\star$) \\
				\cellcolor{gainsboro}\textbf{STANlite} \cite{stanlite_paper,stanlite_code} $^{ap}$ & \cellcolor{gainsboro}Yes  & \cellcolor{gainsboro}N/A & \cellcolor{gainsboro}Yes (A)    		 				 & & \textbf{Teechain} \cite{teechain_paper,teechain_code} $^{ap}$ & Yes  & No (MC) & N/A \\
				\cellcolor{gainsboro}\textbf{StealthDB} \cite{stealthdb_paper,stealthdb_code} $^{a}$ & \cellcolor{gainsboro}Yes  & \cellcolor{gainsboro}Yes & \cellcolor{gainsboro}Yes (B)    	 				 & & \textbf{Town Crier} \cite{towncrier_paper,towncrier_code} $^{a}$ & Yes  & No (\DL) & No (\DL) \\
				
				\cline{1-4}
				\multicolumn{4}{|c|}{\textbf{Applications}}														 & & \textbf{Troxy} \cite{troxy} $^{p}$ & No  & N/A & No (\DL) \\ \cline{1-4}
				\cellcolor{gainsboro}\textbf{BI-SGX} \cite{bisgx_code} $^{a}$ & \cellcolor{gainsboro}Yes  & \cellcolor{gainsboro}No (MC) & \cellcolor{gainsboro}Yes (B) 							 & & \textbf{Twilight} \cite{twilight_paper,twilight_code} $^{a}$ & Yes  & N/A & N/A \\ \cline{6-9}
				\cellcolor{gainsboro}\textbf{CACIC} \cite{cacic_paper,cacic_code} $^{a}$ & \cellcolor{gainsboro}Yes  & \cellcolor{gainsboro}Yes & \cellcolor{gainsboro}Yes (B) 					 & & \multicolumn{4}{c|}{\textbf{Machine Learning}} \\ \cline{6-9}
				\textbf{DEBE} \cite{debe_paper, debe_code} $^{a}$ & Yes  & N/A & N/A 						 	 & & \textbf{Confidential ML} \cite{confidential_ml} $^{a}$ & Yes  & N/A & N/A\\
				\textbf{HySec-Flow} \cite{hysec_paper,hysec_code} $^{a}$ & Yes  & N/A & N/A 					 & & \textbf{DP-GBDT} \cite{dp_gbdt} $^{a}$ & Yes  & No (MC) & N/A \\
				\cellcolor{gainsboro}\textbf{PrivaTube} \cite{privatube} $^{p}$ & \cellcolor{gainsboro}No  & \cellcolor{gainsboro}N/A & \cellcolor{gainsboro}Yes (C) 								 & & \textbf{Plinius} \cite{plinius_paper,plinius_code} $^{a}$ & Yes  & No (MC) & N/A \\
				\textbf{REX} \cite{rex_paper, rex_code} $^{a}$ & Yes  & N/A & N/A 								 & & \textbf{secureTF} \cite{securetf} $^{p}$ & No  & N/A & N/A \\
				\textbf{SGXDedup} \cite{dedup_paper,dedup_code} $^{a}$ & Yes  & N/A & N/A 						 & & \textbf{Secure XGBoost} \cite{xgboost_paper,xgboost_code} $^{a}$ & Yes & N/A & N/A \\
				\textbf{Signal CDS} \cite{signal_blog,signal_code} $^{a}$ & Yes  & N/A & N/A 					 & & \textbf{Slalom} \cite{slalom_paper,slalom_code} $^{ap}$ & Yes  & N/A & N/A \\
				\textbf{SkSES} \cite{skses_paper,skses_code} $^{a}$ & No & N/A & N/A 							 & & \textbf{SOTER} \cite{soter_paper,soter_code} $^{a}$ & Yes  & N/A & N/A \\ \cline{6-9}
				\textbf{SMac} \cite{smac_gen_paper,smac_gen_code} $^{a}$ & Yes  & N/A & N/A 					 & & \multicolumn{4}{c|}{\textbf{Network}} \\ \cline{6-9}
				\textbf{TresorSGX} \cite{tresorsgx_paper,tresorsgx_code} $^{a}$ & Yes  & N/A & N/A 				 & & \textbf{ConsenSGX} \cite{consensgx_paper,consensgx_code} $^{a}$ & No  & N/A & N/A \\
				
				\cline{1-4}
				\multicolumn{4}{|c|}{\textbf{Key + Password Management}} 										 & & \textbf{CYCLOSA} \cite{cyclosa} $^{p}$ & No  & N/A & N/A \\ \cline{1-4}
				\textbf{DelegaTEE} \cite{delegatee} $^{p}$ & No  & No (MC) & N/A 								 & & \textbf{ENDBOX} \cite{endbox} $^{ap}$ & No  & N/A & N/A \\
				\textbf{FeIDo} \cite{feido_paper,feido_code} $^{a}$ & Yes  & N/A & N/A 							 & & \textbf{LightBox} \cite{lightbox_paper,lightbox_code} $^{ap}$ & Yes  & N/A & N/A \\
				\textbf{Keys in Clouds} \cite{keyscloud_paper,keyscloud_code} $^{a}$ & Yes  & No (MC) & N/A		 & & \textbf{SENG} \cite{seng_paper,seng_code} $^{a}$ & Yes  & N/A & N/A \\
				\textbf{SafeKeeper} \cite{safekeeper_paper,safekeeper_code} $^{a}$ & Yes  & No (MC) & N/A 		 & & \textbf{SGX CBR} \cite{sgxcbr} $^{p}$ & No  & N/A & N/A \\
				\cellcolor{gainsboro}\textbf{SGX-KMS} \cite{sgxkms_paper,sgxkms_code} $^{a}$ & \cellcolor{gainsboro}Yes  & \cellcolor{gainsboro}Yes & \cellcolor{gainsboro}Yes (B) 				 & & \textbf{SGX-Tor} \cite{sgxtor_paper,sgxtor_code} $^{ap}$ & Yes  & Yes & N/A \\
				
				\cline{1-4}
				\multicolumn{4}{|c|}{\textbf{Private Search}} 													 & & \textbf{TEE V2V} \cite{v2v_paper,v2v_code} $^{a}$ & No  & N/A & N/A \\ \cline{1-4}
				\textbf{BISEN} \cite{bisen_paper,bisen_code} $^{a}$ & Yes & N/A & N/A 							 & & \textbf{MACSec} \cite{macsec} $^{a}$ & Yes  & No (MC) & N/A \\
				\textbf{DeSearch} \cite{desearch_paper,desearch_code} $^{a}$ & Yes  & N/A & N/A 				 & & \textbf{S-NFV} \cite{snfv} $^{p}$ & No  & N/A & N/A \\
				\textbf{Maiden} \cite{maiden_paper,maiden_code} $^{a}$ & Yes & N/A & N/A 						 & & \textbf{SafeBricks} \cite{safebricks_paper,safebricks_code} $^{ap}$ & Yes  & N/A & N/A \\
				\textbf{POSUP} \cite{posup_paper,posup_code} $^{a}$ & Yes & N/A & N/A 					         & & \textbf{SELIS-PubSub} \cite{pubsub_paper,pubsub_code} $^{p}$ & Yes  & N/A & N/A \\ \cline{6-9}
				\textbf{QShield} \cite{qshield_paper,qshield_code} $^{a}$ & Yes  & N/A & N/A 					 & & \multicolumn{4}{c|}{\textbf{Data Analytics}} \\ \cline{6-9}
				\textbf{Snoopy} \cite{snoopy_paper,snoopy_code} $^{a}$ & Yes  & No (MC) & N/A 					 & & \textbf{Opaque} \cite{opaque_paper,opaque_code} $^{a}$ & Yes  & No (\DL) & No (\DL) \\ 					
				\cline{1-4} \cline{6-9}
			\end{tabular}
		}
		\caption{
			Summary of our analysis of SGX applications. We analysed SGX applications listed in~\cite{sgx_papers} (superscript $p$ next to the citation) and listed in~\cite{awesome_sgx} (superscript $a$ next to the citation). 
			Here, we excluded libraries, runtime frameworks, and projects without documentation. We divide the remaining ones based on their type (Machine Learning, Blockchain, Encrypted Databases, ...). For each application, we report whether the code is available, whether they are vulnerable to rollback attacks, and whether they are vulnerable to cloning attacks (highlighted in gray). In case the application is not vulnerable to a specific attack, we report the countermeasure (MC is monotonic counters, \DL{} is trusted third party). We use N/A in case the attack is not applicable. In case of applications vulnerable to cloning attacks, we categorize the attack type (A, B, C) and provide more details in Appendix~\ref{sec:attacks}. Proof of Luck ($\star$) is not vulnerable to cloning because it books all MCs on the platform at startup; as a result, no clone can be started but this also means that MCs are no longer available for other applications on the same host.
			\label{tab:extended_analysis}
		}
	\end{table*}

	\section{Cloning Attacks on Intel SGX}\label{sec:motivation}
	
	\subsection{Motivation}
	
	Forking attacks against TEEs such as Intel SGX can be mounted either by rolling back the enclave to a previous state or by launching several instances of the victim enclave~\cite{brandenburger17dsn}.
	
	To illustrate how forking attacks based on cloning work, assume an enclave that is not susceptible of rollback attacks---e.g., an enclave that uses monotonic counters to seal its state. We can model the enclave as an automata $E_{ID}$, where $ID$ refers to the identity of the enclave (i.e., MRSIGNER and MRENCLAVE). Upon start, the enclave obtains the initial state $S_{0}$ from the OS and it is ready to process inputs. The enclave moves to the next state $S_j$ as a function $F$ of the current state and the current input $I_j$. For example, without malicious interference, an enclave fed with inputs $I_1$, $I_2$, and $I_3$ (in that order), moves through states $S_1=F(S_0,I_1)$, $S_2=F(S_1,I_2)$, and final state $S_3=F(S_2,I_3)$. Each time the enclave moves to a new state, it seals the new state to disk so to resume from the latest state upon reboot.
	
	To fork the application, the adversary can create two clones, say $E_{ID}$ and $E_{ID}'$, and provide both of them with initial state $S_0$. Next, the OS feeds inputs $I_1$ and $I_2$ to $E_{ID}$ and it feeds $I_3$ to $E_{ID}'$. Thus, enclave $E_{ID}$ moves to state $S_1=F(S_0,I_1)$ and final state $S_2=F(S_1,I_2)$, whereas $E_{ID}'$ move to state $S_3'=F(S_0,I_3)$. The above example implies that a successful forking attack based on cloning requires running multiple instances of the victim enclave \emph{at the same time between two state updates}.
	Running the two instances one at a time does not lead to a fork. To illustrate this, assume $E_{ID}'$ is started \emph{after} that $E_{ID}$ has processed input $I_2$ and sealed state $S_2$. Thus, upon start $E_{ID}'$ fetches the latest state $S_2$ from disk---recall that the application is not susceptible to rollbacks--- obtains input $I_3$ and moves to state $S_3=F(S_2,I_3)$.
	
	Comprehensive solutions to forking attacks rely on a centralized~\cite{DBLP:conf/ccs/DijkRSD07} or distributed TTP~\cite{DBLP:conf/ccs/DijkRSD07,DBLP:conf/ndss/Kaptchuk0M19,brandenburger17dsn,ROTE2017,NARRATOR2022}. For example, the authors of~\cite{brandenburger17dsn} show how to detect forking attacks if clients are mutually trusted---that is, clients themselves act as a distrusted TTP. Solutions like ROTE~\cite{ROTE2017} or NARRATOR~\cite{NARRATOR2022} prevent forking attacks by using a cohort of enclaves---distributed across different hosts---that offer forking prevention to (other) application enclaves. It is interesting to note that solutions like ROTE can be themselves victim of forking attacks by cloning when the cohort of enclaves is being initialized~\cite{NARRATOR2022}. Once the cohort is forked, applications enclaves that use ROTE can be forked.
	ROTE~\cite{ROTE2017} prevents forks of the cohort during initialization by means of a trusted administrator that helps initializing the cohort; NARRATOR removes the need for a centralized TTP---the administrator---by replacing it with a BFT-like blockchain, thereby using a distributed TTP.
	
	This results in the following observation: \emph{some TTP-based solution to forking like NARRATOR needs to use another TTP (i.e., the blockchain) to avoid being forked during its initialization process}. As such, existing solutions are hard to instantiate for most real-world applications. Moreover, trusted parties are hard to find in real-world deployments. Without the aid of a trusted third party, many SGX-based applications mitigate rollback attacks by using TPM's monotonic counters. However, even if rollback attacks are not feasible, an adversary can still clone the victim application in order to mount a forking attack.

	\subsection{Cloning Attacks in the Wild.} We analyzed the security of 72 SGX-based applications against rollback and cloning attacks. Selected applications were taken from curated lists of SGX papers~\cite{awesome_sgx,sgx_papers}. We analyzed the application source-code when available; otherwise we analyzed the description provided in the paper where the proposal was introduced.
	
	Our results are summarized in Table~\ref{tab:extended_analysis}.
	Based on our findings, we draw the following observations:
	\begin{itemize}
		\item Out of the 72 proposals, 14 applications (i.e., roughly 20\%) are vulnerable to forking attacks based on cloning.
		\item 11 of the vulnerable 14 applications do not account for any protection mechanism against forking attacks.
		\item 3 of the 14 vulnerable applications prevent rollback attacks with a monotonic counter; yet, they are vulnerable to forking attacks based on cloning.
		\item 7 of the 14 vulnerable applications do not seal state, and therefore are not vulnerable to rollback attacks per design; however, those applications are vulnerable to cloning.
		\item Out of the 72 proposals, 11 use a TTP to prevent forking attacks. Among these 11 proposals, 9 rely on a decentralized ledger to prevent forking (8 of those are blockchain applications). Finally, 2 applications dismiss rollback attacks by claiming that these attacks can be mitigated by ROTE~\cite{ROTE2017}.
	\end{itemize}	
	
	We categorize the 14 vulnerable applications in three different categories, namely, A, B, and C. Category A mostly consists of in-memory key-value stores (KVS); by cloning the application, the adversary can split the inputs from different clients across multiple KVS instances so that clients have different ``views'' of the store (e.g, updates made by one client to a specific key are not seen by another client). Applications in category B seal state to have it available across restarts; by cloning these applications, the adversary can obtain multiple valid states that can be fed to the enclave when it restarts. Category C mostly consists of applications that leverage an SGX enclave as a proxy to guarantee unlinkability or privacy of client requests; by cloning the application, the adversary can partition the set of clients, thereby reducing the anonymity set for each of the clients. We detail how to mount cloning attacks for each category in Appendix~\ref{sec:attacks}.

	\subsection{Case Study: Cloning attack against BI-SGX}
	\label{sec:bisgx}
	
	As a case-study, we show how to successfully mount a forking attack based on cloning against \textbf{BI-SGX}~\cite{Bisgx}\footnote{\url{https://github.com/hello31337/BI-SGX}}. We chose BI-SGX because (i) its code is open-source, (ii) it was shown to be vulnerable to forking attacks based on rollbacks and a fix based on monotonic counters was proposed~\cite{jangid2021towards}. Our attack against BI-SGX shows that even if applications use monotonic counters to mitigate forking attacks based on rollbacks, they are still vulnerable to forking attacks based on cloning.
	
	\begin{figure}[t]
		\centering
		\includegraphics[width=0.5\textwidth]{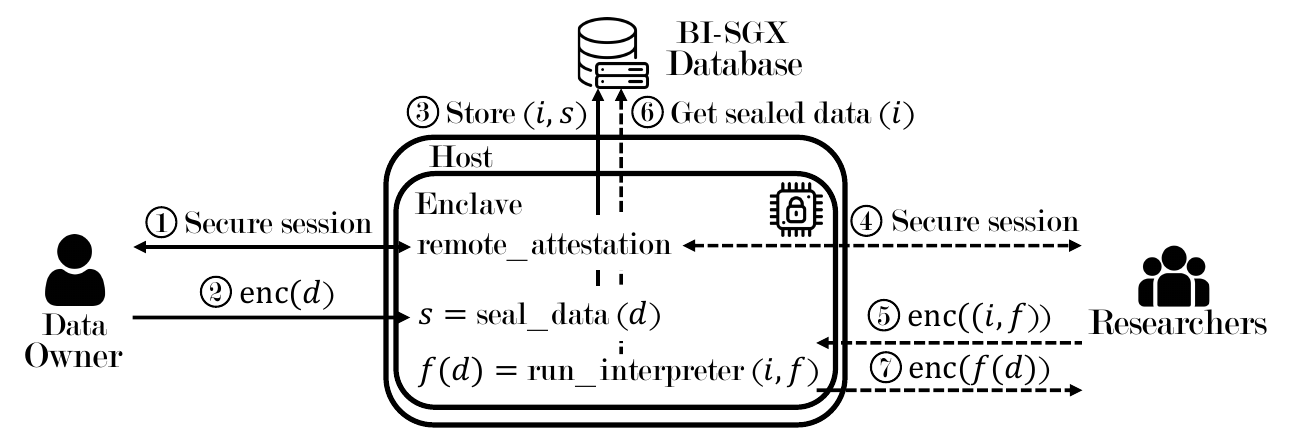}
		\vspace{-2 em}
		\caption{Overview of the BI-SGX enclave and its interactions with Data Owners and Researchers}
		\label{fig_bisgx}
		\vspace{-1.5 em}
	\end{figure}
	
	\vspace{0.25 em}\noindent\textbf{Overview of BI-SGX.} BI-SGX provides secure computation over private data in the cloud by leveraging SGX. As shown in Figure~\ref{fig_bisgx}, a \textit{data-owner} sends to the BI-SGX enclave data $\texttt{d}$ encrypted; the encryption key is agreed between the enclave and the data owner via remote attestation.
	The BI-SGX enclave decrypts the plaintext, seals it, and sends the sealed data (denoted as $\texttt{s}$) to an external database. The database stores $\texttt{s}$  along with an index $\texttt{i}$ as a tuple $\texttt{[i, s]}$.
	Later on, a \textit{researcher} can send requests to the enclave; requests include the index that is used to retrieve data from the database and a description of a function $\texttt{f}$ to be computed over the data. More precisely, a request includes a tuple $\texttt{[i,f]}$; communication is secured with keys agreed between the enclave and the researcher via remote attestation. Once the enclave receives the request, if $\texttt{[i,s]}$ exists in the database, the enclave unseals $\texttt{s}$ to recover data $\texttt{d}$ and returns $\texttt{f(d)}$. Note that the database lies outside of the enclave boundaries. Therefore, it can be under the control of a malicious OS or cloud provider.
	
	\vspace{0.5 em}\noindent\textbf{Rollback Attacks on the early version of BI-SGX.} A system like BI-SGX should offer some state continuity guarantees. More precisely, as stated by Jangid et al.,~\cite{jangid2021towards}, researcher queries containing different indexes should retrieve and process different data items or, the other way around, queries containing the same index should process the same data item.
	Jangid et al.,~\cite{jangid2021towards} used the Tamarin prover to show that BI-SGX could not guarantee such property. Namely, an attacker could feed the enclave with different data even if researchers submit requests with the same index.
	
	To understand how the attack works, we show in Figure \ref{fig_bisgx_func} the pseudocode for the two main functions manipulating the data from the data-owners and researchers perspective, i.e., \texttt{seal\_data} and \texttt{run\_interpreter}, respectively. Note that function \texttt{seal\_data} does not include the index used for data retrieval; the latter is added by the database when it receives the encrypted data for storage. It is straightforward to see how, upon request issued by the BI-SGX enclave to retrieve data item with index $\texttt{i}$, a malicious OS could return any sealed data item; the enclave has no means to tell if the sealed data returned by the OS is the right one.
	
	\begin{figure}[t]
		\begin{procedure*}[nofloat, after={}, width=.475\linewidth, equal height group=TA, size=small,left=0mm, valign=center]{seal\_data}{}
			\hspace{7pt}
			\begin{algorithmic}[1]
				\footnotesize
				\Require Encrypted data $c_{O}$
				\State $d$ = Decrypt $(c_{O})$
				\State $s$ = Seal $(d)$
				\State \textbf{OCALL\_Store $(i,s)$}
			\end{algorithmic}
		\end{procedure*}
		\hfill
		\begin{procedure*}[nofloat, before={}, width=.515\linewidth, equal height group=TA, size=small,left=0mm, top=1mm, valign=center]{run\_interpreter}{}
			\begin{algorithmic}[1]
				\footnotesize
				\Require Encrypted request $c_{R}$
				\Ensure Encrypted result $c_{RES}$
				\State $req$ = Decrypt $(c_{R})$
				\State $i,f$ = BISGX\_main $(req)$
				\State \hspace{3pt} $s$ = \textbf{OCALL\_DB\_get $(i)$}
				\State $d$ = Unseal $(s)$
				\State $c_{RES}$ = Encrypt $(f(d))$
				
			\end{algorithmic}
		\end{procedure*}
		\vspace{-1 em}
		\caption{Pseudocode of the target functions exposed by the enclave as ecalls: seal\_data and run\_interpreter}
		\label{fig_bisgx_func}
		\vspace{-1 em}
	\end{figure}
	
	\begin{figure}[t]
		\begin{procedure*}[nofloat, after={}, width=.475\linewidth, equal height group=AT, size=small,left=0mm, valign=center]{seal\_data}{}
			\hspace{7pt}
			\begin{algorithmic}[1]
				\footnotesize
				\Require Encrypted data $c_{O}$
				\State $d$ = Decrypt $(c_{O})$
				\State \textbf{\textcolor{arsenic}{Increment(MC)}}
				\State \textbf{\textcolor{arsenic}{Read(MC)}}
				\State $s$ = Seal $(d,\textbf{\textcolor{arsenic}{MC}})$
				\State \textbf{OCALL\_Store $(i,s)$}
			\end{algorithmic}
		\end{procedure*}
		\hfill
		\begin{procedure*}[nofloat, before={}, width=.515\linewidth, equal height group=AT, size=small,left=0mm, top=1mm, valign=center]{run\_interpreter}{}
			\begin{algorithmic}[1]
				\footnotesize
				\Require Encrypted request $c_{R}$
				\Ensure Encrypted result $c_{RES}$
				\State $req$ = Decrypt $(c_{R})$
				\State $i,f$ = BISGX\_main $(req)$
				\State \hspace{3pt} $s$ = \textbf{OCALL\_DB\_get $(i)$}
				\State $(d,\textbf{\textcolor{arsenic}{MC}})$ = Unseal $(s)$
				\If {$i == \textbf{\textcolor{arsenic}{MC}}$}
				\State $c_{RES}$ = Encrypt $(f(d))$
				\EndIf
			\end{algorithmic}
		\end{procedure*}
		\vspace{-1 em}
		\caption{Pseudocode of the patched functions from Figure \ref{fig_bisgx_func} using monotonic counters. Changes are highlighted in gray.}
		\label{fig_bisgx_func_pa}
		\vspace{-1.5 em}
	\end{figure}
	
	\vspace{0.5 em}\noindent\textbf{Protecting BI-SGX with Monotonic Counters.}
	The aforementioned vulnerability was reported to the developers of BI-SGX by Jangid et al.,~\cite{jangid2021towards}. The latter also proposed to use monotonic counters (MC) to mitigate this attack. The idea is to seal the index of the data along with the data itself. Hence, when the BI-SGX enclaves requests sealed data with index $\texttt{i}$ and obtains a ciphertext $\texttt{Enc(d,j)}$, it only accepts $\texttt{d}$ as valid if $\texttt{i=j}$. Further, the use of monotonic counters as indexes ensure that not two data items can be stored with the same index. We implemented the fix suggested by~\cite{jangid2021towards} as shown in Figure \ref{fig_bisgx_func_pa}. Here, we use the de-facto ``inc-then store'' mode of monotonic counters to provide security against rollback attacks.
	
\begin{figure}[t]
		\centering
		\includegraphics[width=0.48\textwidth]{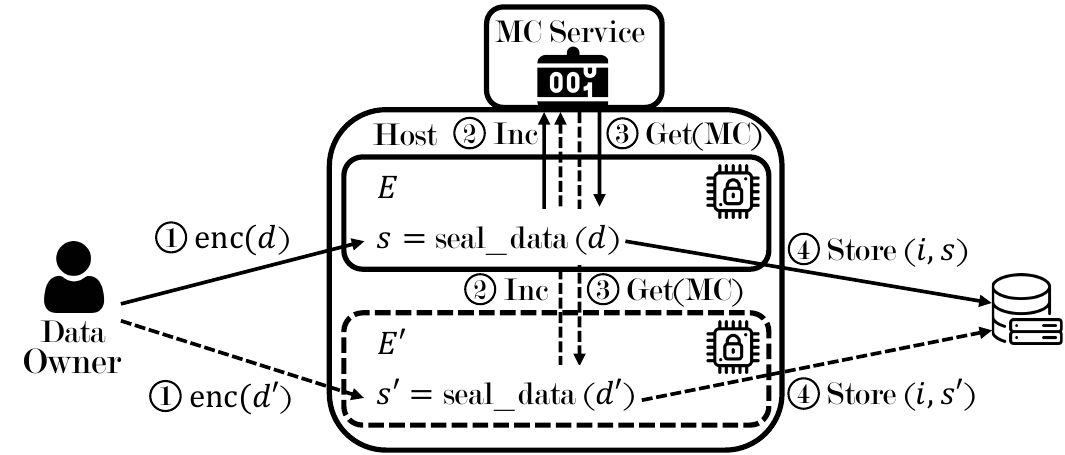}
		\vspace{-1.5 em}
		\caption{Overview of a cloning attack against the fixed version of BI-SGX that uses monotonic counters.}
		\label{fig_bisgx_att}
		\vspace{-1.5 em}
	\end{figure}

	\vspace{0.5 em}\noindent\textbf{Forking the ``fixed'' version of BI-SGX.}
	We argue that this fix is not enough to prevent forks for the BI-SGX enclave. Namely, if there are clones of the enclave running on the system, it is possible to assign the same index to multiple data items. Therefore, when the BI-SGX requests sealed data from the OS, the latter can return one of many valid data items. To carry out this attack, the attacker has to focus on the \textit{data owner} function, i.e. \texttt{seal\_data}. The process is sketched in Figure~\ref{fig_bisgx_att}. The attacker controlling the execution of two BI-SGX enclaves, $E$ and $E'$, has to make sure that both execute \textbf{\textcolor{arsenic}{Increment(MC)}} before allowing them to proceed with \textbf{\textcolor{arsenic}{Read(MC)}}. In a nutshell:
	
	\begin{enumerate}
		\item The adversary starts two BI-SGX enclave instances.
		\item The adversary feeds one data item $\texttt{d}$ to enclave $E$ and another data item $\texttt{d'}$ to enclave $E'$ (as per figure \ref{fig_bisgx_att}). The current value of the counter is MC (cf. Figure~\ref{fig_bisgx_att} stage 1).
		\item The adversary stops the instance that first executes \textbf{\textcolor{arsenic}{Increment(MC)}} until the other one has also executed it. The counter at this state is equal to MC+2. For this proof of concept, we have manually synchronized the execution of both instances, in practice an attacker could use a framework such as SGX-Step \cite{SGXStep17} (cf. Figure \ref{fig_bisgx_att} stage 2).
		\item The adversary allows both instances to proceed. They execute \textbf{\textcolor{arsenic}{Read(MC)}} and get exactly the same value of the counter (MC+2) (cf. Figure \ref{fig_bisgx_att} stage 3).
		\item Instance $E$ seals $\texttt{(d,MC+2)}$ while instance $E'$ seals $\texttt{(d',MC+2)}$. Both ciphertexts are sent to the database. Both ciphertexts are valid for a query from a \textit{researcher} to process data stored at index MC+2, as the BI-SGX enclave only checks if MC in the sealed blob is equal to the index value in the \textit{researcher} request (cf. Figure \ref{fig_bisgx_att} stage 4).
	\end{enumerate}
	
	We note that the adversary is not limited by the number of instances that can be launched at the same time.
	
	We responsibly disclosed this vulnerability to the developers of BI-SGX. They agreed to take into account attacks based on cloning for further releases of BI-SGX. In Section~\ref{sec:eval}, we show how our proposed solution, \our, can efficiently detect any enclave cloning attempts in between the execution of \textbf{\textcolor{arsenic}{Increment(MC)}} and the data sealing phase.

	\section{\our}\label{sec:solution}
	
	\subsection{System \& Threat Model}
	
	Given the observations made in Section~\ref{sec:background} and in Section~\ref{sec:motivation}, we focus on the practical problem of detecting clones on a single platform, in realistic application scenarios where the OS is malicious and the enclave has no access to a trusted third party. As shown in Table~\ref{tab:extended_analysis}, such a setting faithfully mimics most existing deployments.
	
	We consider two enclaves to be clones if (i) they have been loaded with the same binary (hence, they have the same MRSIGNER and MRENCLAVE)\footnote{This also means that each enclave can access data sealed by its clone.}, and (ii) they run at the same time. Condition (i) also implies that clones of an enclave share long-term public keys; condition (ii) is necessary for a successful forking attack as explained in the previous section.
	
	We assume the common threat model for Intel SGX where the hardware is part of the TCB, but the adversary controls privileged software (e.g., the OS) on the host. The goal of our adversary is to run multiple clones on a platform while bypassing the detection mechanism. Similar to~\cite{tsgx17,203672,dejavu17,DBLP:conf/ccs/AlderAKPS19,DBLP:journals/corr/abs-1712-08519,DBLP:conf/acsac/Lang0MLWL22,DBLP:conf/sosp/ChenTZ17,DBLP:conf/ccs/ChenLMZLCW18,DBLP:conf/cloud/TrachFOOBF20}, we consider Denial of Service (DoS) attacks to be out of scope. We note that a malicious OS can anyway DoS a process running on its platform---irrespective of the defense mechanism employed.
	
	\begin{figure}[t!]
		\centering
		\includegraphics[width=0.47\textwidth]{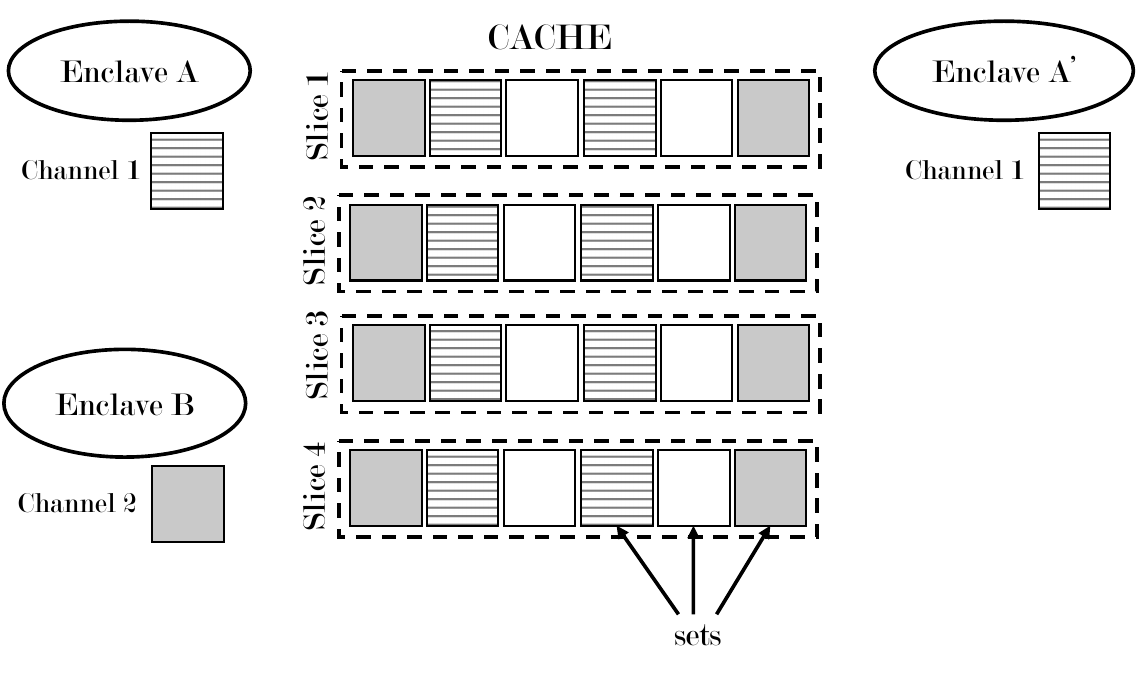}
		\vspace{-2em}
		\caption{Channels in \our. Each enclave uses a group of L3 cache sets to signal its presence and detect the presence of clones. Enclaves with the same binary (Enclaves A and A') use the same cache sets. Enclaves with different binaries (Enclaves A and B or A' and B) use different sets.}\vspace{-1 em}
		\label{fig:newfigure1}
	\end{figure}
	
	\subsection{Overview of \our}
	
	The main intuition behind \our{} is to rely on a covert channel as a signaling mechanism so that each enclave can indicate its presence to (and detect the presence of) clones. Namely, if the enclave instance is truly unique, it will see no response on the channel being monitored. On the other hand, if multiple instances are running, each instance will observe a measurable response in the form of a contention pattern.
	The challenges in using a covert channel as a signaling mechanism for clones lie in how to make communication robust despite (benign) noise due to other applications on the platform and, most importantly, despite a malicious OS that may tamper with the channel so that two clones do not detect each other.
	
	\our{}  undergoes two phases of operation: a preparation phase and a monitoring phase. The preparation phase
	is used to define the ``channel'' to be used for signaling and detection. By channel, we refer to a specific group of cache sets, so that enclaves with the same (resp. different) binary will use the same (resp. a different) channel (cf. Figure~\ref{fig:newfigure1}).
	
	Once the channel has been defined, \our{} builds the eviction sets required to communicate over such (cache-based) channel. During the monitoring phase, \our{} fills the cache sets of its channel with its own data, and continuously measures the time to access such data, in order to detect if it is still cached (cache hit) or if it has been evicted (cache miss). Note that clones will use the same channel (i.e., the same group of cache sets), removing each other's data. The resulting sequence of cache hits and misses is then fed to a classifier whose role is to distinguish whether clones are running on the same host based on the input sequence.
	
\begin{figure}[t!]
		\centering
		\includegraphics[width=0.47\textwidth]{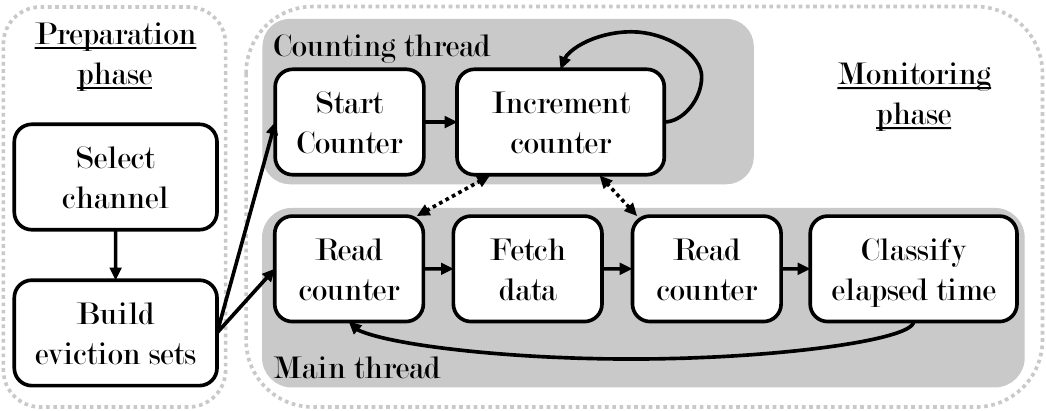}
		\vspace{-1em}
		\caption{Overview of \our.}\vspace{-1 em}
		\label{fig:newfigure2}
	\end{figure}

	From an architectural point of view, \our{} relies on two threads. The main thread measures access time to the cache and runs the classifier in order to detect clones. Recall that SGX 1.0 enclaves have no access to high precision timers API (e.g.,  \emph{rdtsc} and \emph{rdtscp}). Thus, we leverage a second thread that implements a timer by continuously increasing a runtime variable~\cite{armageddon16,MalwareSGX2018}.
	Figure~\ref{fig:newfigure2} summarizes the main execution steps of \our{}. We note that SGX 2.0 allows enclaves to access \emph{rdtsc}, so \our{} could work without the second thread on platforms where SGX 2.0 is available.
	
	Notice that we do not define the specific enclave behavior in case the main thread detects a clone or if a clone raises an alarm, and leave the selection of a suitable choice to the enclave developer. However, it is reasonable to anticipate that the enclave would halt its execution and notify the owner in such cases.
	
	Notice that the cache-based channel used by \our{} is shared with other applications and a potentially malicious OS. That is, any other process may intentionally pollute the channel of an enclave that uses \our. In case the channel is polluted, \our{} experiences a series of cache misses as if a clone were running on the same platform. Hence \our{} detects a clone and raises an alarm (e.g., stops the execution of the enclave). We treat this as a DoS attack and consider DoS attacks as out of scope.
	
	In the following, we provide details on the preparation phase (channel selection and eviction sets) and the monitoring phase.
	
	\subsection{Phase 1: Preparation Phase}
	
	\subsubsection{Channel Selection}
	
	\our{} uses the cache as a channel for an enclave to signal its presence to (and detect the presence of) other enclaves with the same binary. Detection succeeds as long as enclaves with the same binary monitor the same channel, and enclaves with different binaries monitor different channels.
	
	Assuming a typical cache with $s$ slices and 1024 sets per slice, there are 10 bits of a physical address that determine the cache set index (bits 6-15). An enclave only manages 6 of those bits (6-11), but it is unaware of the remaining 4 bits (12-15) that are controlled by the OS. By fixing bits 6-11 of an address, the enclave reduces the possible cache sets where a block of data is being cached within a slice to 16. If all enclaves loaded with the same binary monitor the same 16 cache sets determined by a specific value of bits 6-11, each of them can detect the presence of its clones---despite an adversary that controls the OS and allocates the physical pages of the enclave.
	We provide more details on cache memories and how cache-based covert channels work in Appendix~\ref{sec:memsystem}.
	
	Therefore, \our{} defines a channel as a group of 16 cache sets, in principle allowing for up to 64 concurrent channels.
	In Appendix~\ref{app:minSets}, we show that this choice is optimal, since monitoring less than 16 sets may allow the OS to execute multiple clones of an enclave and evade detection. Note, however, that the channel selected by a given enclave (e.g., by fixing bits 6-11 of the addresses to be monitored) must not be secret and, in particular, security is not affected if the OS knows which channel is being used by an enclave. In a real-world deployment, the OS may even actively help enclave owners in selecting an unused channel prior to attestation; in turn, the enclave owner uses attestation and secret provisioning to instruct the enclave about which channel to use. Note that the OS has no advantage in assigning two different enclaves---loaded with different binaries---to the same channel as this leads to a DoS. In this case, the two enclaves will (mistakenly) detect a clone and take appropriate countermeasures (e.g., stop their execution or report the problem to an external party like the enclave owner). In practice, a malicious OS can easily DoS a process running on its platform---regardless of whether \our{} is used or not.

	\subsubsection{Building Eviction Sets}
	
	\begin{algorithm}[t]
		\caption{Building the eviction sets in \our}
		\label{alg:eviction_sets}
		\begin{algorithmic}[1]
			\footnotesize
			\Require Memory byte array memArr[24MB];
			\Ensure \textit{\textbf{evictionSets[16][SLICES]}}
			\State $spoilerArr[256][LIM] \gets \{\}$ \Comment{LIM depends on memArr size}
			\State $cacheGroups[16][16*LIM] \gets \{\}$
			\State $evictionSets[16][SLICES*WAYS] \gets \{\}$
			\For{$i=0$ \textbf{to} 256}
			\State $cont \gets 0$
			\State test\_address = memArr[i*PAGE\_SIZE + offset];\label{lin:test_address}
			\State spoilerArr[i][cont++] = test\_address;
			\For{$j=(i+1)$ \textbf{to} 24MB; j+=PAGE\_SIZE;}
			\If {\textit{aliasing}(test\_address,memArr[j*PAGE\_SIZE])}
			\State spoilerArr[i][cont++] = memArr[j*PAGE\_SIZE];
			\EndIf
			\EndFor
			\EndFor
			\State {\textit{// Group the addresses with same set number}}
			\For{$i=0$ \textbf{to} 16} \Comment{Reduce before expand}
			\State $cont \gets 0$
			\Comment{it is 16 at the end of each iteration}
			\State test\_array = spoilerArr[i][:];\label{lin:test_set1}
			\State cacheGroups[i][cont++] = test\_array;
			\State {\textit{// Remove used data from the copy array}}
			\State spoilerArrCopy $\gets$ (spoilerArr - cacheGroups)
			\For{$j=i+1$ \textbf{to} 256} \label{lin:cache_groups_first}
			\State remove spoilerArr[j][:] from spoilerArrCopy;
			\If {test\_array is not evicted by spoilerArrCopy}
			\State cacheGroups[i][cont++] = spoilerArr[j][:];
			\State write spoilerArr[j][:] back at spoilerArrCopy;
			\EndIf
			\EndFor
			\For{$j=16$ \textbf{to} 256} \Comment{Find remaining groups}\label{lin:cache_groups}
			\State test\_array = spoilerArr[j][:];\label{lin:test_set2}
			\If {test\_array is evicted by cacheGroups[i][:]}
			\State cacheGroups[i][cont++] = test\_array;	\label{lin:cache_groups_end}
			\EndIf
			\EndFor
			\EndFor
			\For{$i=0$ \textbf{to} 16} \label{lin:reduction}
			\State $evictionSets[i][:] = \textbf{reduce}(cacheGroups[i][:])$
			\EndFor
		\end{algorithmic}
	\end{algorithm}
	
	In order to build eviction sets, the enclave must be aware of the specs of the CPU where it is deployed. This includes the number of slices, the number of sets per slice, and the number of ways per set. Such information must be hardcoded in the enclave. Alternatively, the enclave owner can pass such information to the enclave after the enclave has been deployed and the owner has attested it.
	
	Popular techniques to build eviction sets from within an enclave~\cite{MalwareSGX2018} require that the OS assigns contiguous memory to enclaves. In our settings, a malicious OS may, however, assign non-contiguous memory to the enclave. Therefore, we leverage alternative techniques that rely on false dependencies on load operations which are not under direct control of the OS~\cite{Spoiler2019}.
	We provide additional details on our choice in Appendix~\ref{app:minSets}; more specifically, we use the SATisPy~\cite{satispy} SAT solver to show that a malicious OS may evade detection if evictions sets are built relying on the assumption that enclave memory is contiguous.
	
	We leverage the technique of~\cite{Spoiler2019} to group data whose physical addresses share the last 20 bits and then regroup that data into groups that share the last 16 bits (i.e. groups that share the cache set number). Since 12 out of these 20 bits are controlled by the enclave, we can create $2^8 = 256$ different groups that we call ``spoiler groups''. This step, in turn, ensures that we have enough distinct addresses to build the necessary eviction sets. The spoiler groups are then regrouped into cache groups, and finally, cache groups are reduced and arranged so that all the slices are covered.
	
	The process is summarized in Algorithm~\ref{alg:eviction_sets}. We use an array of 24MB---twice the size of our cache memory---so to ensure that all possible eviction sets can be built. We also point out that when building the ``spoiler groups'' it should be verified that the $test\_address$ (line \ref{lin:test_address}) is not already present in the $spoilerArr$. Similarly, the $test\_array$ should not be part of the $cacheGroups$ (line \ref{lin:test_set1}). These checks have been omitted in the pseudo-code for simplicity and brevity.
	
	The $cacheGroups$ array is filled in two stages. In the first stage (loop at line \ref{lin:cache_groups_first}), a group of arrays or a group of addresses with the same set number that can occupy all the respective slices is obtained. At this point, the data in the $cacheGroups$ could be re-arranged per slices and then reduced to its minimum core (i.e. it should include as many addresses as ways of the cache sets), which is the goal of this algorithm. That is, one could directly execute the steps at line \ref{lin:reduction}. On the other hand, the second stage (lines \ref{lin:cache_groups}-\ref{lin:cache_groups_end}) ensures that the OS has assigned to the enclave the $2^8 = 256$ addresses corresponding to the aforementioned 8 bits of a ``spoiler address''. Besides, the distances between addresses included in the $spoilerArr$ and between the indexes of each $cacheGroup$ show if the memory assigned by the OS is linear and if there are any gaps, i.e., unassigned pages.
	
	We note that by having $256$ different groups of addresses, we ensure that all the possible set numbers are covered. Moreover, by re-grouping those 256 groups into the $16$ groups that share the same cache number while ensuring all the cache slices are covered, we guarantee that \our{} could map any cache location. In case any of the tests fail,  this offers compelling evidence that the OS is manipulating memory to alter the expected view of the memory by \our---in this case, the enclave should refuse to execute. It is worth noting that the value of the offset used in line \ref{lin:test_address} is chosen so that the virtual address of $memArr[offset]$ has its bits 6-11 equal to the selected channel, if the number of cores is not a power of two due to its slice selection function \cite{IrazoquiReverse,YaromReverse}. If, on the contrary, the number of cores is a power of two, the aforementioned slice selection function~\cite{Maurice_2015} makes it possible to use any value for the $offset$, but it should be changed afterwards (e.g during the reduction phase). Finally, the algorithm used to obtain the minimum-size eviction sets from a bigger set of addresses mapping to the same set ($cacheGroups$), could be any of the ones proposed in the literature, e.g.,~\cite{jscache,lastLevelPractical}, that mainly remove elements from the array until it has the same size as ways of the cache, while ensuring it is still able to completely fill the set. In practice, we have taken an approach similar to~\cite{lastLevelPractical}.
	
	\begin{figure}[t]
		\centering
		\includegraphics[width=0.4\textwidth]{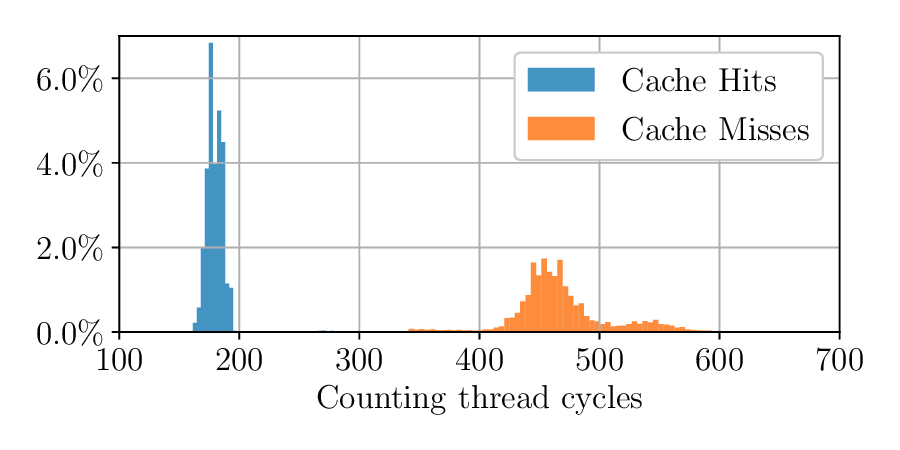}
		\vspace{-1.5 em}
		\caption{Times distribution of cache hits and misses in our test machine measured with the counting thread.}
		\vspace{-0.5 em}
		\label{fig_misses}
	\end{figure}

	\subsection{Phase 2: Monitoring}
	\label{subsec:monitoring}
	
	During the monitoring phase, \our{} reads the data of the sets to be monitored in a loop. Namely, \our{} measures the access times to each of the data blocks in the sets, in order to determine whether they are still cached (hit) or not (miss). The sequences of cache hits or misses---that we refer to as ``observation windows''---are fed to a classification algorithm that decides whether a clone is running on the same host. Like in~\cite{MalwareSGX2018}, we leverage a counting thread to measure access time: we fetch the value of the counter before and after reading an address. If the difference of the two counter values is greater than a pre-defined threshold, we conclude that the data was not cached and treats it as a cache miss; otherwise, we assume a cache hit.
	
	The threshold to distinguish cache hits from misses is machine dependent; it can be pre-computed if the hardware where the enclave is deployed is known a priori. Otherwise, the main thread can compute the threshold by flushing and reloading a block of data (cache miss time), reading again that block of data which will be in the cache (cache hit time), and repeating this process while computing the mean times.
	As shown in Figure~\ref{fig_misses}, cache misses take significant longer than cache hits and can be easily distinguished by observing data access times.
	
	Note that the monitoring and counting threads should run continuously, whenever the enclave is executing a critical piece of code where no clones must be allowed (e.g., between a read and an increment of a monotonic counter).
	If the monitoring/counting thread is interrupted, the obtained measurements will not match the expected ones, i.e., they will differ from the distributions depicted in Figure~\ref{fig_misses}.
	We treat such an event as evidence that the OS is manipulating the enclave with malicious intent and take countermeasures (e.g., halt the execution of the enclave).
	
	Note that before the monitoring phase can actually start, the enclave has to pre-fetch the data to be monitored into the cache to ensure that all the observed cache misses are due to evictions caused by other processes.
	
	We point out that there is no need for an enclave to fill all the ways of the monitored cache sets. In particular, given a $W$-way set-associative cache, clones will evict from cache each other's data---hence, will detect each other---as long as the number of ways filled per cache set, namely $m$, is chosen such that $(W/2) < m \leq W$. Further, if $m=W$, the enclave may detect evictions due to benign applications that happen to use the same cache sets and output a false positive.
	
\begin{figure}[t]
		\centering
		\begin{subfigure}[b]{0.4\columnwidth}
			\centering
			\includegraphics[width=\textwidth]{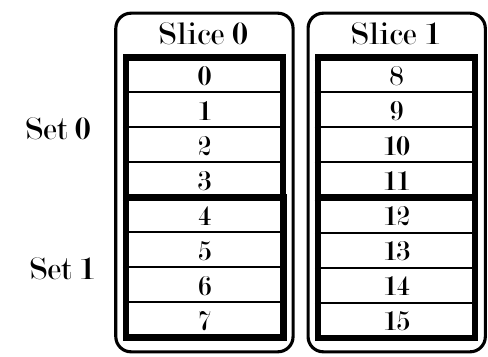}
			\caption{}
			\label{fig:patterns:columnmajor}
		\end{subfigure}
		\hfill
		\begin{subfigure}[b]{0.4\columnwidth}
			\centering
			\includegraphics[width=\textwidth]{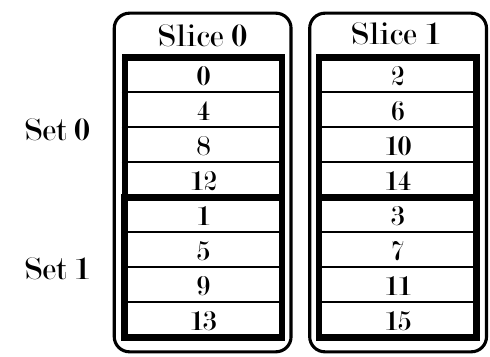}
			\caption{}
			\label{fig:patterns:our}
		\end{subfigure}
		\vspace{-1.5 pt}
		\caption{Different access patterns to the data included in the monitoring set for a 4-Way set associative cache with 2 slices and 2 sets per slice.}
		\label{fig:patterns}
		\vspace{-2 em}
	\end{figure}

	Another important design choice that affects the performance of \our{} is the order in which the enclave loops over the addresses in the monitoring group. In particular, the order impacts the latency in detecting clones. This is exemplified in Figure~\ref{fig:patterns}. Here, we depict a cache with two slices and two sets of 4 ways each. The numbers on each line represents the order in which the enclaves touches the lines in a loop iteration. Assuming the pseudo-LRU eviction policy implemented in Intel processors~\cite{Briongos2019,vila2019cachequery}, data stored at lines touched earlier becomes the candidate to be evicted in case of conflict. If the enclave accesses all lines of a set before moving to the next one (Figure~\ref{fig:patterns:columnmajor}), it might not experience cache misses---thus may not detect a clone---before it loops over all of the address of the monitoring group.
	On the contrary, if the enclave accesses one line per set and then moves to the next set (Figure~\ref{fig:patterns:our}), in the worst case scenario, it will start observing cache misses after looping over one line in each of the sets. In other words, if $m$ stands for the number of ways filled, $c$ for the number of sets in the channel and $s$ for the number of slices in the cache, accessing all the data in the set before moving to the next one might require up to $m*c*s$ accesses prior to detection, whereas accessing one line per set can take up to $c*s$ accesses. In the sequel, we therefore adopt the latter access pattern.
	
	\section{Security Analysis}\label{sec:security}
	
	\noindent \textbf{Knowledge of CPU specifications:} Note that \our{} requires the specifications of the processors where it is to be deployed. In particular, \our{} requires information on the cache, so to build eviction sets. Naturally, a malicious cloud provider may not faithfully report the CPU model where the enclave is going to be deployed. However, we believe that a rational cloud provider has no incentive to provide fabricated information on its CPUs. This is because if  the malicious behavior of the cloud is exposed, its reputation may be severely affected. In a nutshell, \our{} is not designed to counter a malicious cloud provider, but rather an adversary that compromises the OS on the cloud machines.
	
	\vspace{0.5 em} \noindent \textbf{Changing channel assignment:} Recall that the goal of the adversary is to execute two (or multiple) clones---enclaves loaded with the same binary---while evading the clone-detection mechanism. One possible attack strategy is to assign two different channels (i.e., two different groups of cache sets) to two enclave clones. We eliminate this option by ensuring that any two enclaves, loaded with the same binary, monitor the same group of cache sets. In particular, if all enclaves with the same binary fix bits 6-11 of the addresses to be monitored, each of those addresses can only be mapped to one out of 16 cache sets. By monitoring all of the 16 cache sets, we guarantee that two clones cannot be assigned to different channels. Note that monitoring less than 16 cache sets---out of those determined by fixing bits 6-11 of an address---may allow  the adversary to evade the detection mechanism. In particular, we used a SAT-solver (SATisPy \cite{satispy}, which in turn is a wrapper of MiniSAT \cite{minisat}) to simulate memory mapping and to show that, if less than 16 cache sets are monitored, the OS can find multiple mappings that effectively assign clones to different channels. We provide more details on this in Appendix~\ref{app:minSets}.
	
	\vspace{0.5 em} \noindent \textbf{Side-stepping the enclave:} Alternatively, the adversary might leverage the ability to control the execution of the enclave at instruction level, e.g., by using frameworks such as SGX-Step~\cite{SGXStep17}. By choosing which of the clones is making progress, one or few instructions at a time, the adversary may prevent one enclave instance from detecting the presence of the other. We argue that such strategy is not viable because the cache as a covert channel allows two enclaves to detect each other, even if they are not running at the same time. Take, for example, the BI-SGX enclave described in Section~\ref{sec:motivation}. The enclave uses monotonic counters and a forking attack requires two clones, say $E$ and $E'$, such that the following instructions are executed in a sequence: (a) $E$ calls \texttt{Increment(MC)}, (b) $E'$ calls \texttt{Increment(MC)}, (c) $E$ calls \texttt{Read(MC)}, and finally (d) $E'$ calls \texttt{Read(MC)}. The outcome is two sealed data items, one from $E$ and the other from $E'$, with the same value of the monotonic counter. This attack can be mitigated by using \our. In particular, if $E$ runs first, it writes its fingerprint to the cache. Next $E'$ runs and overwrites with its own fingerprint what enclave $E$ had written into the cache. Finally, $E$ resumes, detects that its fingerprint into the cache was overwritten and determines that a clone is running.
	Once a clone has been detected, the enclave could take appropriate countermeasures (e.g., refuse to seal data). Notice that other interruption strategies, besides single-stepping the enclave, could be used by the adversary. For instance, the adversary might try to infrequently interrupt either clones in an attempt to prevent detection. While such attacks could result in a false positive (raising an alarm by \our), it remains unclear whether \our{} can comprehensively detect all such attack strategies.
	
	\vspace{0.5 em} \noindent  \textbf{Polluting the channel:} Note that ``polluting'' the cache-based channel is not a viable option for a malicious OS. If the OS deliberately touches the cache lines used by \our{}, the detection mechanism (wrongly) infers that a clone is running thereby generating a false positive. Upon detection, the enclave may, e.g., halt its execution but no fork would take place. We confirm this by experiments in Section~\ref{sec:eval}.
	
	\vspace{0.5 em} \noindent  \textbf{Slowing down threads:} Further, the OS may as well try to make a cache miss look like a hit so that the enclave running \our{} fails to detect its clone. To do so, a malicious OS needs to slow down the counting thread while the main thread measures access times to its eviction set.
	The OS can achieve this by scheduling the counting thread on a core along with other applications. This strategy would slow down the counting thread and result in anomalous readings by the main thread. Here, the main thread reads the counter and computes an elapsed time value that does not match the elapsed time of a cache miss nor it matched the elapsed time of a cache miss. The current version of \our{} does not address such attack. However, we believe that it could be addressed by having the main thread raising an alarm every time it detects an anomalous reading of the counter. We have empirically verified this by scheduling threads on the same core where the counting thread was running and the main thread witnessed no increases of the counter variable. Similarly, if the adversary slows down the main thread, the corresponding AEX could be detected by monitoring the SSA area as done in previous work~\cite{varys18}; once the thread resumes and detects the asynchronous exit, it could raise an alarm.
	
	\vspace{0.5 em} \noindent \textbf{Modifying core frequency:} Another approach to make a cache miss look like a hit would be to change the frequency of the different cores available. Concretely, the adversary may run the counting thread on a slower core and the main thread on a faster one. We note that there is no SGX-enabled processor with per-core frequency scaling; this feature seems to be available only on some HPC processors that do not feature SGX~\cite{7284406,Schone_2019}. Hence, if the OS changes the frequency of a core in an SGX-capable processor, it would cause a frequency change on all other cores~\cite{6008552,9200659}. We have empirically verified this in our platform.  Even assuming future processors with SGX and per-core frequency scaling~\cite{xeonscalable}, some time elapses between the instant when the OS makes a frequency change request until this change is effective. As reported in~\cite{7284406,Schone_2019}, this time interval amounts to roughly 500 $\mu$s; in contrast a cache miss only takes around 0.15 $\mu$s. Thus, adding a periodic re-calibration phase where the main thread measures the time of a cache miss, may prevent the OS from scaling the frequency. In particular, if the re-calibration phase occurs every 500 $\mu$s, frequency scaling by the OS could be spotted. As an alternative strategy, the OS may configure core frequencies in advance, and then move the counting thread across cores. Again this could be spotted with periodic re-calibration.
	
	\vspace{0.5 em} \noindent \textbf{Changing memory mapping:} A malicious OS may change the physical to virtual mapping by leveraging its ability to control some of the bits of an address that determine the cache set (bits 12-15). We note that the enclave fixes bits 6-11 and monitors all sets corresponding to all configurations of the remaining 4 bits. As an example, fixed bits 6-11 as 010101, then \our{} monitors the sets given as XXXX010101 where XXXX ranges from 0000 to 1111. In case the OS  changes the mapping between a virtual address and a physical address (e.g., by swapping pages using the EWB instruction) an address would move from one of the sets monitored by \our{} to another set that is also monitored by \our{}. In this case, \our{} may end up polluting its own cache sets. If the number of addresses that underwent a change set is high, \our{} would mistakenly detect a clone and raise an alarm. This is another case of false positive and, as mentioned before, false positives are not in the attacker's best interest. Changes to the mapping between physical and virtual addresses may also be carried differently, e.g., by adding and removing pages via EDMM (available for SGX 2.0). While we could not verify this strategy on SGX 2.0, we speculate, however, that such changes made to the page mappings using EDMM are likely to trigger a notification before becoming effective, which, in turn, can be detected by \our{} (see~\cite{McKeenAACJLR16}  page 3, Section 3.1 for more details).

	\section{Other Applications of \our{}}\label{sec:multiple}
	
	\subsection{Extending \our{} to detect $N>1$ Clones}
	
	So far, we have described how \our{} is used to allow the execution of a single legitimate instance of an application enclave, and to detect whenever the number of cloned instances is two or greater. In what follows, we show how \our{} can allow the execution of $N\geq 2$ legitimate instances and detect whenever the number of instances reaches $N+1$ or above. This could be useful in application scenarios where more than one enclave instance may be honestly needed for redundancy or increased throughput~\cite{replicatee2019}.
	
	As mentioned earlier, allowing $N=1$ instances requires that an enclave monitors at least $W/2+1$ of the available ways in a set. To generalize to $N\geq 2$, we tune the number $m$ of monitored ways in a set such that (i) as long as $N$ clones are running, none of them evicts data pushed to cache by the other instances, and (ii) if $N+1$ clones are running, the enclaves will witness evictions of their data from the cache. In other words, we set $m$ such that $m\leq W/N$ and $m>W/(N+1)$. For example, if the number of ways in a cache set is $W=16$ and we wish to allow up to $N=2$ concurrent instances, we set $m=6$. As a result, two instances will fill at most $12$ ways of a cache set, whereas a third instance will fill up all the lines and evict data of the first two. Assuming a 16-way last-level cache, Table~\ref{table_nenclaves} shows possible values of $m$ for the different values of $N$. Note that for some values of $N$, no $m$ satisfies the constraints defined above. For example, no integer $m$ satisfies $W/(N+1)<m\leq W/N$ for $W=16$ and $N=7$.
	
	As a by-product, we argue that \our{} can also estimate the number of clones running on a machine by dynamically tuning the number of monitored ways per cache set as shown in Table~\ref{table_nenclaves}. For example, the enclave starts monitoring $m=12$ ways per set to check for the presence of a clone; in case the classifier infers the presence of another instance, the enclave decreases $m$ to $8$ to spot whether two or more other instances are running. If, given the new value of $m$, the classifier outputs that there are no clones, the enclave concludes there was only one other instance running in the system. Otherwise, if the classifier infers again the presence of clones, the enclave decreases again $m$ to $5$. Again, if the classifier outputs that there are no clones, the enclave concludes there were only two instances running in the system. On the contrary, if the classifier keeps detecting clones, the enclave decreases $m$ once again. This process is repeated until no clones are detected, which would then allow the enclaves to estimate the number of clones running on the platform.
	
	\begin{table}[!t]
		\centering
		\scalebox{0.97}{\begin{tabular}{|c|c|c|c|c|c|c|c|}
				\hline
				\# of allowed instances ($N$) & 1 & 2 & 3 & 4 & 5 & 8 & 16 \\
				\hline
				\# of ways to be monitored ($m$) & 9-16 & 6-8 & 5 & 4 & 3 & 2 & 1 \\
				\hline
		\end{tabular}}
		\caption{Number of ways that have to be monitored per cache set to allow different numbers of legitimate  enclaves, given a 16-way set associative cache.\label{table_nenclaves}}
		\vspace{-2 em}
	\end{table}
	
	\subsection{Using \our{} to Enhance NARRATOR and ROTE}
	\label{sec:fix_others}
	
	As mentioned earlier, solutions against forking attacks like ROTE~\cite{ROTE2017} and NARRATOR~\cite{NARRATOR2022} use a cohort of enclaves hosted in different platforms that act as a (distributed) TTP and provide forking prevention to other application enclaves. Both solutions require yet another TTP to avoid being forked at system initialization, since a fork on the cohort allows the adversary to fork application enclaves.
	
	ROTE~\cite{ROTE2017} uses a trusted administrator that provisions secret keys to the cohort members; it uses the linkable attestation mechanism of Intel SGX to ensure that at most one enclave per platform is provisioned.
	On the other hand, NARRATOR~\cite{NARRATOR2022} replaces the centralized administrator with a distributed TTP, i.e., a blockchain.
	In particular, enclaves use the hash of their sealing key (derived from the platform private key and the enclave identity) as a platform ID. An enclave being initialized as a cohort member checks if its platform ID has been written on the ledger; if not, the enclave writes the ID to the ledger and joins the cohort. Otherwise, the enclave assumes that a clone on the same platform has already joined the cohort and refuses to proceed.
	
	\our{} can be leveraged to remove the need of a TTP (be it a trusted administrator or a blockchain) during the initialization phases of ROTE or NARRATOR.
	In a nutshell, an enclave being initialized to join the cohort uses \our{} to ensure that no clone on the same platform is also being initialized.

	\begin{table*}[!htbp]
		\centering
		\setlength\tabcolsep{1pt}
		\noindent\scalebox{0.99}{
			\begin{tabular}{cc|c|c|c|c|c|c|c|c|c|c|c|c|c|c|c|c|c|c|}
				\cline{3-20}
				&&\multicolumn{6}{c|}{\textbf{Threshold}}&\multicolumn{6}{c|}{\textbf{Naive Bayes}}&\multicolumn{6}{c|}{\textbf{Neural Network}} \\ \cline{3-20}
				&& w=1&w=4&w=16&w=64&w=256&w=1024&w=1&w=4&w=16&w=64&w=256&w=1024&w=1&w=4&w=16&w=64&w=256&w=1024  \\ \hline
				\multicolumn{1}{|c|}{\multirow{3}{*}{Baseline}}&m=9 & 0.113 & 0.087 & 0.051 & 0.035 & 0.023 & 0.012
				& 0.102 & 0.078	& 0.039 & 0.021 & 0.022 & 0.004
				&0.103 & 0.068 & 0.027 & 0.009 & 0.007 & 0.001
				\\ \cline{2-20}
				\multicolumn{1}{|c|}{}&m=12 & 0.097 & 0.024 & 0.002 & 0.001 &0.001 & 0.000
				&0.098 & 0.045 & 0.009 & 0.009 & 0.005 & 0.004
				&0.098 & 0.030 & 0.004 & 0.003 & 0.003 & 0.000
				\\ \cline{2-20}
				\multicolumn{1}{|c|}{} &m=16 &  0.158 &  0.107 & 0.064 & 0.021 &0.012 & 0.001
				& 0.113 & 0.103 & 0.021 & 0.011 & 0.003 & 0.006
				& 0.114 & 0.102 & 0.021 & 0.008 & 0.002 & 0.001
				\\ \hline
				\multicolumn{1}{|c|}{\multirow{3}{*}{x265}}&m=9 & 0.189 &  0.108 & 0.048 & 0.007 & 0.002 & 0.001
				&0.182 & 0.119 & 0.061 & 0.034 & 0.019 & 0.015
				&0.189 & 0.108 & 0.049 & 0.038 & 0.014 & 0.009
				\\ \cline{2-20}
				\multicolumn{1}{|c|}{} &m=12 &0.082 & 0.043 & 0.021 & 0.021 & 0.003 & 0.002
				&0.091 & 0.045 & 0.027 & 0.021 & 0.012 & 0.003
				&0.092 & 0.039 &0.016 & 0.010 & 0.006 & 0.004
				\\ \cline{2-20}
				\multicolumn{1}{|c|}{} &m=16  &0.198 & 0.156 & 0.153 & 0.102 & 0.012 & 0.009
				&0.210 & 0.178 & 0.134 & 0.098 & 0.041 & 0.009
				&0.211 & 0.189 & 0.123 & 0.085 & 0.054 & 0.023
				\\ \hline
				\multicolumn{1}{|c|}{\multirow{3}{*}{sql}} &m=9 & 0.208 & 0.098 & 0.053 & 0.041 & 0.010 & 0.005
				& 0.212 & 0.102 & 0.056 & 0.021 & 0.011 & 0.008
				&0.199 & 0.191 & 0.042 & 0.012 & 0.007 & 0.001
				\\ \cline{2-20}
				\multicolumn{1}{|c|}{} &m=12 & 0.043 & 0.021 & 0.019 & 0.018 & 0.010 & 0.006
				& 0.045 & 0.023 &0.018 & 0.008 &0.006 &0.004
				&0.045 & 0.023 &0.018 & 0.009 &0.006 & 0.003
				\\ \cline{2-20}
				\multicolumn{1}{|c|}{} &m=16 & 0.198 & 0.134 & 0.124 & 0.098 & 0.052 & 0.018
				& 0.187 & 0.152 & 0.146 & 0.127 & 0.078 & 0.012
				& 0.187 & 0.160 & 0.129 & 0.085 & 0.013 & 0.013
				\\ \hline
				\multicolumn{1}{|c|}{\multirow{3}{*}{opencv}} &m=9 & 0.223 & 0.102 & 0.072 & 0.065 & 0.043 & 0.021
				& 0.161 & 0.126 & 0.087 & 0.081 & 0.047 & 0.030
				& 0.161 & 0.125 & 0.086 & 0.075 & 0.039 & 0.027
				\\ \cline{2-20}
				\multicolumn{1}{|c|}{} &m=12 & 0.091 & 0.065 & 0.058 & 0.045 & 0.032 & 0.012
				&0.090 & 0.062 & 0.054 & 0.042 & 0.031 & 0.011
				&0.090 & 0.062 & 0.054 & 0.039 & 0.021 & 0.010
				\\ \cline{2-20}
				\multicolumn{1}{|c|}{} &m=16&0.320 & 0.213 & 0.211 & 0.193 & 0.072 & 0.034
				&0.310 & 0.223 & 0.132 & 0.092 & 0.089 & 0.032
				&0.310 & 0.218 & 0.081 & 0.070 & 0.034 & 0.021
				\\ \hline
				\multicolumn{1}{|c|}{\multirow{3}{*}{gcc}} &m=9 & 0.163 & 0.124 & 0.073 & 0.043 & 0.032 & 0.028
				&0.159 & 0.099 & 0.050 & 0.032 & 0.017 & 0.017
				&0.159 & 0.099 & 0.056 & 0.024 & 0.018 & 0.017
				\\ \cline{2-20}
				\multicolumn{1}{|c|}{} &m=12 & 0.068 &0.031 & 0.019 & 0.009 & 0.006 & 0.005
				&0.069 & 0.026 &0.018 & 0.013 & 0.013 &0.007
				&0.069 & 0.026 &0.017 & 0.012 & 0.013 &0.010
				\\ \cline{2-20}
				\multicolumn{1}{|c|}{} &m=16& 0.190 & 0.176 & 0.132 & 0.102 & 0.087 & 0.054
				& 0.192 & 0.162 & 0.112 & 0.096 & 0.087 & 0.017
				& 0.191 & 0.161 & 0.103 & 0.087 & 0.068 & 0.036
				\\ \hline
				\multicolumn{1}{|c|}{\multirow{3}{*}{cloud}} &m=9 & 0.067 & 0.059 & 0.031 & 0.023 & 0.017 & 0.010
				& 0.079 & 0.045 & 0.031 &  0.032 & 0.020 & 0.010
				& 0.081 & 0.043 & 0.025 & 0.021 & 0.009 &0.006
				\\ \cline{2-20}
				\multicolumn{1}{|c|}{} &m=12 & 0.100 & 0.071 & 0.047 & 0.018 & 0.007 & 0.003
				& 0.091 & 0.040 & 0.032 & 0.012 & 0.012 & 0.002
				& 0.083 & 0.032 & 0.023 & 0.009 & 0.004 & 0.003
				\\ \cline{2-20}
				\multicolumn{1}{|c|}{} &m=16&0.142 &0.090 &0.061 & 0.011 & 0.009 & 0.009
				& 0.109 &0.089 & 0.056 & 0.043 & 0.021 &0.013
				& 0.102 &0.056 & 0.049 & 0.037 & 0.013 & 0.006
				\\ \hline
		\end{tabular}}
		\vspace{0.5em}
		\caption{False negative rates for various detection algorithms for different values of $w$ and $m$. ``Baseline'' refers to the scenario where no background application is running, whereas the others refer to different applications running in the background.}
		\label{tab:initial_results1}
		\setlength\tabcolsep{1pt}
		\noindent\scalebox{0.99}{
			\begin{tabular}{cc|c|c|c|c|c|c|c|c|c|c|c|c|c|c|c|c|c|c|}
				\cline{3-20}
				&&\multicolumn{6}{c|}{\textbf{Threshold}}&\multicolumn{6}{c|}{\textbf{Naive Bayes}}&\multicolumn{6}{c|}{\textbf{Neural Network}} \\ \cline{3-20}
				&& w=1&w=4&w=16&w=64&w=256&w=1024&w=1&w=4&w=16&w=64&w=256&w=1024&w=1&w=4&w=16&w=64&w=256&w=1024  \\ \hline
				\multicolumn{1}{|c|}{\multirow{3}{*}{Baseline}}&m=9 & 0.120	& 0.055 & 0.024 & 0.008 & 0.007 & 0.004
				& 0.134 & 0.048 & 0.027 & 0.019 & 0.010 & 0.008
				&0.132 & 0.059 & 0.027 & 0.009 & 0.007 & 0.001
				\\ \cline{2-20}
				\multicolumn{1}{|c|}{}&m=12 & 0.090 & 0.047 & 0.010 & 0.005 & 0.003 & 0.002
				&0.089 & 0.041 & 0.009 & 0.009 & 0.011 & 0.006
				&0.089 & 0.032 & 0.004 & 0.003 & 0.003 & 0.002
				\\ \cline{2-20}
				\multicolumn{1}{|c|}{} &m=16 &  0.189 & 0.103 & 0.045 & 0.023 & 0.020 & 0.019
				& 0.253 & 0.107 & 0.019 & 0.015 & 0.007 & 0.008
				& 0.252 & 0.111 & 0.027 & 0.010 & 0.010 & 0.003
				\\ \hline
				\multicolumn{1}{|c|}{\multirow{3}{*}{x265}}&m=9 & 0.214 & 0.099 & 0.059 & 0.006 & 0.004 & 0.001
				&0.224 & 0.117 & 0.059 & 0.032 & 0.019 & 0.015
				&0.214 & 0.101 & 0.047 & 0.028 & 0.012 & 0.009
				\\ \cline{2-20}
				\multicolumn{1}{|c|}{} &m=12 & 0.106 & 0.039 & 0.033 & 0.031 & 0.015 & 0.010
				&0.095 & 0.037 & 0.031 & 0.033 & 0.018 & 0.013
				&0.094 & 0.043 & 0.039 & 0.016 & 0.012 & 0.006
				\\ \cline{2-20}
				\multicolumn{1}{|c|}{} &m=16  &0.317 & 0.263 & 0.172 & 0.143 & 0.028 & 0.027
				&0.297 & 0.230 & 0.218 & 0.105 & 0.071 & 0.027
				&0.296 & 0.229 & 0.201 & 0.089 & 0.067 & 0.054
				\\ \hline
				\multicolumn{1}{|c|}{\multirow{3}{*}{sql}} &m=9 & 0.055 & 0.052 & 0.038 & 0.022 & 0.010 & 0.005
				& 0.085 & 0.056 & 0.027 & 0.003 & 0.005 & 0.002
				& 0.071 & 0.073 & 0.038 & 0.004 & 0.003 & 0.003
				\\ \cline{2-20}
				\multicolumn{1}{|c|}{} &m=12 & 0.068 & 0.033 & 0.025 & 0.036 & 0.016 & 0.010
				& 0.066 & 0.031 & 0.028 & 0.016 & 0.014 & 0.006
				&0.041 & 0.031 & 0.026 & 0.021 & 0.014 & 0.009
				\\ \cline{2-20}
				\multicolumn{1}{|c|}{} &m=16 & 0.173 & 0.181 & 0.176 & 0.098 & 0.037 & 0.022
				& 0.189 & 0.169 & 0.167 & 0.122 & 0.054 & 0.026
				& 0.189 & 0.148 & 0.157 & 0.089 & 0.069 & 0.038
				\\ \hline
				\multicolumn{1}{|c|}{\multirow{3}{*}{opencv}} &m=9 & 0.065 & 0.050 & 0.044 & 0.040 & 0.026 & 0.005
				& 0.124 & 0.092 & 0.063 & 0.053 & 0.043 & 0.028
				& 0.121 & 0.096 & 0.043 & 0.029 & 0.018 & 0.015
				\\ \cline{2-20}
				\multicolumn{1}{|c|}{} &m=12 & 0.104 & 0.076 & 0.069 & 0.039 & 0.034 & 0.020
				&0.106 & 0.079 & 0.075 & 0.046 & 0.041 & 0.021
				&0.106 & 0.079 & 0.071 & 0.027 & 0.033 & 0.016
				\\ \cline{2-20}
				\multicolumn{1}{|c|}{} &m=16&0.350 & 0.244 & 0.012 & 0.021 & 0.083 & 0.040
				&0.370 & 0.228 & 0.132 & 0.160 & 0.085 & 0.063
				&0.370 & 0.236 & 0.077 & 0.066 & 0.055 & 0.037
				\\ \hline
				\multicolumn{1}{|c|}{\multirow{3}{*}{gcc}} &m=9 & 0.132 & 0.051 & 0.016 & 0.033 & 0.024 & 0.018
				&0.138 & 0.071 & 0.050 & 0.012 & 0.017 & 0.017
				&0.138 & 0.071 & 0.027 & 0.004 & 0.016 & 0.009
				\\ \cline{2-20}
				\multicolumn{1}{|c|}{} &m=12 & 0.070 & 0.023 & 0.021 & 0.023 & 0.024 & 0.013
				&0.069 & 0.034 & 0.030 & 0.067 & 0.021 & 0.009
				&0.069 & 0.028 & 0.023 & 0.026 & 0.021 & 0.012
				\\ \cline{2-20}
				\multicolumn{1}{|c|}{} &m=16& 0.192 & 0.143 & 0.118 & 0.080 & 0.083 & 0.062
				& 0.190 & 0.164 & 0.132 & 0.100 & 0.061 & 0.013
				& 0.191 & 0.163 & 0.107 & 0.078 & 0.068 & 0.036
				\\ \hline
				\multicolumn{1}{|c|}{\multirow{3}{*}{cloud}} &m=9 & 0.067 & 0.024 & 0.017 & 0.015 & 0.007 & 0.010
				& 0.053 & 0.020 & 0.001 & 0.024 & 0.008 & 0.006
				& 0.051 & 0.022 & 0.009 & 0.011 & 0.005 & 0.004
				\\ \cline{2-20}
				\multicolumn{1}{|c|}{} &m=12 & 0.087 & 0.062 & 0.047 & 0.022 & 0.013 & 0.011
				& 0.098 & 0.080 & 0.055 & 0.016 & 0.014 & 0.016
				& 0.107 & 0.089 & 0.067 & 0.015 & 0.012 & 0.013
				\\ \cline{2-20}
				\multicolumn{1}{|c|}{} &m=16&0.147 & 0.077 & 0.070 & 0.027 & 0.017 & 0.015
				& 0.191 & 0.078 & 0.075 & 0.066 & 0.035 & 0.031
				& 0.200 & 0.106 & 0.081 & 0.069 & 0.027 & 0.014
				\\ \hline
		\end{tabular}}
		\vspace{0.5em}
		\caption{False positive rates for various detection algorithms for different values of $w$ and $m$. ``Baseline'' refers to the scenario where no background application is running, whereas the others refer to different applications running in the background.}
		\label{tab:initial_results2}
		\vspace{-2 em}
	\end{table*}

	\section{Implementation and Evaluation}\label{sec:eval}
	
	\subsection{Implementation Setup}
	We implemented a prototype of \our, including the code for the creation of the eviction sets (and some tests to ensure they have been properly built) and the counting thread that serves as a clock. Our implementation accounts for approximately 800 LoC.
	
	We deployed the prototype on a Xeon E-2176G (12 vCores at 3.70GHz, 64 GB RAM, and a 12 MB 16-way cache). To assess the performance of \our, we evaluated the impact on performance of (i) the choice of classification algorithm used to infer the presence of a clone given a sequence of cache hits and misses, (ii) the number of ways per set to be monitored $m$, and (iii) the size of the observation window $w$.
	
	We evaluate performance in an ideal scenario where no other application apart from the enclave (and possibly its clone) is running, as well as in a more realistic scenario where background processes---taken from the Phoronix benchmark suite~\cite{phoronix}---are running on the host at the same time. In scenarios featuring background processes, we run as many instances of the benchmark as needed to reach a total CPU usage close to 100\%. For each configuration of parameters and background process, we collected 100.000 samples while the enclave and a clone are running, and the same number of samples while the enclave is running without clones. We labeled these samples accordingly and obtained multiple datasets of 200.000 samples per scenario.

	\subsection{Evaluation Results}
	
	\begin{figure*}[ht]
		\centering
		\begin{subfigure}[b]{\picscalingvalue\textwidth}
			\centering
			\includegraphics[trim=0.1cm 0.1cm 0cm 0cm, clip, width=0.87\textwidth]{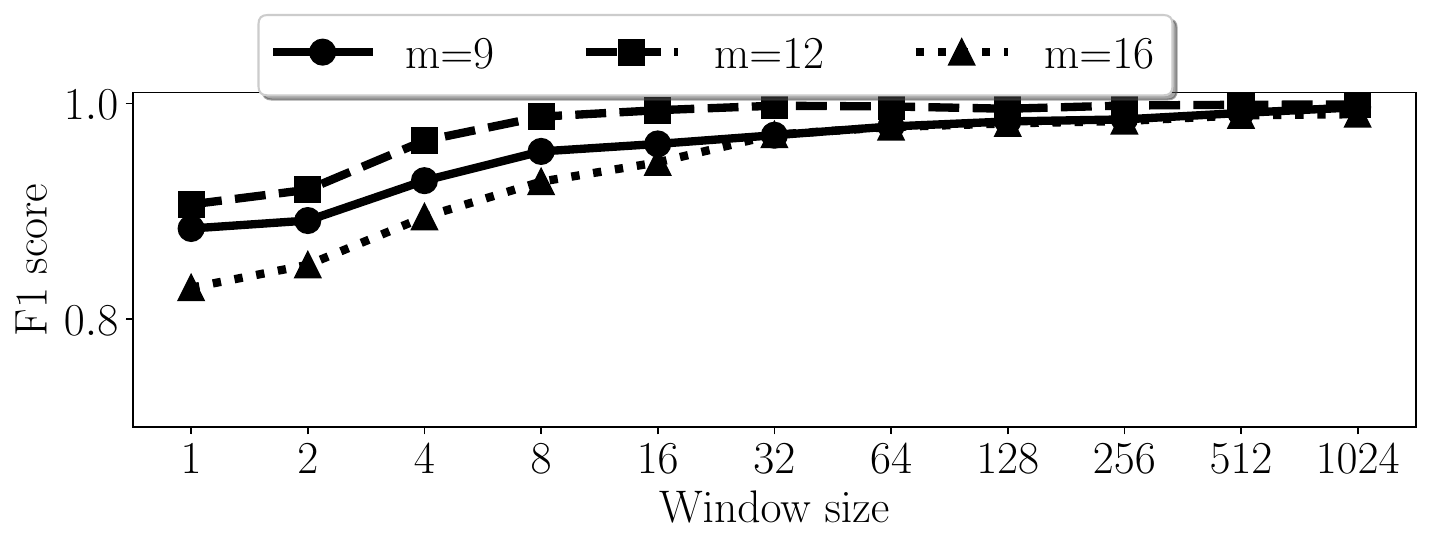}
			\caption{Threshold}
			\label{fig:thbase}
		\end{subfigure}
		\hfill
		\begin{subfigure}[b]{\picscalingvalue\textwidth}
			\centering
			\includegraphics[trim=0.1cm 0.1cm 0cm 0cm, clip, width=0.87\textwidth]{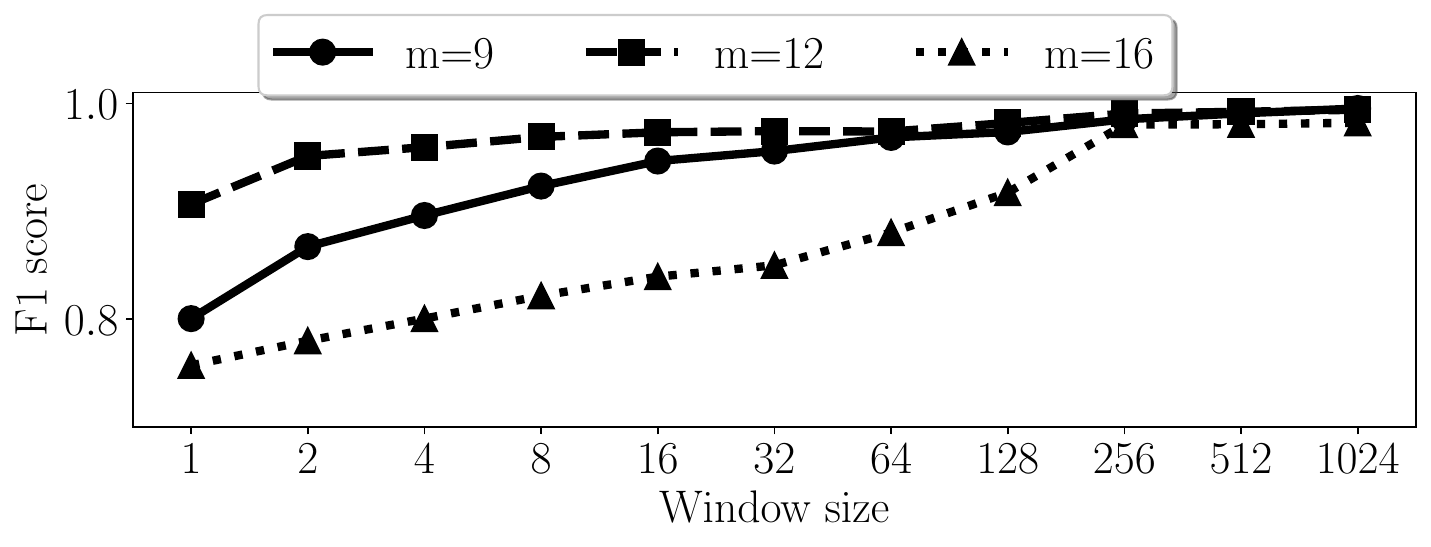}
			\caption{Threshold (x265)}
			\label{fig:thbase_x265}
		\end{subfigure}
		\begin{subfigure}[b]{\picscalingvalue\textwidth}
			\centering
			\includegraphics[trim=0.1cm 0.1cm 0cm 0cm, clip, width=0.87\textwidth]{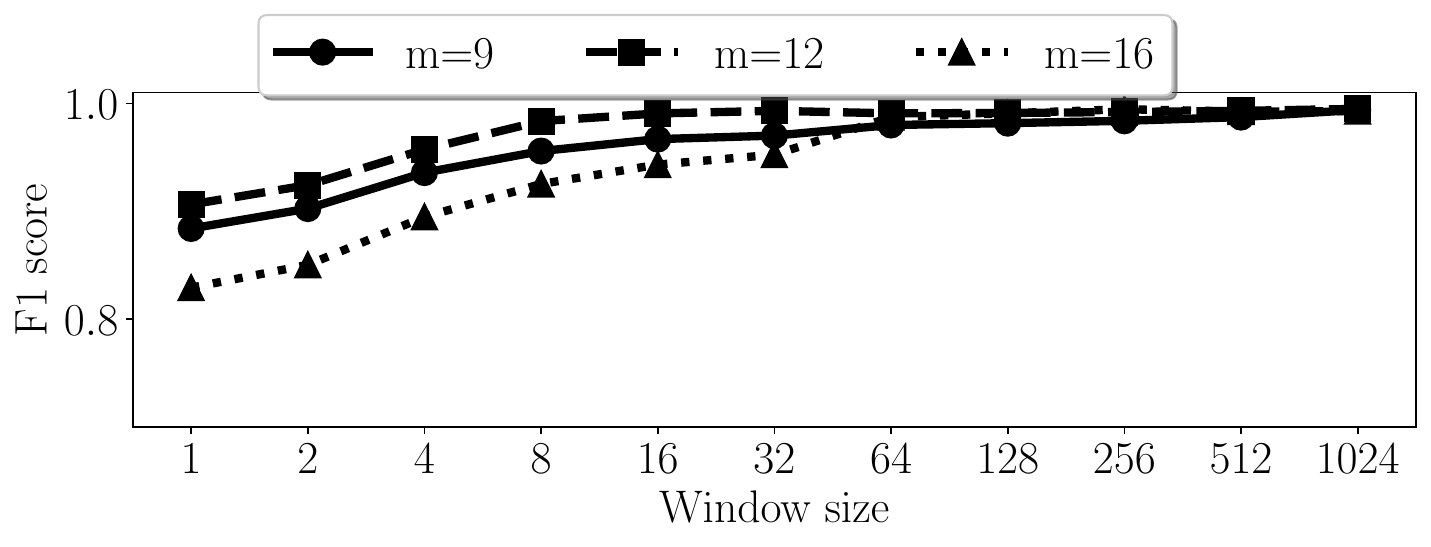}
			\caption{Gaussian Naive Bayes}
			\label{fig:bayes}
		\end{subfigure}
		\hfill
		\begin{subfigure}[b]{\picscalingvalue\textwidth}
			\centering
			\includegraphics[trim=0.1cm 0.1cm 0cm 0cm, clip, width=0.87\textwidth]{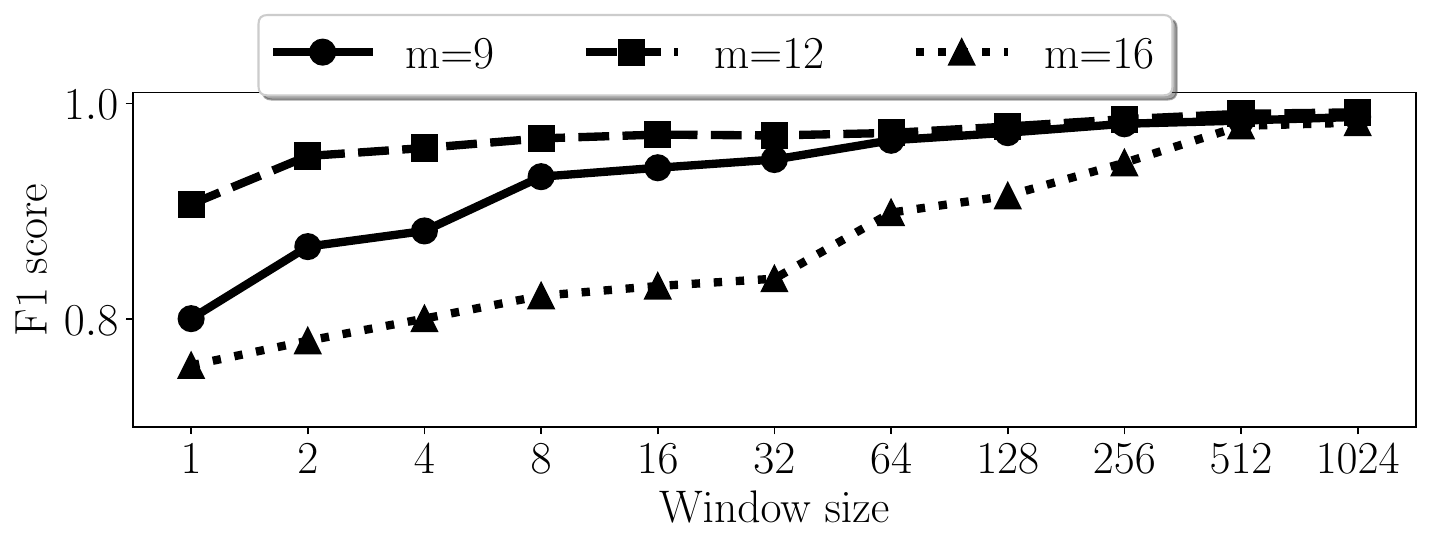}
			\caption{Gaussian Naive Bayes (x265)}
			\label{fig:bayes_x265}
		\end{subfigure}
		\begin{subfigure}[b]{\picscalingvalue\textwidth}
			\centering
			\includegraphics[trim=0.1cm 0.1cm 0cm 0cm, clip, width=0.87\textwidth]{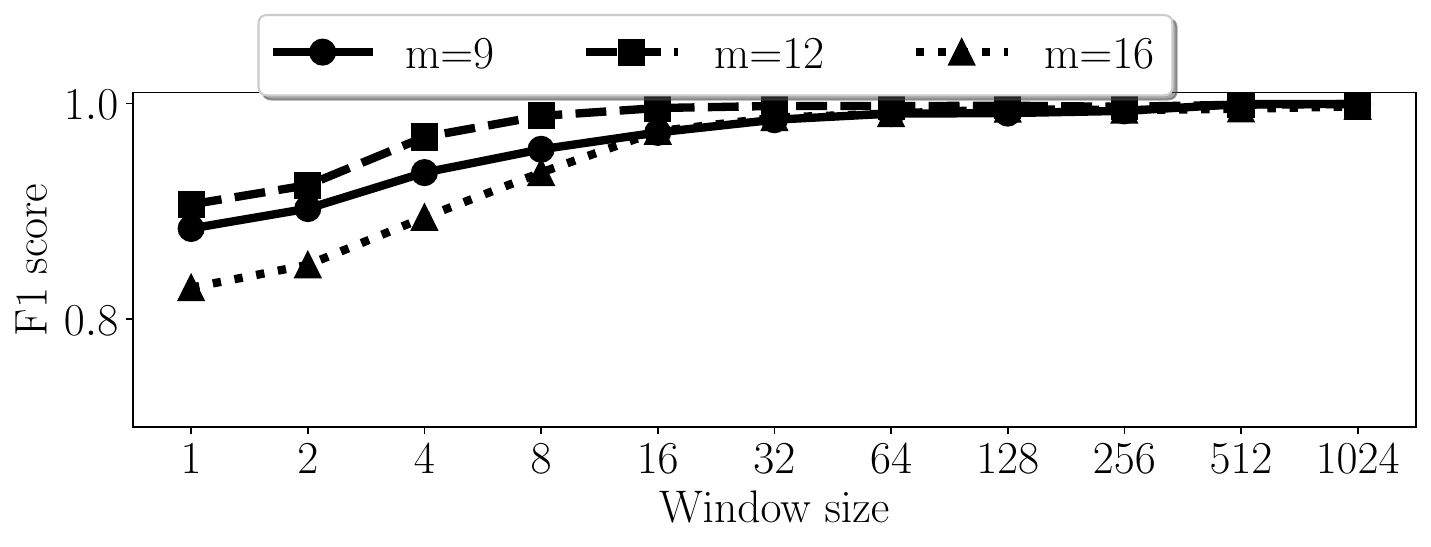}
			\caption{Neural Network}
			\label{fig:network}
		\end{subfigure}
		\hfill
		\begin{subfigure}[b]{\picscalingvalue\textwidth}
			\centering
			\includegraphics[trim=0.1cm 0.1cm 0cm 0cm, clip, width=0.87\textwidth]{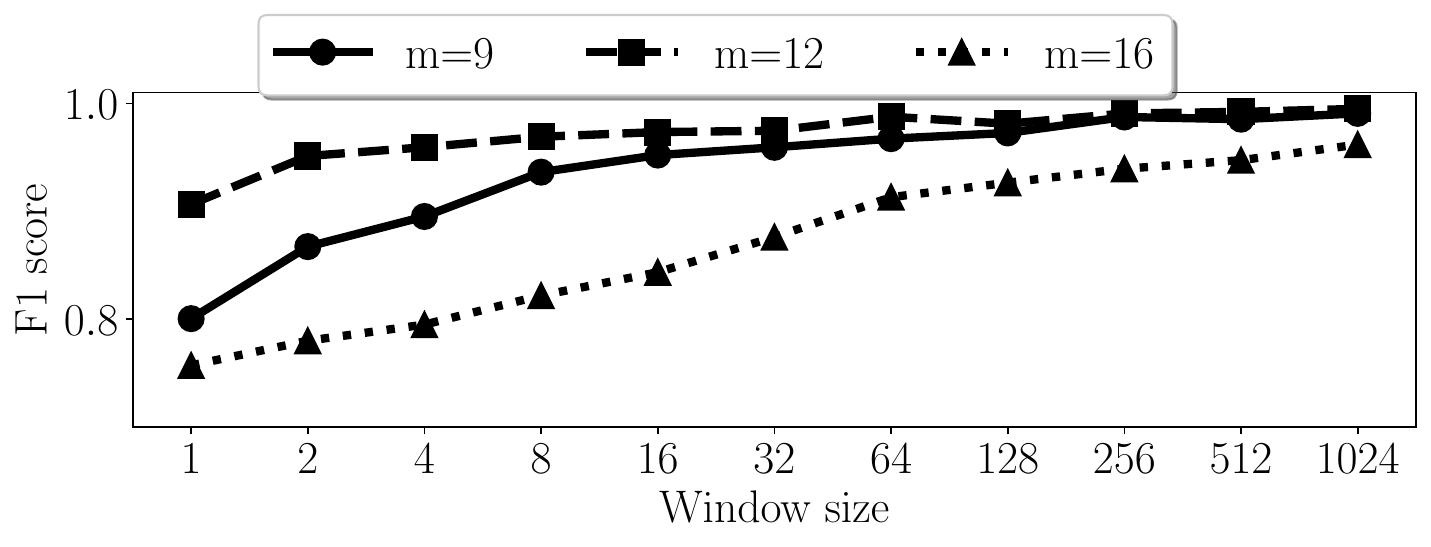}
			\caption{Neural Network (x265)}
			\label{fig:network_x265}
		\end{subfigure}
		\vspace{-1 em}
		\caption{F1 score of various detection algorithms for different values of $w$ and $m$. Figures on the left show the performance of \our{} with no other application; figures on the right show the performance when \texttt{x265 video encoder} runs in the background.}
		\label{fig:comparisson_algorithm_noisy}
		\vspace{-1.5 em}
	\end{figure*}
	
	We assess the performance of \our{} by means of F1 score. We additionally report in Tables~\ref{tab:initial_results1} and~\ref{tab:initial_results2} the associated false positive and negative rates for each experiment. Each data point represents the mean of 10-fold cross-validation.
	
	\vspace{0.5 em}\noindent\textbf{Choice of the Detection Algorithm.} We evaluate the performance of different detection algorithms in inferring the presence of a clone, given a sequence of cache hits and misses. In particular, we considered a number of classifiers included in Scikit-learn ~\cite{scikit-learn} as well as a simple threshold-based algorithm. For the latter, the threshold $t$ of cache misses for the detector to report a clone is selected empirically, as the one that allows to obtain the best performance.
	
	Figure~\ref{fig:comparisson_algorithm_noisy} compares the performance of various detection algorithms for $w\in[1,2024]$ and for $m \in\{9,12,16\}$ both in the ideal scenario with no background processes and in a realistic scenario where processes are running in background. For the latter scenario, we use \texttt{x265 video encoder}---the application with the most intensive memory use among the ones we have tested from the Phoronix benchmark suite---as the background process.
	
	The comparison between the plots on the left side of Figure~\ref{fig:comparisson_algorithm_noisy} (with no background process) and the ones on the right side of the same figure (with \texttt{x265 video encoder} running in parallel) allows us to assess the impact of background processes on the performance.
	
	\begin{figure*}[tbp]
		\centering
		\begin{subfigure}[b]{0.84\textwidth}
			\centering
			\includegraphics[trim=0.1cm 0.1cm 0cm 0cm, clip, height=4cm]{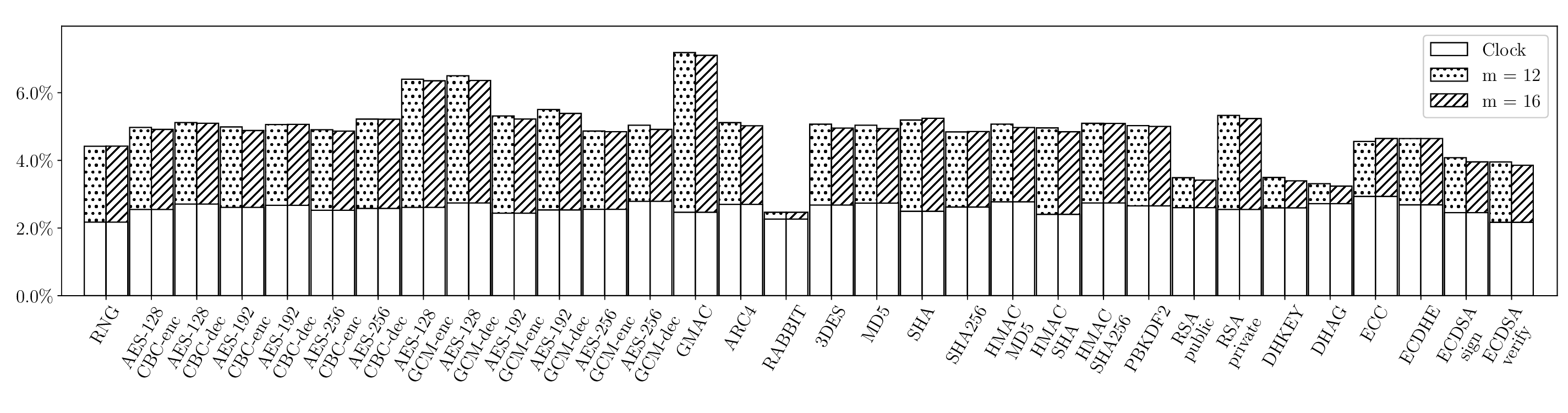}
			\caption{WolfSSL}
			\label{fig:1a}
		\end{subfigure}
		\hfill
		\begin{subfigure}[b]{0.14\textwidth}
			\centering
			\includegraphics[trim=0.1cm 0.1cm 0cm 0cm, clip, height=4cm]{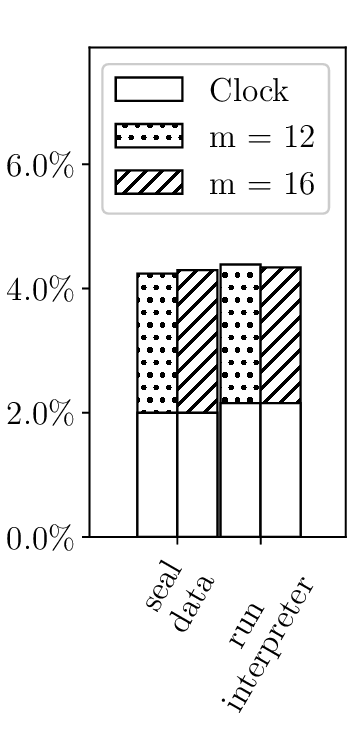}
			\caption{BI-SGX}
			\label{fig:2a}
		\end{subfigure}
		\vspace{-1 em}
		\caption{Mean penalty (in achieved throughput) due to \our{} for different WolfSSL applications (a) and BI-SGX (b). Here $w=32$.}
		\label{fig_performance1}
		\vspace{-1 em}
	\end{figure*}
	
	\begin{figure*}[h]
		\centering
		\begin{subfigure}[b]{\picscalingvalue\textwidth}
			\centering
			\includegraphics[trim=0.1cm 0.1cm 0cm 0cm, clip, width=0.87\textwidth]{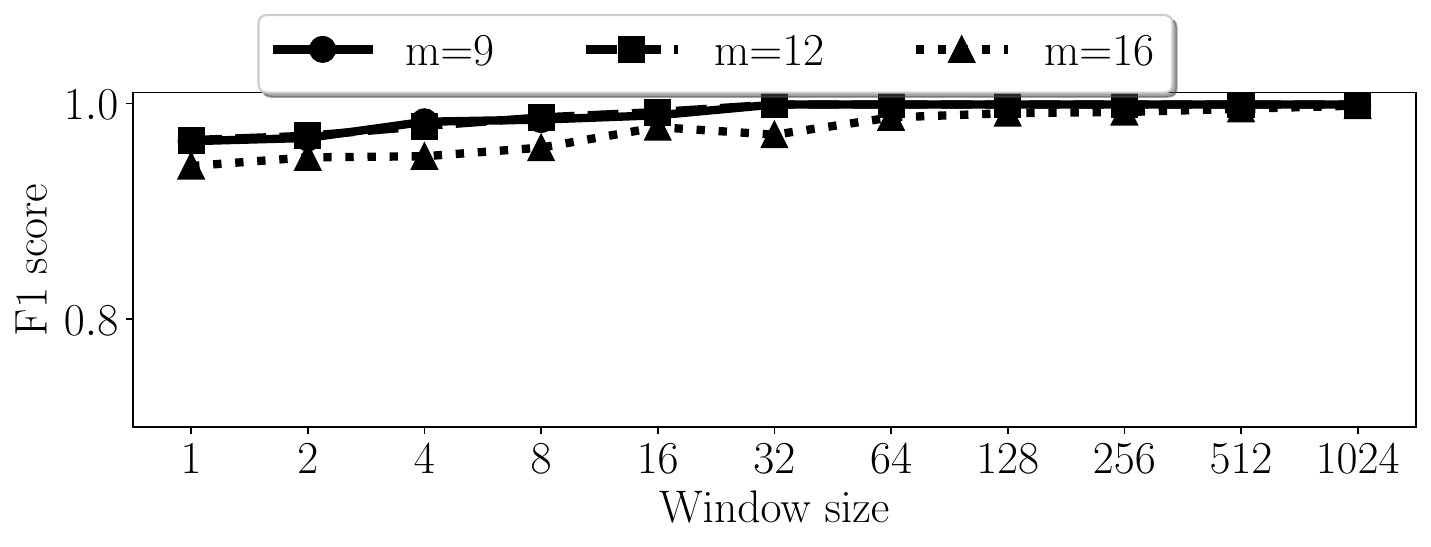}
			\vspace{-0.5 em}
			\caption{Threshold (no background process)}
			\label{fig:1b}
		\end{subfigure}
		\hfill
		\begin{subfigure}[b]{\picscalingvalue\textwidth}
			\centering
			\includegraphics[trim=0.1cm 0.1cm 0cm 0cm, clip, width=0.87\textwidth]{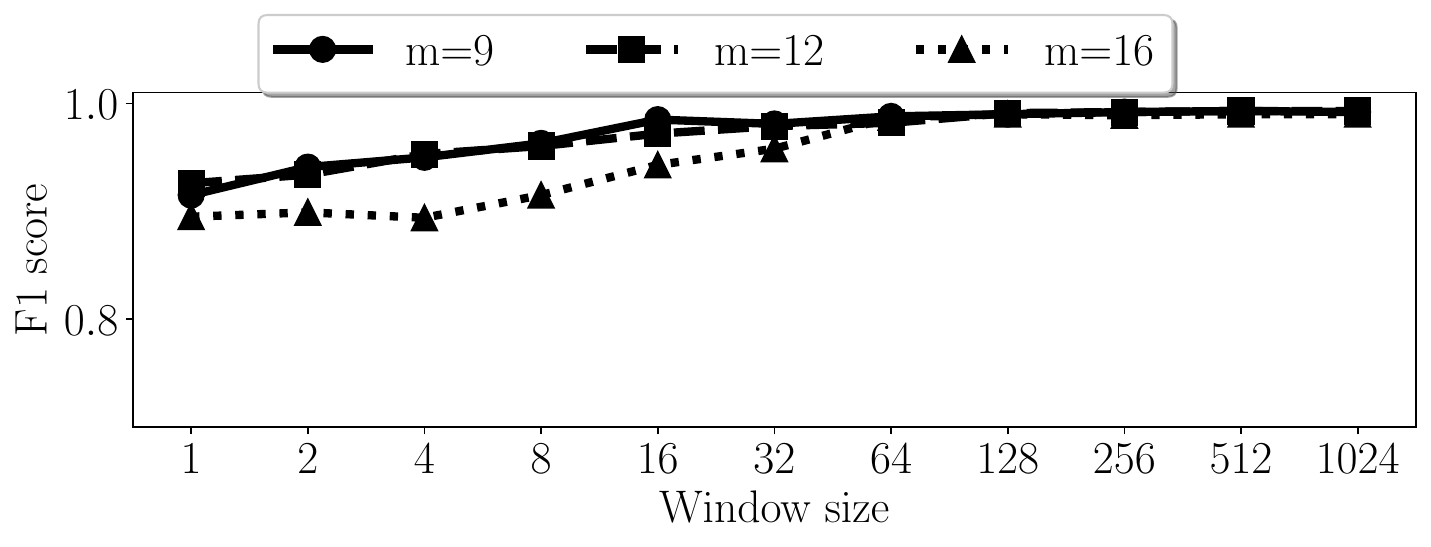}
			\vspace{-0.5 em}
			\caption{Threshold (x265)}
			\label{fig:2b}
		\end{subfigure}
		\vspace{-.5 em}
		\caption{F1 score of the threshold based detection algorithm for different values of $w$ and $m$ for attacks against BI-SGX. Figure on the left show the performance of \our{} with no other application; figure on the right show the performance when \texttt{x265 video encoder} runs in the background.}
		\label{fig:securing_bisgx}
		\vspace{-1 em}
	\end{figure*}
	
	\begin{figure}[h]
		\centering
		\includegraphics[width=0.47\textwidth]{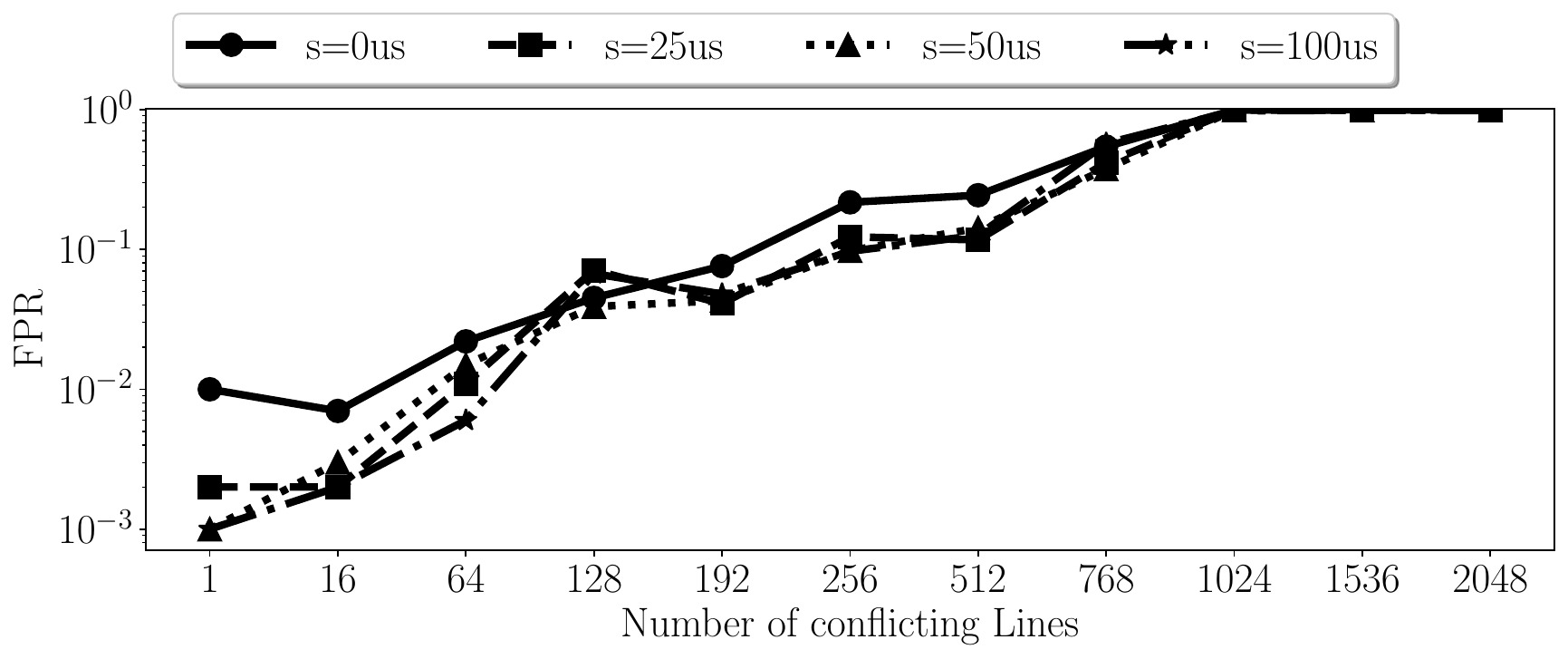}
		\vspace{-1 em}
		\caption{False positive rate when a malicious OS injects noise by placing into the cache based covert channel different amount of cache lines. $s$ refers to the waiting interval between injections of noise}\vspace{-1 em}
		\label{fig:noisefigure1}
	\end{figure}

	We note that the threshold-based algorithm is among the ones with higher F1 scores, for most configuration of $w$ and $m$. Indeed, the threshold-based detector emerges as the most suitable choice---owing to its simplicity, small code-size, and F1 score (and its associated false positive/negative rates).
	
	In Appendix~\ref{app:noise} we provide additional results with alternative detection algorithms and background applications of the Phoronix benchmark suite.
	
	\vspace{0.5 em}\noindent \textbf{Impact of observation window size $w$.} Figure~\ref{fig:comparisson_algorithm_noisy} also shows the impact of the size of the observation window $w$ on the F1 score. Clearly, increasing $w$ leads to better performance. In particular, a small observation window may only account for cache misses due to benign applications running on the same host, and may cause false positives. For example, by using the threshold-based classifier with
	$w=1$, the F1 score for $m=9$, $m=12$, and $m=16$ is 0.884, 0.906, and 0.829, respectively in an ideal scenario; when \texttt{x265 video encoder} is running in the background, F1 scores are 0.801, 0.907, and 0.757 for $m=9$, $m=12$, and $m=16$, respectively.
	By increasing $w$, classification becomes more robust: with $w=1024$, F1 score is 0.996 ($m=9$), 0.999 ($m=12$), and 0.990 ($m=16$) in the scenario where no application is running in the background and reaches 0.999 ($m=9$), 0.994 ($m=12$), and 0.982 ($m=16$) when \texttt{x265 video encoder} runs in the background.
	
	We also note that $w$ has a direct impact on detection latency, since it determines the time to fill the observation window with cache hits and misses---before the window is fed to the classifier. For instance, given that measuring a cache miss on the test machine takes approximately 450 cycles, setting $w=256$ results in detection latency of roughly 115k cycles. To put this number in context, computing an RSA-2048 signature with \texttt{openssl} on the test machine requires 17390k cycles.
	
	\vspace{0.5 em}\noindent\textbf{Impact of number of ways $m$.} As expected, the detector performance in a scenario with no background processes is only marginally affected by $m$---because no other process is polluting the cache. In scenarios with background processes, the impact of $m$ on the F1 score is more prominent since those processes may be polluting the monitored lines and may be causing false positives.
	
	Indeed, $m=16$ resulted in the highest number of false positives for most of the configurations tested (cf. Table~\ref{tab:initial_results2}). On the other hand, performance difference between $m=9$ and $m=12$, depends on the process running in the background. For instance, when $w=256$, as shown in Tables~\ref{tab:initial_results1} and~\ref{tab:initial_results2}, the number of false positives is generally higher for $m=12$ and the number of false negatives is generally higher for $m=9$. Since we consider false negatives more damaging than false positives, we opt for $m=12$.
	
	\vspace{0.5 em}\noindent\textbf{Impact of malicious noise on detection.} As discussed in Section~\ref{sec:security}, a malicious OS might artificially add noise to the channel. We have tested such scenario with the following experiment. We run \our{} (with the threshold-based classifier, $w=64$ and $m=12$) while increasing both the number of lines in the channel polluted by the OS, as well as the frequency with which the OS pollutes those lines. The number of lines polluted by the OS ranged from 1 to 2048 (i.e., from one to all lines of the cache sets monitored by \our); the OS injected noise in intervals of 0, 25, 50, and 100 $\mu$s. Figure~\ref{fig:noisefigure1} shows the impact of such strategy on the false positive rate. Our results show that if the adversary can pollute more than 768 cache lines, \our{} always results in a false positive. Conversely, when the adversary pollutes less than 192 cache lines, the resulting FPR is very low. On the other hand, our experiments show that this strategy does not impact the false negative rate of \our{} (it consistently remains between 0 and 0.009).
	
	\vspace{0.5 em}\noindent\textbf{Performance overhead for WolfSSL.} We use applications of WolfSSL~\cite{Wolfssl}---a suite of cryptographic applications ported to SGX---as exemplary applications to assess the overhead of \our. For each application in the WolfSSL benchmark, we run the vanilla version as baseline and compare its throughput with the one of the same application when enhanced with \our.
	
	Figure~\ref{fig_performance1} (a) depicts the performance penalty incurred for each application in WolfSSL, normalized with respect to the baseline. Each data point is averaged over 100 independent runs. Here, ``clock'' refers to the performance of the application instrumented with \our{} but with only the counting thread running, whereas ``m=12'' and ``m=16'' refer to the performance of the application when both the counting and main threads of \our{} are running. The mean performance penalty across all applications of the WolfSSL benchmark is 2.58 $\pm$ 0.17 \% if just the counting thread is running, and 4.82 $\pm$ 0.91\% and 4.88 $\pm$ 0.90\% if the countermeasure is running with 12 and 16 monitored ways, respectively. We conclude that parameter $m$ has little effect on the overhead and that the performance penalty due \our{} can be tolerated by most applications.
	
	\vspace{0.5 em}\noindent\textbf{Evaluating \our{} when used with BI-SGX.} We now evaluate the performance penalty incurred by BI-SGX~\cite{Bisgx} when \our{} is used to detect attacks. Figure~\ref{fig_performance1} (b) shows the penalty for the two main functions of BI-SGX, namely \texttt{seal\_data} and \texttt{run\_interpreter} (Figure~\ref{fig_bisgx_func_pa}).
	We measure the time for each function to execute with input data comprised of 5000 characters. The performance penalty is normalized with respect to the baseline (i.e., BI-SGX without \our) and we report the average over 100 runs. The mean performance penalty was measured to be 1.99 $\pm$ 2.15\% if just the counting thread is running, and 4.24 $\pm$ 4.39\% and 4.30 $\pm$ 4.33\% if \our{} is running and monitoring 12 or 16 ways, respectively. In Figure~\ref{fig:securing_bisgx}, we asses the performance of \our{} in detecting clones of BI-SGX. We use the threshold-based detection algorithms for $w\in[1,1024]$ and for $m \in\{9,12,16\}$, both in the ideal scenario with no background processes and in a realistic scenario where a background process (\texttt{x265 video encoder}) runs in the background. We collect samples for 10,000 executions of BI-SGX running in a benign setting and while carrying out the attack described in Section~\ref{sec:bisgx}, respectively. Figure~\ref{fig:securing_bisgx} shows that even with background noise, the F1 score reaches 0.999 for $w \geq 64$, with a false positive rate of 0.0015 and a false negative rate of 0.0004.
	
	\begin{figure*}[t]
		\centering
		\begin{subfigure}[b]{0.77\textwidth}
			\centering
			\includegraphics[trim=0cm 0.25cm 0cm 0cm, clip,width=\textwidth]{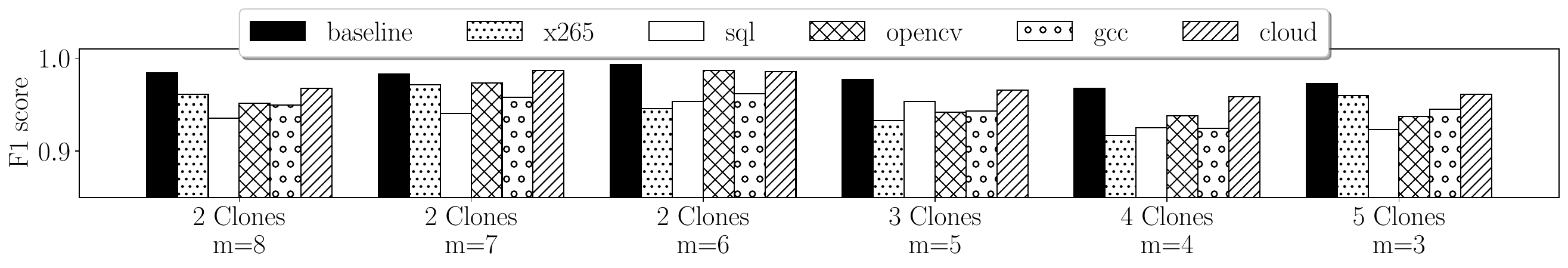}
			\caption{$w$=16 for the case of 2 clones and $w$=32 for the remaining cases}
			\label{fig:multiple_clones_w16}
		\end{subfigure}
		\hfill
		\begin{subfigure}[b]{0.77\textwidth}
			\centering
			\includegraphics[trim=0cm 0.25cm 0cm 0cm, clip,width=\textwidth]{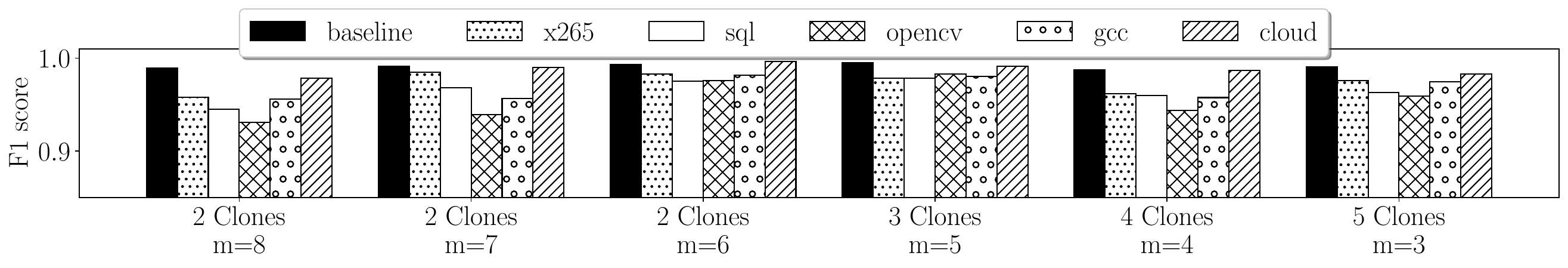}
			\caption{ $w$=256}
			\label{fig:multiple_clones_w256}
		\end{subfigure}
		\vspace{-0.5em}
		\caption{F1 scores of the threshold-based detection algorithm for the detection of  $N+1>2$ clones.
			\label{fig:multiple_clones}}
		\vspace{-1em}
	\end{figure*}
	
	\noindent\textbf{Detection of multiple clones.} We now focus on assessing the effectiveness of \our{} in detecting malicious clones when $N\geq 2$ legitimate instances are allowed to run and an alarm should be raised as soon as $N+1$ instances are running. That is, for each value of $N\in\{2,3,4,5\}$ we run either $N$ or $N+1$ instances and assess whether \our{} can distinguish the two scenarios.
	
	Figure~\ref{fig:multiple_clones} shows the result for this set of experiments. The gray-colored bars show the baseline performance, i.e., when no other application is running on the machine, apart from the $N$ enclave instances (and possibly an additional clone); colored bars depict the results when other processes are running in the background. Note that $m$ is chosen according to Table~\ref{table_nenclaves} and $w$ is set to either $16$, $32$, or $256$.
	
	In Figure~\ref{fig:multiple_clones_w16}, we set $w=16$ for $N=2$ and $w=32$ for $N\in\{3,4,5\}$ because, for $N\in\{3,4,5\}$, we witnessed larger increase of the F1 score when doubling the size of the observation window from $w=16$ to $w=32$. Concretely, for $N\in\{3,4,5\}$, increasing $w$ to $32$ improves the F1 scores by 0.0661, 0.0751, and 0.0718, respectively. Figure~\ref{fig:multiple_clones_w16} shows that the F1 score does not fall below 0.9168 in any of the considered scenarios. When $N=2$, the F1 score ranges between 0.983 and 0.993 with no other applications running in the background; in case of background noise, the F1 score varies between 0.946 and 0.987. Similarly, we measured F1 values of 0.977, 0.967, 0.973 for $N\in\{3,4,5\}$ and observed that it falls to 0.917 in the worst case for  $N=4$ while running \texttt{x265 video encoder} in the background.
	
	Figure~\ref{fig:multiple_clones_w256} shows results with a larger window ($w=256$). As discussed above, increasing $w$ impacts detection latency; however, the performance gain justifies the choice of a larger observation window in most cases.
	When no applications are running in the background, the F1 score ranges between 0.989 and 0.993 for $N=2$ and reaches 0.995, 0.988, and 0.991 for $N\in\{3,4,5\}$, respectively. In scenarios where applications are running in the background, F1 drops as low as 0.931 for $N=2$ and $m=8$ when the clones are running along \texttt{OpenCV}.
	
	Figure \ref{fig:multiple_clones} shows results up to $N=5$, since the number of enclaves using \our{} on a given host at a given time is bounded by the numbers of available cores. In particular, this limitation comes from the registers used by the counting thread. Each core has only one such register, and it is shared by its two hyper-threads. As a consequence, as soon as the number of running enclaves using \our{} is greater than the number of physical cores, multiple context switches are required. The result is an increase in the number of asynchronous exits or a counter that does not increase between consecutive reads. Both cases should be labeled as an anomaly and appropriate countermeasure should be taken.

	\section{Concluding Remarks}
	\label{conclusion}

	In this work, we addressed the problem of forking attacks against Intel SGX by cloning the victim enclave. We analyzed 72 SGX-based applications and found that roughly 20\% are vulnerable to such attacks, including those that rely on monotonic counters to prevent forking attacks based on rollbacks. A comprehensive solution to forking attacks requires a trusted third party that, unfortunately, are hard to find in real-world deployments.
	
	To address this problem, we introduced \our, the first practical clone detection mechanism for SGX enclaves that does not rely on a trusted third party. We analyzed the security of \our{} and showed that a malicious OS cannot bypass it to spawn clones without detection. We implemented \our{} and evaluated its performance in existing SGX applications and under various realistic workloads. Our evaluation results show that \our{} achieves high accuracy in detecting clones, only incurs a marginal performance overheard, and adds up to 800 LoC to the TCB.
	
	\section*{Acknowledgments}
	This work is partly funded by the Deutsche Forschungsgemeinschaft
	(DFG, German Research Foundation) under Germany’s
	Excellence Strategy - EXC 2092 CASA - 390781972,
	and the
	European Union’s Horizon 2020 research and innovation programme (SPATIAL, Grant Agreement No. 101021808). Views
	and opinions expressed are however those of the authors
	only and do not necessarily reflect those of the European
	Union. Neither the European Union nor the granting
	authority can be held responsible for them. 
	
	\bibliographystyle{ACM-Reference-Format}
	\bibliography{references}
	
	\appendix
	
	\section*{Appendix}
	\renewcommand\floatpagefraction{0.1}
	\thispagestyle{empty}

	\section{Additional Background}
	
	\begin{figure}[t]
		\centering
		\includegraphics[width=0.45\textwidth]{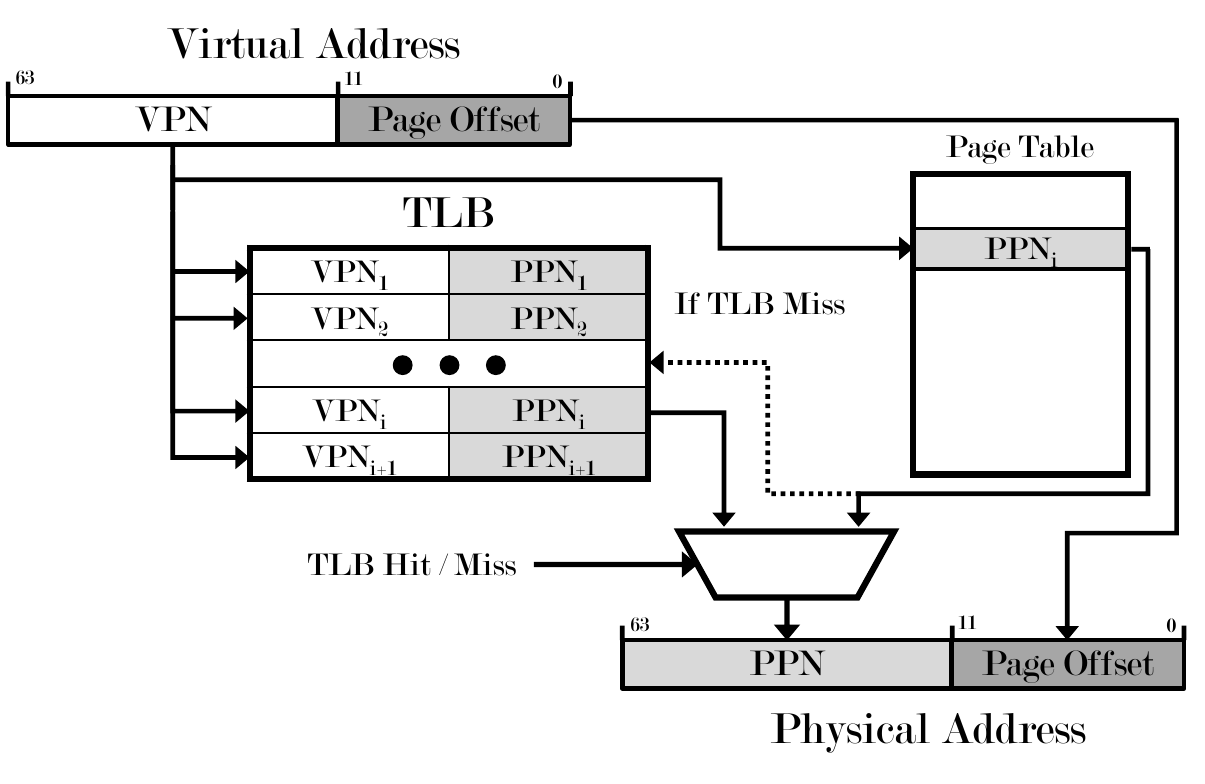}
		\caption{Virtual to physical address translation. If a TLB miss occurs, a page walk is triggered and the translation is retrieved from the page table.}
		\vspace{-1.5 em}
		\label{fig_trans}
	\end{figure}
	
	\subsection{Cache-based covert channels}\label{sec:memsystem}
	
	Covert channels refer to communication channels that were not intended nor designed to transmit information. In particular, hardware-based covert channels constitute an effective means to establish communication between two processes that share the hardware~\cite{7056069}. Various micro-architectural elements can be used as a covert channel, e.g. the TLBs~\cite{tlbSecrets}, the memory bus~\cite{180210} or the DRAM~\cite{Drama2016} but, among all of them, cache memories have been the most widely studied~\cite{lastLevelPractical,SgxPectre19,flushreload14,cross2015,7163049}. In what follows, we briefly discuss the operation of virtual memory and caches, before providing details on cache-based covert channels as they are relevant to \our.
	
	\vspace{0.5 em}\noindent\textbf{Virtual Memory and Cache Hierarchy.} The virtual memory system is one main constituent of the memory system. In a nutshell, the virtual address space of each process is allocated in pages of typically 4KB of size, and each is stored physically in the DRAM by the OS. The OS determines the mapping between virtual and physical pages by managing a structure called ``page table''. Moreover, the hardware caches the mapping information of the page table in the so-called Translation Lookaside Buffer (TLB), so subsequent translations of the same virtual address are faster. Figure~\ref{fig_trans} shows how a virtual address is translated to a physical one by means of the page table. A virtual address is made of a Virtual Page Number (VPN) and a page offset (12-bits long if 4KB pages are used). The VPN is used to address the page table and obtain the corresponding Physical Page Number (PPN); the latter is concatenated with the page offset to obtain the physical address corresponding to the virtual address.
	
	Cache memories are several orders of magnitude faster than the main memory and speed up memory access by keeping copies of recently used data. In Intel SGX, caches are shared between enclaves and the untrusted software. Modern processors include hierarchically organized caches: L1 and L2 caches are private to each core whereas L3 cache (also known as last level cache or LLC) is shared among all the cores. Traditionally, Intel has built processors with inclusive last-level caches: data in the core-private low-level caches is also present in the shared L3 cache. The size and the latency of the caches increase as their distance to the processor grows. Indeed, by measuring the time it takes to load a piece of data, it is possible to determine whether it was located in the cache hierarchy (cache hit) or if it has been fetched from main memory (cache miss).
	
\begin{figure}[t]
		\centering
		\includegraphics[width=0.45\textwidth]{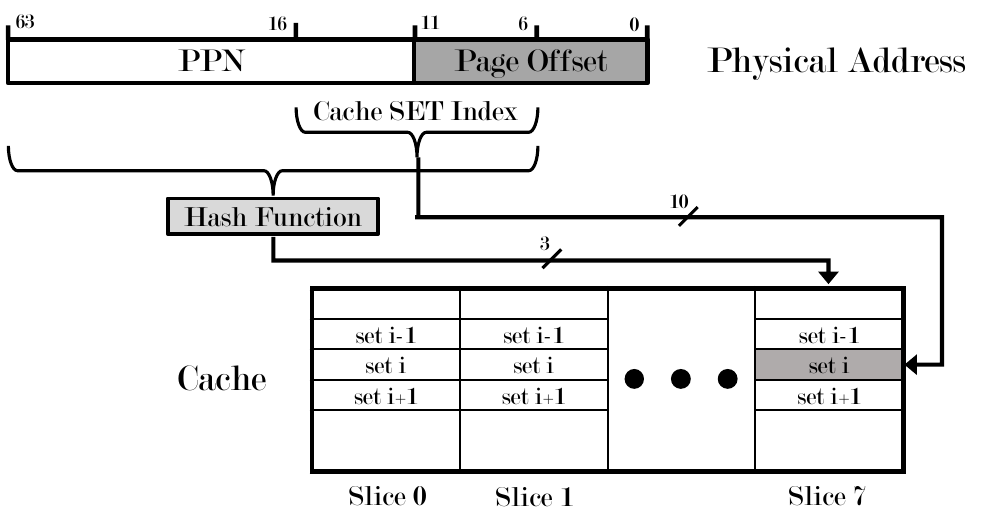}
		\caption{Representation of the cache set and slice addressing function for a cache with 8 slices and 1024 sets per slice.
		}
		\label{fig_trans2}
		\vspace{-1.5 em}
	\end{figure}

	Intel processors include $W$-way set-associative caches. This type of cache is organized into multiple sets ($S$), each containing $W$ lines of usually 64 bytes of data. Additionally, sets can be grouped into slices. Figure~\ref{fig_trans2} shows how the hardware determines the cache location of a memory block from its physical address. The least significant bits (line offset) locate the data within the cache line, i.e., 6 bits for a 64 byte line; the following $\log_2(S)$ bits (cache set index) select the set; the slice number is determined by computing a hash function on some bits of the physical address; the remaining bits (tag) determine whether the data is cached or not. The newest generations of Intel processors include slices of 1024 sets, therefore requiring 10 bits for addressing a cache set. That is, bits 0-5 are used for the cache line, and bits 6-15 for the cache set index. If we consider 4KB pages, the application only knows 6 of the bits that determine the cache set number (bits 6-11), whereas the remaining bits (12-15) are defined by the physical address that, in turn, is chosen by the OS.
	
	The order in which the elements of a cache set are removed from the cache and replaced with new data is determined by insertion/replacement policies~\cite{Briongos2019,Abel19,vila2019cachequery}. In particular, in the case of the L3 cache, the Quad-Age Replacement Policy assigns ages ranging from 0 to 3 to the cache-lines which are updated upon accesses to the L3 cache; in the event of a cache miss the first block whose age is equal to 3 is replaced.
	
	\vspace{0.5 em}\noindent\textbf{Cache-based Covert Channels.} Cache-based covert channels exploit the noticeable timing difference between cache hits (the data is fetched from the cache) and cache misses (data has to be retrieved from main memory). In a nutshell, assuming a sender and a receiver, the receiver first sets the cache into a known state (e.g., by forcing some data into the cache), then the sender would either remove that data (e.g., by either flushing the cache or forcing some new data into the cache so to ensure that the original data from the receiver is removed from the cache) to send a ``1'', or do nothing to send a ``0''. Finally, the receiver checks (e.g., by probing the cache) whether the cache has changed from its original state.
	
	Using the L3 cache as a covert channel requires building eviction sets, which are groups of memory addresses mapping to a particular set and slice~\cite{vila2018theory}. These are the main building blocks for Prime+Probe attacks~\cite{lastLevelPractical,DanielCachezoom,MalwareSGX2018} or for achieving fast evictions required to carry out rowhammer attacks~\cite{gruss2016rowhammer}. For regular applications, large memory pages of 2MBs ease the creation of eviction sets. Unfortunately, large pages are not available for Intel SGX enclaves. However, Islam et al.~\cite{Spoiler2019} have shown how to leverage the processor load and store operations to tell whether two addresses share up to 20 bits; this technique can be used to build eviction sets even in case large pages are not available. 
	
	\subsection{DRAM}
	
	We now provide additional background information on the DRAM and how it can be used to infer information on the physical memory allocated to an enclave~\cite{MalwareSGX2018,malware_extended}. This information will be particularly useful for Appendix~\ref{app:minSets}.
	
	The DRAM, sometimes referred to as main memory, is organized hierarchically: the memory arrays arranged in rows and columns form banks which, in turn, form bank groups in the case of DDR4.  A set of banks composes a rank (typically 8 banks on DDR3 and 16 banks on DDR4). Each memory cell is uniquely addressed with the information referring to its channel, DIMM, rank, bank, row and column. The CPU maps physical memory addresses to cells by means of an (officially) undocumented function. Previous work has reverse-engineered the function~\cite{Drama2016} and showed that two alternating accesses to the DRAM mapping to the same bank (i.e., they have the same value of BA0, BA1, BG0 and BG1, rank, and channel) but that have different row indexes (row conflict) would take considerably longer than if they had the same row index (row hit). By timing memory accesses, a process can infer relations among some bits of its physical address space.
	
\begin{figure}[t!]
		\centering
		\includegraphics[trim=0cm 0cm 0cm 0cm, clip, width=0.45\textwidth]{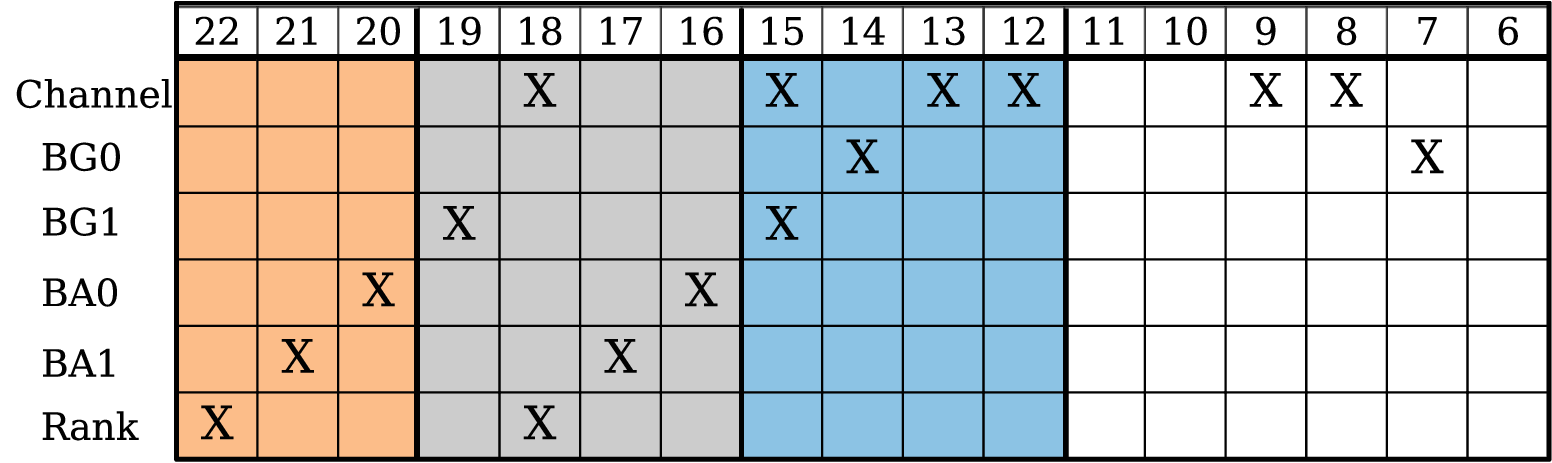}
		\caption{DRAM mapping function of our test machine (Xeon E-2176G). Each output bit value (row) is computed XOR-ing the bits of the physical address (columns) marked with an $X$.}
		\label{fig_dram_function}
	\end{figure}

	We reverse-engineered the mapping function for our Xeon E-2176G (2x32GB DDR4 RAM) as shown in~\cite{Drama2016}. The function is depicted in Figure~\ref{fig_dram_function}. In a nutshell, given a physical address as $a_{n-1},a_{n-2},...,a_{0}$, our memory controller computes its channel as $a_{18} \oplus a_{15} \oplus a_{13} \oplus a_{12} \oplus a_{9} \oplus a_{8}$; the bank values BG0, BG1, BA0, and BA1 as $a_{19} \oplus a_{15}$, $a_{20} \oplus a_{16}$, $a_{21} \oplus a_{17}$, and $a_{22} \oplus a_{18}$, respectively; the rank as $a_{22} \oplus a_{18}$. Finally, the row index uses bits 18 and above from the physical address (assuming a typical row size of 8KB).
	
	Note that Figure~\ref{fig_dram_function} does not show bits 0 to 5 since they are used neither to address the DRAM nor the cache sets. We use different background colors to distinguish between regions: the region between bit 6 and bit 11 represents the bits of the cache set index that are controlled by the enclave; the region between bit 12 and bit 15 (blue) corresponds to the remaining bits of the cache set index under control of the OS. Additionally, the region between bits 16 and 19 (gray) shows the bits used by Spoiler~\cite{Spoiler2019} that are not part of the cache set index; finally, the region that goes from bit 20 to bit 22 are the bits exclusively used by the DRAM mapping function.

	\begin{algorithm}[t!]
		\caption{Encoding of the conditions for the SAT Solver}
		\footnotesize
		\begin{algorithmic}[1]
			\Require Set of conditions on arr[N][B]
			\Ensure \textit{\textbf{Mapping that fulfills the conditions}}
			\State{\textit{// Setup phase (assign known values)}}
			\State c1=(), c2=(),c3=(), c4=(), c5=() \Comment{Initialize conditions}
			\For{$i=0$ \textbf{to} N}
			\State arr[i][0:11] = i mod $2^{12}$ (4096)
			\EndFor
			\State {\textit{// Condition 1}}
			\For{$i=0$ \textbf{to} N}
			\For{$j=i$ \textbf{to} N}
			\State c1 \&= (at least one bit of arr[i][:] different from arr[j][:])
			\EndFor
			\EndFor
			\State {\textit{// Condition 2}}
			\For{$i=0$ \textbf{to} $2^{20}$}
			\For{$j=i+256$ \textbf{to} N; $j+=2^{20}$;}
			\State c3 \&= (arr[i][0:19] equal to arr[j][0:19])
			\EndFor
			\EndFor
			\State {\textit{// Condition 3}}
			\For{$i=0$ \textbf{to} $2^{16}$}
			\For{$j=i+16$ \textbf{to} N; $j+=2^{16}$;}
			\State c3 \&= (arr[i][0:15] equal to arr[j][0:15])
			\EndFor
			\EndFor
			\State {\textit{// Condition 4}}
			\For{$i=0$ \textbf{to} $2^{6}$} \Comment{i $\equiv$ \{channel,BG0,BG1,BA0,BA1,Rank\}}
			\State $list \gets ()$
			\For{$j=0$ \textbf{to} N}
			\State $channel_j = j_{18} \oplus j_{15} \oplus j_{13} \oplus j_{12} \oplus j_{9} \oplus a_{8}$;
			\State $BG0_j = j_{19} \oplus j_{15}$
			\State $BG1_j = j_{20} \oplus j_{16}$
			\State $BA0_j = j_{21} \oplus j_{17}$
			\State $BA1_j = j_{22} \oplus j_{18}$
			\State $Rank_j = j_{22} \oplus j_{18}$;
			\If{i == \{$channel_j,BG0_j,BG1_j,BA0_j,BA1_j,$\\$Rank$\}}
			\State Add j to $list$
			\EndIf
			\EndFor
			\ForAll{k in list}
			\ForAll{l in list such  l!= k}
			\If{$k_{18:B}!=l_{18:B}$}\Comment{Row bits}
			\State c4 \&= arr[k][:] conflicts with arr[l][:]
			\Else
			\State c4 \&= arr[k][:] \Comment{No conflict}
			\EndIf
			\EndFor
			\EndFor
			\EndFor
			\State {\textit{// Condition 5}}
			\State c5 \&= arr[0][:] has row conflict with arr[N-1][:])
			\For{$i=0$ \textbf{to} N-1}
			\If i mod $2^{22}$ != $(\mathtt{0x7FFFFF})$
			\State c5 \&= arr[i][:] has no conflict with arr[i+1][:])
			\Else
			\State c5 \&= arr[i][:] conflicts with arr[i+1][:])
			\EndIf
			\EndFor
			\State \textbf{Expression to satisfy = c1 \& c2 \& c3 \& c4 \& c5}
		\end{algorithmic}
		\label{alg:satsolver}
	\end{algorithm}
	
	\section{Linear memory mapping and physical addresses}\label{app:minSets}
	
	To build eviction sets for \our{}, we opted not to rely on the technique proposed in~\cite{MalwareSGX2018,malware_extended} due to its reliance on an OS that assigns linear memory to enclaves.
	We now provide more details on this choice. In particular, we show that a malicious OS can assign to an enclave a non-linear memory region that ``looks'' linear from the enclave perspective. As a consequence, if eviction sets are built by assuming linear memory, a malicious OS can deceive two enclave clones to monitor different cache sets---so that they would not detect each other.
	
	Recall that~\cite{MalwareSGX2018} builds evictions sets by relying on the assumption that the physical memory assigned to an enclave is linear. In particular, if the memory is linear, given a virtual address $VA_i$ and its corresponding physical address $PA_i$, the physical address of another virtual address $VA_j$ is computed as $PA_j = PA_i + (VA_j - VA_i)$. As a consequence, given any two addresses $VA_i,VA_j$ such that $VA_i[15-0]=VA_j[15-0]$, then $PA_i[15-0]=PA_j[15-0]$ and the two addresses share the same cache set. In this setting, it is straightforward to find an eviction set for a specific cache set, if the value of $PA_i[15-0]$ is known. In case the assigned memory is linear, the enclave can find out bits 15-0 of a physical address $PA_i$ corresponding to a virtual address $VA_i$ by exploiting the DRAM mapping function. In particular, a process witnesses a DRAM row conflict between two consecutive virtual addresses $VA_i$ and $VA_i+1$ only if their physical addresses have their last 22 bits equal to all ones and all zeros, respectively~\cite{MalwareSGX2018,malware_extended}. Thus, a process can cycle through its virtual memory to find two virtual addresses $VA_i$ and $VA_i+1$ such that their corresponding physical addresses have 
	$PA_i[15-0]=\mathtt{0xFFFF}$ and $PA_{i+1}[15-0]=\mathtt{0x0000}$, respectively. Once those physical addresses have been found, either can be used as a starting address to build an eviction set.
	
	In our scenario, the OS is however considered malicious and has an obvious advantage in assigning non-linear memory to the victim enclave. In particular, assigning non-linear memory to two enclave clones may deceive them to monitor different cache sets---and therefore not detect each other---as follows. Given two enclave instances $E1$ and $E2$ the OS assigns the physical memory of $E1$ such that, when $E1$ performs the above DRAM test, it finds two consecutive physical addresses such that $PA_i[15-0]=\mathtt{0xFFFF}$ and $PA_{i+1}[15-0]=\mathtt{0x0000}$. Assume $E1$ uses $PA_{i+1}$ as the starting point to build an eviction set so that the monitored cache set has cache set number $0$ (because $PA_i[15-6]$ is all zeros). Now the OS ``swaps'' the two addresses for $E2$ so that, when running the above test, $E2$ finds $PA_i[15-0]=\mathtt{0x0000}$ and $PA_{i+1}[15-0]=\mathtt{0xFFFF}$. Thus, if $E2$ uses $PA_{i+1}$ as the starting point to build an eviction set, it ends up monitoring cache set number $1023$  (because $PA_i[15-6]$ is all ones) and the two enclaves will not detect each other.
	
	Since we cannot trust the OS to assign contiguous linear memory to enclaves, building eviction sets using~\cite{MalwareSGX2018} requires checking that the memory assigned to the enclave is indeed linear. To do so the enclave could leverage known techniques to infer information of the physical memory corresponding to its virtual memory region. In particular, (i) speculative loads~\cite{Spoiler2019} allow the enclave to check whether two addresses share the least significant 20 bits, (ii) cache memory allows the enclave to verify if groups of addresses share the least significant 16 bits, and (iii) the DRAM provides information about the bits of the physical addresses by measuring row hits and conflicts---as explained earlier.
	
	Given the above techniques, we define a set of conditions that a linear memory arrangement must meet:
	
	\begin{description}
		\item[Condition 1] All physical address, to which virtual addresses are mapped, must be different.
		\item[Condition 2] Given two virtual addresses sharing the last 20 bits, we must observe speculative load hazards~\cite{Spoiler2019}.
		\item[Condition 3] Given two virtual addresses sharing bits 6-15, they must be assigned to the same cache set.
		\item[Condition 4] Given any virtual address $VA_i$ for which we assume its physical address is known, if we calculate its corresponding channel, BG0, BG1, BA0, BA1, Rank and row value we must observe row conflicts with any other $VA_j$ whose physical address (obtained assuming there is linear memory) that maps to the same channel, BG0, BG1, BA0, BA1 and Rank while having a different row value. Analogously, we must observe row hits if the row value is the same.
		\item[Condition 5] For any $VA_i$ and $VA_j$ such that $VA_j=VA_i+1$, we should observe a row conflict if and only if $PA_i$ has its 22 least significant bits equal to 1 and
		consequently $PA_j$ has its 22 least significant bits equal to 0 (e.g., $VA_i=\mathtt{0x3FFFFF}$ and $VA_j=\mathtt{0x400000}$)
	\end{description}
	
	Ideally, if an enclave checked the conditions above and if a linear memory arrangement were the only arrangement to fulfill such conditions, we could build eviction sets as in~\cite{MalwareSGX2018}, while being absolutely sure about the value of the bits 12-15.
	
	To verify the existence of alternative (i.e., non-linear) memory assignments, we formulate this problem as a Boolean satisfiability problem (SAT).
	We encode each bit of the physical addresses of a memory range as a condition and then use a SAT solver to figure out the possible values of such physical addresses that satisfy these conditions. Concretely, we used SATisPy~\cite{satispy}, which is a wrapper of the popular MiniSAT \cite{minisat}. For reproducibility purposes, the encoding of the problem is given in Algorithm~\ref{alg:satsolver}. SATisPy provided multiple non-linear memory arrangements that satisfied all of the conditions. We conclude that building eviction sets by using the technique of~\cite{MalwareSGX2018} may not be suitable for \our, as the OS may provide a page alignment that would help it to place two clone enclaves on two different channels and evade detection. For this reason, \our{} builds eviction sets by using a technique derived from ~\cite{Spoiler2019}. Also, because of a malicious OS, \our{} may not be able to figure out bits 12-15 of one of its physical addresses and, therefore, cannot be sure of its cache set index. For this reason, \our{} uses the 16 possible cache sets as a covert channel---fixing bits 6-11 of the addresses of the eviction sets---and monitors all of them to make sure that the OS cannot evade detection.

	The code in Algorithm~\ref{alg:satsolver} uses an array $arr[N][B]$ that mimics the region of the virtual memory we want to study in the sense that it encodes N different addresses of B bits. The actual content of each location of the array represents the physical address, whereas the location itself represents the virtual address. In particular, we used a memory region that comprises 8MB because the DRAM function uses up to the 23rd bit. The conditions are coded as if $arr[0][:]$ was going to have its content equal to 0, i.e., they assume that the physical address of $arr[0][:]$ is $\mathtt{0x000000}$.

	\section{Additional experiments}\label{app:noise}
	
	\begin{table*}[h]
		\centering
		\setlength\tabcolsep{1pt}
		\scalebox{0.99}{\begin{tabular}{cc|c|c|c|c|c|c|c|c|c|c|c|c|c|c|c|c|c|c|}
				\cline{3-20}
				&&\multicolumn{6}{c|}{\textbf{Logistic regression}}&\multicolumn{6}{c|}{\textbf{Linear Discriminant Analysis}}&\multicolumn{6}{c|}{\textbf{KNN}} \\ \cline{3-20}
				&& w=1&w=4&w=16&w=64&w=256&w=1024&w=1&w=4&w=16&w=64&w=256&w=1024&w=1&w=4&w=16&w=64&w=256&w=1024  \\ \hline
				\multicolumn{1}{|c|}{\multirow{3}{*}{Baseline}}&m=9&0.884&0.936&0.973&0.992 &0.991& 0.986
				&0.884&0.936&0.957&0.979 & 0.983&0.993
				& 0.885&0.925&0.961& 0.954& 0.973&0.989
				\\ \cline{2-20}
				\multicolumn{1}{|c|}{}&m=12 &0.906&0.965&0.996&0.977&0.983&0.990
				&0.906&0.966&0.985& 0.986& 0.991& 0.993
				&0.907&0.968&0.992& 0.993& 0.992& 0.993
				\\ \cline{2-20}
				\multicolumn{1}{|c|}{} &m=16 &0.829&0.894&0.967& 0.977&0.984 &0.985
				&0.829&0.894&0.957&0.977 &0.977 &0.978
				&0.746&0.885&0.965& 0.977&0.980 &0.986
				\\ \hline
				\multicolumn{1}{|c|}{\multirow{3}{*}{x265}}&m=9&0.801&0.870&0.912& 0.922& 0.951&0.961
				&0.801&0.870&0.888& 0.900&0.952 &0.957
				&0.719&0.878&0.925& 0.931& 0.954&0.959
				\\ \cline{2-20}
				\multicolumn{1}{|c|}{} &m=12 &0.907&0.959&0.972& 0.973 &0.974 &0.983
				&0.907&0.953&0.971&0.971 & 0.971&0.972
				& 0.363&0.957&0.976&0.978 & 0.979&0.984
				\\ \cline{2-20}
				\multicolumn{1}{|c|}{} &m=16 &0.757&0.801&0.829& 0.901&0.952 &0.959
				&0.757&0.801&0.827&0.897 & 0.912&0.937
				& 0.227&0.780&0.849&0.892 & 0.914&0.939
				\\ \hline
				\multicolumn{1}{|c|}{\multirow{3}{*}{sql}} &m=9&0.894&0.919&0.938&0.944 &0.973 &0.989
				&0.894&0.908&0.914&0.936 &0.959 &0.971
				&0.893&0.923&0.940& 0.941&0.944 &0.939
				\\ \cline{2-20}
				\multicolumn{1}{|c|}{} &m=12&0.945&0.971&0.978&0.984 & 0.990&0.994
				&0.945&0.969&0.978& 0.980& 0.975&0.987
				&0.945&0.973&0.978& 0.983&0.987 &0.987
				\\ \cline{2-20}
				\multicolumn{1}{|c|}{} &m=16&0.812&0.841&0.849& 0.875& 0.932&0.980
				& 0.812&0.839&0.848&0.870 & 0.924& 0.978
				& 0.321&0.849&0.853&0.886 &0.905 &0.934
				\\ \hline
				\multicolumn{1}{|c|}{\multirow{3}{*}{opencv}}&m=9&0.856&0.888&0.925&0.935 &0.947 &0.955
				& 0.855&0.889&0.924& 0.945&0.956 &0.965
				& 0.855&0.888&0.934& 0.946 &0.964 &0.967
				\\ \cline{2-20}
				\multicolumn{1}{|c|}{}&m=12&0.903&0.930&0.937&0.956 &0.963 &0.974
				&0.903&0.930&0.936& 0.945& 0.955&0.963
				& 0.903&0.930&0.938& 0.949& 0.952&0.961
				\\ \cline{2-20}
				\multicolumn{1}{|c|}{}&m=16&0.670&0.775&0.876&0.900 & 0.911&0.917
				& 0.670&0.775&0.868&0.905 & 0.910&0.915
				& 0.670&0.775&0.921& 0.925&0.923 &0.911
				\\ \hline
				\multicolumn{1}{|c|}{\multirow{3}{*}{gcc}} &m=9&0.850&0.909&0.933& 0.938& 0.960&0.979
				& 0.850&0.909&0.917& 0.937& 0.939&0.946
				& 0.848&0.907&0.938& 0.954& 0.961&0.962
				\\ \cline{2-20}
				\multicolumn{1}{|c|}{} &m=12&0.931&0.973&0.980& 0.982& 0.985&0.987
				&0.931&0.967&0.980&0.983 & 0.987&0.989
				& 0.931&0.970&0.979&0.979 &0.982 &0.978
				\\ \cline{2-20}
				\multicolumn{1}{|c|}{} &m=16&0.809&0.838&0.873& 0.887&0.923 &0.933
				& 0.809&0.839&0.850&0.864 &0.906 &0.923
				& 0.559&0.835&0.870&0.900 &0.933 &0.943
				\\ \hline
				\multicolumn{1}{|c|}{\multirow{3}{*}{cloud}}   &m=9&0.933&0.967&0.979& 0.987& 0.990&0.990
				& 0.933&0.955&0.955& 0.959& 0.963&0.971
				& 0.932&0.970&0.974& 0.985&0.986 &0.988
				\\ \cline{2-20}
				\multicolumn{1}{|c|}{} &m=12&0.906&0.941&0.953& 0.964& 0.987 & 0.990
				&0.906&0.929&0.929& 0.9423& 0.954&0.973
				& 0.901&0.942&0.944&0.959 & 0.983&0.989
				\\ \cline{2-20}
				\multicolumn{1}{|c|}{} &m=16&0.856&0.916&0.932&0.939 & 0.941&0.949
				& 0.856&0.908&0.932& 0.946& 0.957&0.972
				& 0.256&0.918&0.934& 0.943&0.951 &0.966
				\\ \hline
				\cline{3-20}
				&&\multicolumn{6}{c|}{\textbf{Decision Tree}}&\multicolumn{6}{c|}{\textbf{Random Forest}}&\multicolumn{6}{c|}{\textbf{SVM}} \\ \cline{3-20}
				&& w=1&w=4&w=16&w=64&w=256&w=1024&w=1&w=4&w=16&w=64&w=256&w=1024&w=1&w=4&w=16&w=64&w=256&w=1024  \\ \hline
				\multicolumn{1}{|c|}{\multirow{3}{*}{Baseline}}&m=9 &0.884&0.936&0.982&0.989 & 0.988&0.990
				& 0.884&0.936&0.963& 0.973& 0.979&0.982
				&0.886&0.931&0.977&979 &0.980 &0.981
				\\ \cline{2-20}
				\multicolumn{1}{|c|}{}&m=12 &0.906&0.970&0.998&0.992&0.993 &0.991
				&0.906&0.970&0.995& 0.996& 0.994&0.993
				&0.907&0.969&0.993& 0.993& 0.993&0.993
				\\ \cline{2-20}
				\multicolumn{1}{|c|}{} &m=16 &0.829&0.895&0.980&0.981 &0.989 &0.992
				&0.829&0.895&0.956&0.978 & 0.985&0.991
				&0.830&0.901&0.963& 0.979&0.978 &0.986
				\\ \hline
				\multicolumn{1}{|c|}{\multirow{3}{*}{x265}}&m=9&0.801&0.895&0.948&0.966&0.970&0.973
				& 0.801&0.895&0.946& 0.957& 0.963&0.973
				& 0.798&0.894&0.952& 0.966 & 0.974&0.980
				\\ \cline{2-20}
				\multicolumn{1}{|c|}{} &m=12 &0.907&0.960&0.971& 0.971&0.954 & 0.982
				&0.907&0.960&0.972& 0.978& 0.983&0.992
				& 0.906&0.959&0.972&0.970 &0.973 &0.980
				\\ \cline{2-20}
				\multicolumn{1}{|c|}{} &m=16 &0.757&0.792&0.848& 0.912&0.944 &0.951
				& 0.757&0.792&0.841& 0.916& 0.952&0.969
				& 0.755&0.784&0.845& 0.920 & 0.931&0.920
				\\ \hline
				\multicolumn{1}{|c|}{\multirow{3}{*}{sql}} &m=9&0.894&0.923&0.964& 0.969 & 0.972&0.972
				&0.894&0.923&0.951&0.968 & 0.954&0.962
				& 0.893&0.924&0.961& 0.983& 0.990&0.992
				\\ \cline{2-20}
				\multicolumn{1}{|c|}{}&m=12&0.945&0.973&0.977&0.978 &0.980 &0.981
				& 0.945&0.973&0.978& 0.980&0.981 &0.983
				& 0.945&0.971&0.974& 0.968&0.978 &0.979
				\\ \cline{2-20}
				\multicolumn{1}{|c|}{} &m=16&0.812&0.846&0.859&0.899 &0.927 &0.952
				&0.812&0.846&0.853& 0.938& 0.941&0.944
				&0.811&0.846&0.854& 0.875&0.955 &0.978
				\\ \hline
				\multicolumn{1}{|c|}{\multirow{3}{*}{opencv}}&m=9 &0.855&0.888&0.933& 0.959&0.960 &0.963
				&0.855&0.888&0.918& 0.931 &0.960 &0.967
				&0.855&0.889&0.931&0.943&0.935 &0.939
				\\ \cline{2-20}
				\multicolumn{1}{|c|}{} &m=12&0.903&0.930&0.934& 0.943&0.955 &0.969
				& 0.903&0.930&0.937& 0.954&0.959 &0.973
				& 0.903&0.928&0.939&0.960 &0.967 &0.969
				\\ \cline{2-20}
				\multicolumn{1}{|c|}{} &m=16 &0.670&0.775&0.957& 0.947& 0.971&0.970
				& 0.670&0.775&0.868& 0.911& 0.927&0.963
				& 0.667&0.774&0.890&0.917 &0.963&0.971
				\\ \hline
				\multicolumn{1}{|c|}{\multirow{3}{*}{gcc}} &m=9 &0.850&0.914&0.951&0.958 & 0.965&0.973
				& 0.850&0.914&0.952&0.962 &0.965 &0.969
				& 0.849&0.915&0.957& 0.962& 0.962&0.973
				\\ \cline{2-20}
				\multicolumn{1}{|c|}{} &m=12&0.931&0.973&0.975& 0.981&0.982 &0.983
				& 0.931&0.973&0.979&0.980 & 0.982&0.982
				& 0.931&0.971&0.979& 0.981& 0.986&0.986
				\\ \cline{2-20}
				\multicolumn{1}{|c|}{} &m=16&0.809&0.839&0.896&0.917 &0.926 &0.937
				& 0.809&0.839&0.869& 0.920& 0.946&0.972
				& 0.806&0.841&0.890&0.915 & 0.915&0.920
				\\ \hline
				\multicolumn{1}{|c|}{\multirow{3}{*}{cloud}} &m=9&0.933&0.967&0.984& 0.986& 0.989&0.990
				&0.933&0.967&0.977& 0.986& 0.987&0.987
				& 0.933&0.966&0.984& 0.986& 0.989&0.992
				\\ \cline{2-20}
				\multicolumn{1}{|c|}{} &m=12&0.906&0.941&0.960&0.982 & 0.983&0.983
				&0.906&0.941&0.952& 0.972& 0.980&0.9832
				& 0.907&0.943&0.957& 0.977& 0.993&0.996
				\\ \cline{2-20}
				\multicolumn{1}{|c|}{} &m=16&0.856&0.921&0.944& 0.951 & 0.966&0.972
				&0.856&0.921&0.934& 0.941&0.948 &0.967
				& 0.857&0.920&0.935& 0.966& 0.987 & 0.989
				\\ \hline
		\end{tabular}}
		\caption{F1 score of various detection algorithms for different values of $w$ and $m$. Baseline refers to the scenario where no background applications are running, whereas the others refer to different applications running in the background simultaneously with the clone detector.}
		\label{tab:complete_data}
		\vspace{-1 em}
	\end{table*} 
	
	In Table~\ref{tab:complete_data}, we report for completeness, the evaluation results when \our{} is equipped with other detection algorithms; Table~\ref{tab:complete_data} also shows the performance of \our{} when different applications of the Phoronix benchmark suite~\cite{phoronix} are running in the background. Our results reveal a performance degradation when $m=16$---that is, when monitoring all of the lines in the 16 cache sets chosen
	as the covert channel---and an application is running in background. Finally, we note that there is no clear choice between $m=9$ and $m=12$.

	\section{Alternative Architectures}\label{sec:alternative}
	
	\noindent\textbf{Other TEEs.} \our{} has been mostly designed for Intel SGX. Nevertheless, it also finds direct applicability in a number of other TEE instantiations. In what follows, we discuss how \our{} can be adapted to other TEE architectures.
	
	Recall that the main requirement for \our{} to work is the existence of a resource shared among clones that can be used as a covert channel. We argue that if a platform is vulnerable to covert-channels, this vulnerability could be turned into a feature to detect clones. Previous work has shown a number of micro-architectural features on different platforms that could be used to setup side- or covert-channels (see ~\cite{DBLP:journals/jce/GeYCH18} for a survey). In the following, we provide more details on ARM TrustZone and AMD SEV---arguably the most popular TEEs along with Intel SGX.
	
	ARM TrustZone~\cite{10.1145/3291047} is the TEE instantiation proposed by ARM and distinguishes between two protection domains dubbed secure world (SW) and normal world (NW). Previous work has shown how normal world software can instruct the OS in the trusted world to launch multiple instances of the same process~\cite{KwonSCLP20,DBLP:conf/dais/AmacherS19}. Hence, applications running in the secure world, could leverage a clone-detection mechanism like \our, by exploiting one of the available covert-channel~\cite{9152801,armageddon16,10.1145/3274694.3274704}. Notice, however, that details of ARM TrustZone vary across implementations and the design of a particular anti-cloning mechanism for ARM TrustZone should take into account the specific feature of the underlying ARM platform.
	
	Similarly, AMD Secure Encrypted Virtualization (SEV)~\cite{AMDsev} aims to protect virtual machines from untrusted cloud providers or hypervisors. 
	In a nutshell, SEV includes an encryption engine that automatically encrypts and decrypts data in main memory to ensure that it is not readable by other software on the same platform. According to its documentation, SEV does not offer any mechanism to prevent a malicious hypervisor from launching several clones of a victim VM. Further, unless the VM owner sets its VM as non-migratable (by setting the \texttt{NOSEND} bit in guest policy), the hypervisor can clone enclaves across platforms, thereby making the problem of detecting clones even tougher to solve. Assuming that migration is disabled---thus the adversary can only clone the victim VM on the same platform--- a VM could detect its clones on the same machine by using a covert channel as shown by \our. We note AMD processors are vulnerable to side- and covert-channels~\cite{10.1145/3320269.3384746}, and also the official AMD documentation acknowledges that SEV VMs are vulnerable to Prime+Probe attacks~\cite{AMDwhitepaper}. We also note that the way AMD SEV partition caches among guests---guests sharing the same Address Space Identifier (ASID) are assigned to the same cache sets---may facilitate the creation and improve the throughput of the covert channel between clones; this is because two different VMs (with different ASIDs) will not use the same cache sets, hence they will not pollute each-other's cache.
	
	\vspace{0.5 em}\noindent\textbf{Non-inclusive caches.} \our{} requires the enclave to build eviction sets on the host, despite a potentially malicious OS. We have shown how to do so on a CPU architecture with inclusive caches, since this is the cache architecture available on all SGX-enabled CPUs to date. Nevertheless, Intel has recently announced that SGX will also be available on next-generation Xeon processors~\cite{intelproducts}, which most likely will feature non-inclusive caches and increase the size of the EPC. In order to use \our{} on such processors, we could leverage techniques to build eviction sets on architectures with non-inclusive caches as proposed by Yan et al.~\cite{8835325}. In particular, one could build eviction sets by targeting cache directories rather than the cache itself.
	
	\vspace{0.5 em}\noindent\textbf{Cache Allocation Technology.} Cache Allocation Technology (CAT)~\cite{intelcat} allows the OS or any privileged software to define multiple partitions in the LLC. In a nutshell, CAT introduces an intermediate abstraction, so-called Class of Service (CLOS), into which applications can be grouped. The CLOS is in turn associated with a capacity bitmask (CBMs), which indicates the fraction of the LLC that can actually be used by the given CLOS~\cite{7446102}.
	
	If future SGX-enabled processors were to support CAT\footnote{Notice that CAT is not available on SGX-enabled processors~\cite{intelrdtcat}.}, \our{} could easily detect if it is only assigned to one partition of the cache since, in this case, it will fail to build eviction sets. In this case, the application may take appropriate countermeasures, e.g., refuse to execute. In principle, a malicious OS could try to use CAT dynamically: it could let \our{} build eviction sets while accessing the whole cache, and later on restrict each clone to a different cache partition. We speculate that such a strategy would cause false positives (i.e., each enclave would signal the presence of clones) as the enclave will evict its own data when trying to fill the monitored sets. This is due to the fact that some of the data will not fit in the newly assigned sets or slices (physical addresses do not change and we monitor all the slices and expand through multiple sets).

	\newcommand{\catA}{{FIm}}
	\newcommand{\catB}{{ForKVS}}
	\newcommand{\catC}{{BUG}}
	
	\section{Cloning Attacks}\label{sec:attacks}
	
	We have examined 72 application based on Intel SGX with respect to forking attacks (see Section~\ref{sec:motivation}). Among the examined applications, 14 were susceptible to cloning-based forking attacks. In what follows, we describe in details how to mount cloning attacks on these applications. We observe that cloning-based attacks can be grouped into three broad categories. To ease the presentation, we describe each of the attack categories and, for each category, we  show how the attack can mounted against an exemplary applications (chosen from the 14 vulnerable applications).
	
	\subsection{\catA ~-- Forking In-memory Key-value Stores}
	We start by describing cloning attacks on in-memory key-value stores (KVS), dubbed \catA.
	A KVS exposes \texttt{PUT} and \texttt{GET} interfaces to clients. At any time, there is at most one value per key; further, when a \texttt{GET} request is issued for a given key $k$, the latest value that was written to the database for that key is returned.
	A forking attack against a KVS may break these security guarantees.
	
	In-memory KVSs that use Intel SGX, e.g., Aria \cite{aria}, Enclage \cite{enclage}, STANLite \cite{stanlite_paper}, ObliDB \cite{oblidb_paper}, and Avocado \cite{avocado_paper}, are designed to store data larger than EPC memory (limited to 128 MB). Hence, they seal data to persistent memory but keep meta-data in runtime memory to ensure integrity and rollback protection. Since meta-data is not sealed to persistent storage, it is lost if the enclave terminates.
	
	\subsubsection{Attack overview}
	
\begin{figure} [!t]
		\centering 
		\includegraphics[width=0.41\textwidth]{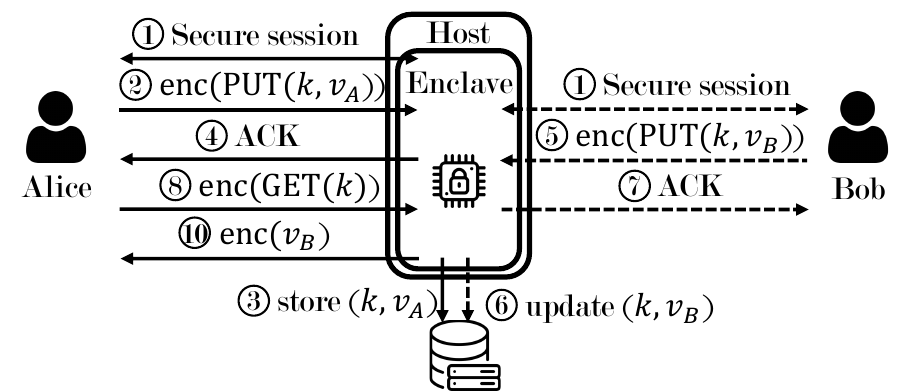}
		\caption{Overview of an SGX-backed in-memory key-value store if it operates in a benign setting.}
		\label{fig:fim_1}
	\end{figure}

	In benign settings, assume a server running an enclave-backed KVS that two clients, $A$ and $B$, can access as depicted in Figure \ref{fig:fim_1}.
	First, both clients attest the enclave and establish a session key to encrypt their messages.
	All subsequent messages are encrypted using the session key.
	In a benign setting, $A$ sends a \texttt{PUT} request to post the key-value (KV) pair $(k, v_A)$ to the storage.
	The enclave recognizes that the key $k$ does not exist, creates a new entry with the pair $(k, v_A)$, and returns an $ACK$ message.
	Afterward, $B$ sends a \texttt{PUT} request to post the KV pair $(k, v_B)$.
	At this time, the enclave recognizes that the key $k$ exists and updates the value to $v_B$ before returning an $ACK$ message.
	If $A$ later requests the value associated with $k$ from the KVS, it receives the value $v_B$.
	
\begin{figure} [!t]
		\centering
		\includegraphics[width=0.42\textwidth]{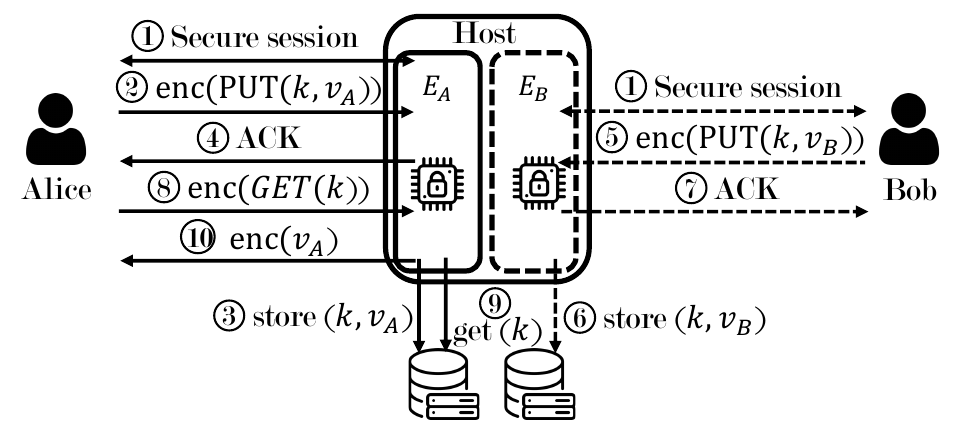}
		\caption{Overview of a generic \catA ~attack on an SGX-backed in-memory key-value store.}
		\label{fig:fim_2}
	\end{figure}

	In an adversarial setting (cf. Figure \ref{fig:fim_2}), the adversary can provide two different KVS instances to $A$ and $B$.
	Even if clients attest the enclave where the KV instance is running, they cannot tell whether they are communicating with the same instance or not.
	
	The adversary launches two enclave instances, $E_A$ and $E_B$, and connects each client to one instance.
	Each client attests the connected enclave.
	Assume the same sequence of requests as in the previous setting.
	First, $A$ sends a \texttt{PUT} request to post the KV pair $(k, v_A)$ to the storage.
	$E_A$ recognizes that $k$ does not exist in the associated storage, creates a new entry with the pair $(k, v_A)$, and returns an $ACK$ message.
	Afterward, client $B$ sends a \texttt{PUT} request to post the KV pair $(k, v_B)$.
	$E_B$ does not find an entry for $k$ in its copy of the storage and creates a new entry for the pair $(k, v_B)$.
	Both clients receive an $ACK$ reporting the correct execution of their request.
	However, if client $A$ later requests the value associated with $k$, $E_A$ returns $v_A$ which is the latest value it has seen---this is however different from the newest value in the system.
	
	We note that the above issue can be fixed if clients are mutually trusted and each client shares its view of the KVS with the others~\cite{brandenburger17dsn}. Essentially, the set of client acts as a distributed TTP. Unfortunately, assuming a mutually trusted set of clients limits the scenarios where the KVS can be used.
	
	\subsubsection{Concrete example: \catA ~Attack against Aria}
	
	As an example, we now show how to mount a \catA ~attack against \textbf{Aria} \cite{aria}.
	Aria provides an in-memory KVS in the cloud.
	Each entry is protected against rollback attacks by means of  MC.
	For confidentiality, the entries are encrypted with AES (in CTR mode), where the counter value is set to be the current MC value of the entry.
	The enclave generates a pseudo-random key at initialization and uses the same key for encrypting all data.
	Additionally, each entry contains a MAC over the encrypted data for integrity protection.
	The integrity of the MCs is guaranteed by a Merkle tree structure over all MCs.
	The enclave exclusively stores the Merkle root in its runtime memory.
	Additionally, it stores all recently used MCs in its local cache.
	The cached counters can be used to decrypt entries directly without verifying the Merkle root, thus reducing latency.
	
\begin{figure}
		\centering
		\includegraphics[width=0.45\textwidth]{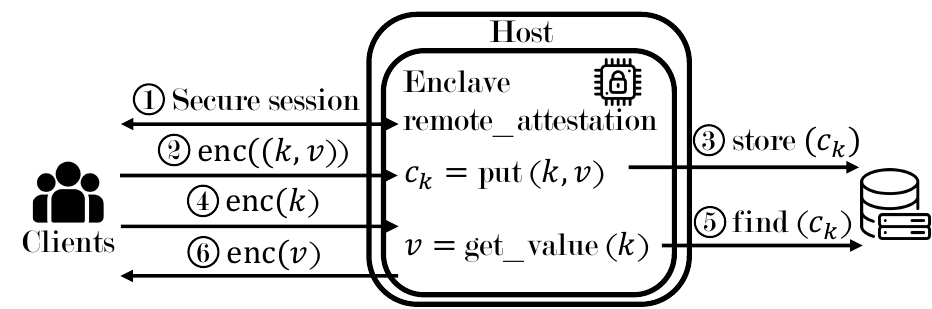}
		\caption{Overview of the main functions exposed by the Aria enclave and its interactions with clients.}
		\label{fig:aria_1}
	\end{figure}

	As depicted in Figure \ref{fig:aria_1}, the client first attests the enclave and establishes a secure session key.
	The client sends a \texttt{PUT} request for $(k, v)$, encrypted with the session key.
	The enclave decrypts the message and checks if $k$ exists in storage.
	If so, it updates the corresponding counter and encrypted KV pair.
	Otherwise, it assigns the key a free counter and stores it in the database.
	Later, the client can access $v$ by sending an encrypted \texttt{GET} request for $k$.
	The enclave verifies the counter integrity and decrypts entries until it finds the requested KV pair.
	Finally, it returns $v$ through the secure channel.
	
	Assume now a malicious host and two clients, $A$ and $B$ who share access to the same KVS, e.g. for customer records.
	
\begin{figure}
		\centering
		\includegraphics[width=0.39\textwidth]{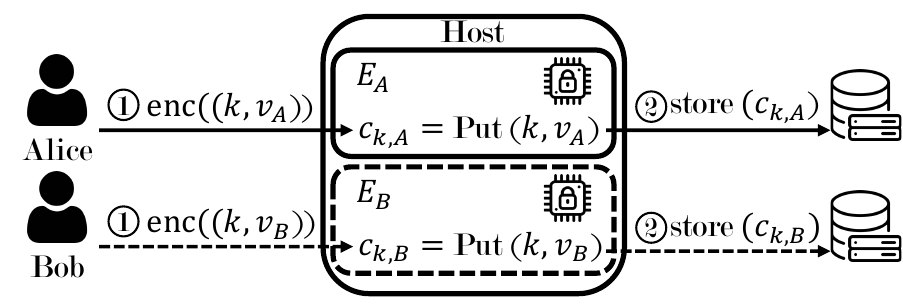}
		\caption{Overview of a cloning attack against Aria enclaves.}
		\label{fig:aria_2}
		\vspace{-1em}
	\end{figure}

	As shown in Figure \ref{fig:aria_2}, one can mount \catA ~attacks on Aria as follows:
	
	\begin{itemize}
		\item The adversary starts two Aria enclave instances, $E_A$ and $E_B$.
		\item The adversary connects $A$ to $E_A$, and $B$ to $E_B$.
		\item The clients attest the enclaves and establish secure communication sessions.
		\item The clients send encrypted \texttt{PUT} requests for $(k, v_A)$ and $(k, v_B)$ to $E_A$ and $E_B$, respectively.
		\item $E_A$ and $E_B$ decrypt the requests and create/update the corresponding encrypted entries in the their storage instances. $A$ and $B$ cannot determine if they are communicating with the same instance.
	\end{itemize}
	
	Hence, \catA ~violates the consistency of Aria by cloning the enclave.
	The adversary is not limited by the number of enclaves and can run arbitrarily many instances.
	
	\subsection{\catB ~-- Forking persistent Key-value Stores}
	
	We now describe cloning attacks on persistent KVSs, dubbed \catB.
	Persistent KVSs that are susceptible to cloning attacks are EnclaveCache \cite{enclavecache}, NeXUS \cite{nexus_paper}, StealthDB \cite{stealthdb_paper}, ShieldStore \cite{shieldstore_paper}, SGX-KMS \cite{sgxkms_paper}, and CACIC \cite{cacic_paper}.
	In contrast to in-memory KVSs, persistent KVS use sealing to make meta-data available across reboots.
	
	\subsubsection{Attack overview}
	
\begin{figure}[t]
		\centering
		\includegraphics[width=0.26\textwidth]{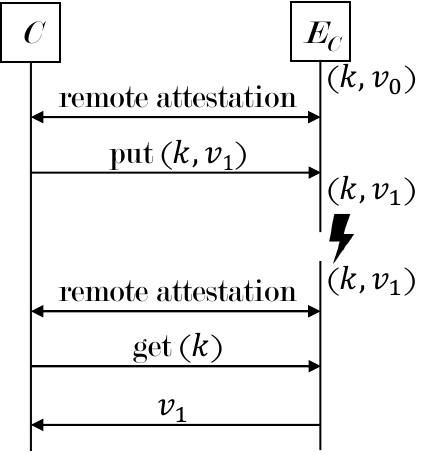}
		\caption{Overview of the interaction of a persistent key-value store with a client in a benign setting.}
		\label{fig:forkvs_1}
	\end{figure}

	Assume a server running an enclave-backed KVS that stores a KV pair $(k, v_0)$ when a client, $C$, connects to the system.
	First, $C$ attests the enclave $E_C$ and establishes a session key.
	All subsequent messages are encrypted using the session key.
	In a benign setting (cf. Figure \ref{fig:forkvs_1}), $C$ sends a \texttt{PUT} request to update the value associated with $k$ to $v_1$.
	$E_C$ updates the KV pair in its storage.
	Afterward, it creates a snapshot, sealing the meta-data with the incremented MC value $ctr_i + 1$ and incrementing the MC value $ctr_{i+1} \leftarrow ctr_{i} + 1$.
	Later, $E$ crashes and needs to restart.
	It successfully verifies the MC value in the sealed data and restores the KVS.
	$C$ must attest the restarted enclave instance and establish new session keys.
	When $C$ requests the value associated with $k$, the KVS correctly returns the latest value, $v_1$.
	
\begin{figure}[t]
		\centering
		\includegraphics[width=0.35\textwidth]{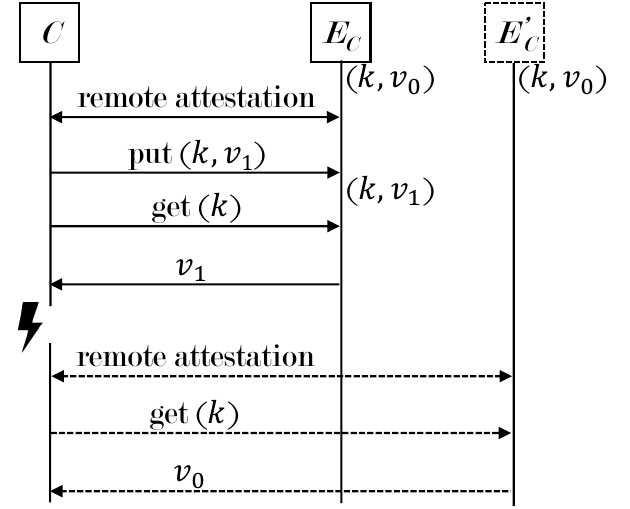}
		\caption{Overview of a generic \catB ~attack against persistent key-value stores.}
		\label{fig:forkvs_2}
	\end{figure}

	In an adversarial setting (cf. Figure \ref{fig:forkvs_2}), the adversary can provide two different views of the same KVS instance to $C$.
	The adversary launches two enclave instances, $E_C$ and $E'_C$.
	Both instances have the same initial state storing the KV pair $(k, v_0)$.
	First, the adversary connects $C$ to enclave $E_C$, and the value is updated to $v_1$.
	If the client requests the value associated to $k$, the enclave correctly provides with $v_1$.
	Afterward, the adversary connects $C$ to the second instance, $E'_C$.
	$C$ assumes the enclave has crashed and successfully attests $E'_C$.
	However, when requesting the value associated with $k$, $E'_C$ returns $v_0$, the latest state it stores.
	Consequently, the same key is associated with different values in different enclave instances.
	The same attack holds if multiple clients use the KVS instead of $C$ connecting to the KVS in different sessions.
	
	Cloning attacks on in-memory KVSs are limited to providing two instances of a KVS.
	They do not share entries unless the same data is provided to both instances in different sessions.
	\catB ~is more powerful: two instances of the KVS share common data that has been sealed by the first instance before the second instance starts.
	Therefore, \catB ~can have the same effect as rollback attacks, even if classical rollback attacks are not possible.
	
	Notice that BI-SGX can be seen as an instantiation of a persistent KVS. We therefore refer the reader to Section \ref{sec:bisgx} as a concrete example of \catB.

	\subsection{\catC ~-- Breaking Unlinkability Guarantees}
	
	We now describe cloning attacks on SGX proxies, dubbed \catC.
	Applications affected by \catC, i.e., X-Search \cite{xsearch_paper} and PrivaTube \cite{privatube}, provide unlinkability by leveraging an SGX-backed proxy.
	The proxy receives encrypted requests and obfuscates them, e.g., by adding fake requests, to ensure that an adversary accessing the service cannot link the plaintext requests to individual clients.
	
	\subsubsection{Attack overview}
	
	By cloning the enclave, an adversary can break the unlinkability and link a request to a specific user or at least reduce the anonymity set.
	We now describe the generic cloning attack for breaking unlinkability guarantees of SGX-backed proxies, considering an honest setting first.
	
\begin{figure}
		\centering
		\includegraphics[width=0.42\textwidth]{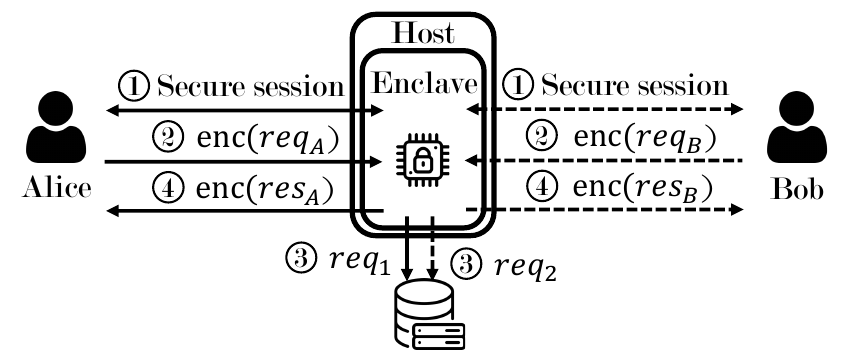}
		\caption{Overview of an SGX-backed proxy in a benign setting.}
		\label{fig:bug_1}
	\end{figure}

	Assume a server running an enclave-backed proxy that receives requests from two clients, $A$ and $B$, as depicted in Figure \ref{fig:bug_1}.
	First, both clients attest the enclave and establish session keys.
	In a benign setting, clients send requests $req_A$ and $req_B$, encrypted with the session key.
	The enclave decrypts the requests and forwards two (decrypted) requests, $req_1$ and $req_2$, without knowing the identity of the issuer.
	Afterward, the proxy maps the responses to the client requests, encrypts and forwards them to the corresponding client.
	The server cannot distinguish if $A$ send $req_1$ or $req_2$.
	The anonymity set increases with the number of clients simultaneously connected to the enclave.
	
\begin{figure}
		\centering
		\includegraphics[width=0.42\textwidth]{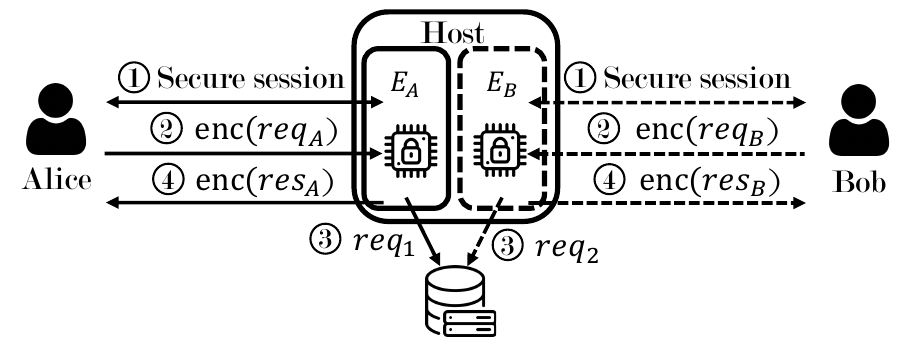}
		\caption{Overview of a generic \catC ~attack against an SGX-backed proxy.}
		\label{fig:bug_2}
	\end{figure}

	In an adversarial setting (cf. Figure \ref{fig:bug_2}), the adversary can recover the assignment and break the unlinkability guarantee.
	The adversary starts two proxy enclaves, $E_A$ and $E_B$, and connects clients $A$ and $B$ to one instance each.
	The clients attest the connected enclave and send the encrypted requests.
	The adversary observes which enclave forwards the request to the server, e.g., $E_A$ sends the request $req_1$.
	The adversary connected $A$ to $E_A$, thus can infer that $A$ sent $req_1$.
	Linking the decrypted requests to the clients, the \catC ~attack breaks the unlinkability guarantee.
	
	\subsubsection{Concrete example: \catC ~Attack against PrivaTube proxies}
	
	As an example, we now show how to mount a \catC ~attack against \textbf{PrivaTube} \cite{privatube}.
	PrivaTube is a distributed Video on Demand system leveraging fake requests and SGX enclaves to ensure the unlinkability of requests to individual users.
	Requests for video segments can be served by video servers and assisting platforms.
	Assisting platforms are other users that requested a specific video segment in the past and can provide other users with this segment.
	Each peer in the system hosts an enclave, an HTTP proxy, to break the link between clients and requests.
	
\begin{figure}[t]
		\centering
		\includegraphics[width=0.50\textwidth]{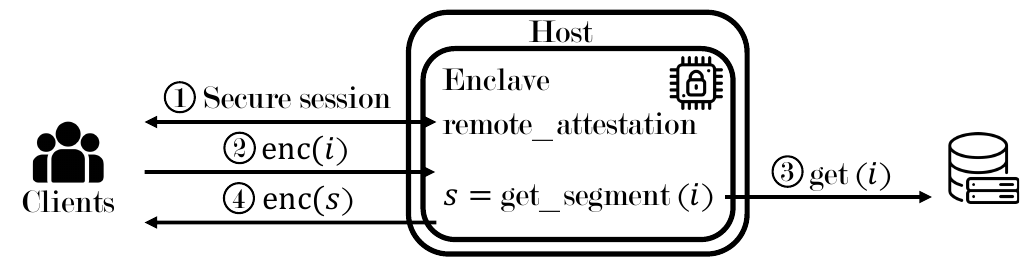}
		\caption{Overview of the functions exposed by a PrivaTube proxy and its interaction with clients.}
		\label{fig:privatube_1}
	\end{figure}

	As shown in Figure \ref{fig:privatube_1}, a client attests the proxy enclave and sends an encrypted request for a video segment with the ID $i$.
	The enclave decrypts the segment ID and requests the video segment from the peer's video database.
	It encrypts the received segment $\texttt{s}$ and sends it to the client.
	PrivaTube assumes that each video server serves multiple requests simultaneously, thus preventing the precise assignment of users to requested video segments.
	
\begin{figure}[t]
		\centering
		\includegraphics[width=0.50\textwidth]{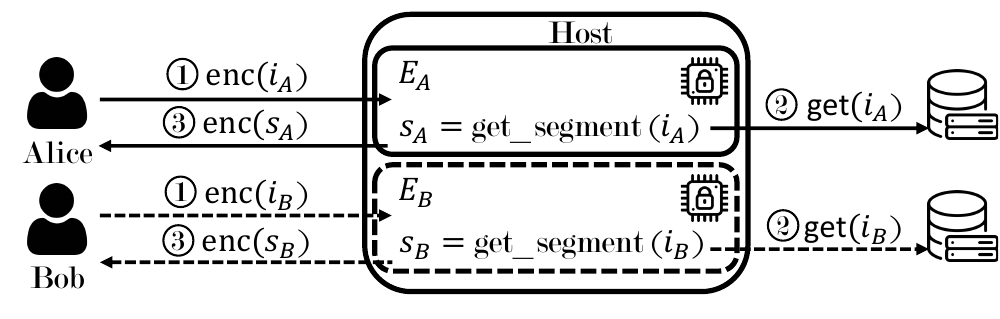}
		\caption{Overview of a cloning attack against a PrivaTube proxy.}
		\label{fig:privatube_2}
	\end{figure}

	Assume now a malicious video server and two users, $A$ and $B$.
	As shown in Figure \ref{fig:privatube_2}, one can mount a \catC ~attack on PrivaTube proxies as follows:
	
	\begin{itemize}
		\item The adversary starts two proxy enclave instances, $E_A$ and $E_B$.
		\item The adversary connects $A$ to $E_A$, and $B$ to $E_B$.
		\item The clients attest the enclaves and establish secure communication sessions.
		\item The clients send encrypted requests for $i_A$ and $i_B$ to $E_A$ and $E_B$, respectively.
		\item The adversary observes the decrypted requests for $i_A$ and $i_B$ to the database, issued by $E_A$ and $E_B$, respectively.
		\item Knowing $A$ is connected to $E_A$ and $B$ is connected to $E_B$, the adversary can recover that $A$ requested the video segment $i_A$ and $B$ requested $i_B$.
		Both requests are served correctly.
		Further, $A$ and $B$ cannot determine that they are connected to different proxies.
	\end{itemize}
	
	The adversary is not limited by the amount of enclaves it can execute at the same time.
	For every client requesting the video server, the adversary can start a new enclave, precisely recovering the assignment of requested video segments to clients.
	Here, the unlinkability guarantee is broken.
	
	Notice that \catC{} attacks can be mounted against other proposals (not present in the lists we have analyzed) that use Intel SGX to cloak client requests. For example, Prochlo~\cite{prochlo} and subsequent work (e.g.,~\cite{orshuffle}) provide real-world privacy-preserving analytic frameworks to collect anonymous statistics. They use Intel SGX to shuffle client inputs so to break the link between a data item and the identity of the client where data originates. In a nutshell, an external observer cannot tell the client that sent a given input, among a set of $k$ clients. We note that, by cloning the shuffler enclave, an adversary can split the set of clients in disjoint subsets, so that two clients of two different subsets send their inputs to two different instances of the shuffler enclave. As a result, each client input is shuffled with less than $k$ data items and matching an input to its client becomes an easier task.

\end{document}